\documentclass[preprint,prd,nofootinbib,tightenlines,groupedaddress,superscriptaddress,amsmath,amssymb]{revtex4}

\usepackage{graphicx}
\usepackage[hypertex]{hyperref}
\usepackage{color}

\def\sla#1{\rlap/#1}

\begin{document}
\title{One loop radiative correction to Kaluza-Klein masses  \\
in $S^2/Z_2$ universal extra dimensional model}
\preprint{OCU-PHYS 397, STUPP-14-216}
\pacs{
}
\author{Nobuhito Maru}
\email{nmaru@sci.osaka-cu.ac.jp}
\affiliation{Department of Mathematics and Physics, Osaka City University, Osaka, 558-8585, Japan \vspace{2.5mm}} 
\author{Takaaki Nomura}
\email{nomura@mail.ncku.edu.tw}
\affiliation{Department of Physics, National Cheng-Kung University, Tainan 701, Taiwan \vspace{2.5mm}} 
\author{Joe Sato}
\email{joe@phy.saitama-u.ac.jp}
\affiliation{Department of Physics, Saitama University, Shimo-Okubo, Sakura-ku, Saitama 355-8570, Japan \vspace{2.5mm}} 

\date{\today}

\begin{abstract}
We investigate a radiative correction to the masses of Kaluza-Klein(KK) modes in a universal extra dimensional model which are defined on a six-dimensional spacetime
with extra space as a two-sphere orbifold $S^2/Z_2$.
We first define the Feynman rules which are necessary for the calculation.
%
We then calculate the one-loop diagrams which contribute to the radiative corrections to the KK masses, and obtain one-loop corrections to masses for fermions, gauge bosons and scalar bosons.
We estimate the one-loop corrections to KK masses for the first KK modes of standard model particles as a function of momentum cut-off scale, 
and we determine the lightest KK particle which would be a promising candidate of a dark matter.  
\end{abstract}
\maketitle

\section{Introduction} 

The Standard Model(SM) has been well established. It has indeed passed test of the accelerator experiments.  It
is, however, not a satisfactory theory for all the physicists.  There
seems to be several flaws, e.g., the hierarchy
problems, no candidate of dark matter, and so on.
With the fact that relic abundance of the dark matter is well explained by
weakly interacting massive particle, 
these problems strongly indicate
a new physics beyond the SM at TeV scale. 

There are many candidates of such models, say, models with
supersymmetry, little higgs, extra dimensions, and so on. 
Since the Large Hadron Collider experiment is now operating, 
which will explore the physics at TeV scale, 
it is urgent to investigate possible models at that scale.

Among these, the idea of Universal Extra Dimensional(UED)
model is very interesting \cite{Appelquist:2000nn,Antoniadis:1990ew}.
The minimal version of UED has recently been studied very extensively.
It is a model with one extra dimension defined on an orbifold $S^1/Z_2$.
This orbifold is given by identifying the extra spatial coordinate $y$
with $-y$ and hence there are fixed points $y=0,\pi$.  By this
identification chiral fermions are obtained.  It is shown that this
model is free from the current experimental constraints if the scale of
extra dimension $1/R$, which is the inverse of the compactification
radius $R$, is larger than 400 GeV \cite{Appelquist:2000nn,
Agashe:2001ra}.  The dark matter can be explained by the first or second
Kaluza-Klein (KK) mode \cite{Cheng:2002ej}, which is often the first KK
photon, and this model can be discriminated from other models
\cite{Datta:2005zs}. This model can also give plausible explanations for
SM neutrino masses which are embedded in extended models
\cite{Matsumoto:2006bf}.

In contrast with the UED models in five dimensions,
UED models in six dimensions have interesting properties which would explain some problems in the SM.
For example, in six dimensions, the
number of generations of quarks and leptons is derived by anomaly
cancellations \cite{Dobrescu:2001ae} and the proton stability is
guaranteed by a discrete symmetry of a subgroup of six dimensional Lorentz symmetry
\cite{Appelquist:2001mj}.  Candidate of UED models
in six dimensional model are the one with extra dimensions of $T^2$, 
a torus\cite{Appelquist:2000nn},
or the one with $S^2$, a sphere\cite{Maru:2009wu,Dohi:2010vc}.
In these two classes of models, the latter is quite new and
its phenomenology has been studied recently in Refs.~\cite{Maru:2009cu, Nishiwaki:2011vi, Nishiwaki:2011gk, Kakuda:2013kba}. 

Although we do not have experimental hints of existence of extra dimensions presently, it is meaningful to explore non-minimal UED models.
There are many kind of UED models and we should know the properties for different models to test them in experiments.
For example, we can explore preferred KK mass scale by estimating relic density of dark matter for each UED models using KK mass spectrum and couplings, 
and it will suggest what values of KK masses we can expect  to discover at collider experiments. 
Thus it is important to investigate properties of each UED models such as mass spectrum and couplings
to examine the validity of the UED in experiments.
%
%
Furthermore, $S^2$ has the structure of the simplest coset space SU(2)/U(1) and it has been applied in Coset Space Unification scenario~\cite{Kapetanakis:1992hf, Nomura:2008sx} to construct unified models.
Thus it is worth investigating the radiative corrections to mass spectrum of the KK modes even apart from UED models.

To study it in detail, first of all we have to calculate
the quantum correction to mass spectrum.
It is well known that at tree level all the particles in the same
Kaluza-Klein (KK) mode are degenerate in mass and therefore
it is impossible to predict even a decay mode.
It is hence inevitable to calculate the quantum correction to their mass
to find its phenominological consequences, say, collider 
signatures. This small correction relative
to their tree mass is crucial since mass differences by it
is ``infinitely'' large campared with it at tree level,
and hence it determines physics.
In five-dimensional model, radiative corrections to KK masses and couplings are discussed in \cite{Bauman:2011xf} where
the importance of choice of regulator is taken into account~\cite{Bauman:2007rt}.
For UED model, the aurthors of \cite{Cheng:2002iz} have calculated the radiative corrections in $S^1$ and $T^2$.
Also the running Higgs self couplings and Yukawa couplings are discussed in \cite{Ohlsson:2012hi} for $T^2$ UED and in \cite{Kakuda:2013kba} for $S^2$-based UED models.
However there is no calculation of the corrected KK masses for the $S^2 $ model\cite{Maru:2009wu}.


In models with two sphere, fermions cannot be
massless because of the positive curvature and hence they have a mass of
O($1/R$) \cite{Lichnerowicz:1964zz,A.A.Abrikosov}.  We cannot overcome
the theorem simply by the orbifolding of the extra spaces.  In usual
cases, we have no massless fermion on the curved space with positive
curvature, but we know a mechanism to obtain a massless fermion on that
space by introducing a nontrivial background gauge field
\cite{Horvath:1977st,RandjbarDaemi:1982hi}.
The nontrivial background gauge field can cancel the spin connection
term in the covariant derivative.  As a result, a massless fermion
naturally appears.  Furthermore, we note that the background gauge field
configuration is energetically favorable since the background gauge
kinetic energy lowers a total energy.  In order to realize chiral
fermions, the orbifolding is required, for instance.
Unfortunately it makes the calculation of the quantum correction
very difficult.

In this paper, we study a new type of UED with $S^2/Z_2$ extra dimensions.
We treat this theory as cut-off theory.
We show feynmann rules for it and calculate the quantum correction
as a function of the cut-off. By this we show the mass spectrum
for the theory and offer a basics for studying phenominology
such as LHC physics.


The paper is organized as follows.  In Section \ref{S2UED}, we
recapitulate the model\cite{Maru:2009wu}.
We then specify the Feynman rules for propagators and vertices on the six dimensional spacetime with $S^2/Z_2$ extra space.  
In Section \ref{Sec:OneLoop}, we discuss the one loop calculation for KK mass correction and derive a formulas to estimate corrected KK masses.
In Section \ref{Sec:correctedMasses}, we estimate the corrected first KK masses for each SM particles and determine the lightest KK particle of the model.
Section \ref{summary} is devoted to the summary and discussions. 


\section{$S^2/Z_2$ UED model \label{S2UED}}

In this section, we first review the universal extra dimensions defined on the six dimensional spacetime which has extra space as two-sphere orbifold $S^2/Z_2$~\cite{Maru:2009wu}.
We then define Feynman rules relevant to our calculation.

\subsection{Structure of the model}
The model is defined on the six-dimensional spacetime $M^6$ which has extra dimensional space compactified as two-sphere orbifold $S^2/Z_2$.
The coordinate of $M^6$ is denoted by $X^M = (x^\mu, y^\theta=\theta,y^\phi=\phi)$, where $x^\mu$ and $\{ \theta, \phi \}$ are the $M^4$ coordinates and the $S^2$ spherical coordinates, respectively.
The orbifold is defined by identifying the point $(\theta, \phi)$ with $(\pi-\theta,-\phi)$.

The spacetime index $M$ runs over $\mu \in \{0,1,2,3\}$ and $\alpha \in \{\theta,\phi \}$.
The metric of $M^6$ can be written as 
\begin{equation}
g_{MN} = \begin{pmatrix} \eta_{\mu \nu} & 0 \\ 0 & - g_{\alpha \beta} \end{pmatrix},
\end{equation}
where $\eta_{\mu \nu}={\rm diag}(1,-1,-1,-1)$ and $g_{\alpha \beta} = {\rm diag}(R^2,R^2 \sin^2 \theta)$ are metric of $M^4$ and $S^2/Z_2$ with a radius $R$, respectively, 
with radius of $S^2/Z_2$ as $R$.

We introduce a gauge field $A_M(x,y)=(A_\mu(x,y),A_\alpha(x,y))$, SO(1,5) chiral fermions $\Psi_{\pm}(x,y)$, and complex scalar field $H(x,y)$ as the SM Higgs field. 
The chiral fermion is defined by the action of SO(1,5) chiral operator $\Gamma_7=\gamma_5 \otimes \sigma_3$, where $\sigma_i(i=1,2,3)$ are Pauli matrices and $\gamma_5$ is SO(1,3) chiral operator, such that
\begin{equation}
\Gamma_7 \Psi_{\pm}(x,y) =  \pm \Psi_{\pm}(x,y), 
\end{equation}
so that the chiral projection operator is given by $\Gamma_\pm = (1\pm \Gamma_7)/2$.
The boundary conditions for each field can be defined as 
\begin{align}
\label{BCF}
\Psi^{(\pm \gamma_5)}_\pm(x,\pi-\theta,-\phi) & = \pm \Upsilon_5 \Psi^{(\pm \gamma_5)}_\pm(x,\theta,\phi), \\
\label{BCA}
A_\mu(x,\pi-\theta,-\phi) & = A_\mu(x,\theta,\phi), \\
\label{BCex}
A_\alpha (x,\pi-\theta,-\phi) & = -A_\alpha(x,\theta,\phi), \\
\label{BCH}
H (x,\pi-\theta,-\phi) & = H (x,\theta,\phi), 
\end{align}
where $\Upsilon_5 = \gamma_5 \otimes I_2$ with $I_2$ being $2 \times 2$ identity, requiring the invariance of an action in six-dimensions under the $Z_2$ transformation.
%

\begin{table}[b] \vspace{1ex}
\begin{tabular}{|c||c|c|c|c|c|} \hline
 & $Q(x,y)$ & $U(x,y)$ & $D(x,y)$ & $L(x,y)$ & $E(x,y)$ \\ \hline
(SU(3),SU(2))(U(1)$_Y$,U(1)$_X$) & $(3,2)(\frac{1}{6},\frac{1}{2})$ & $(3,1)(\frac{3}{2},\frac{1}{2})$ & $(3,1)(-\frac{1}{3},\frac{1}{2})$ & $(1,2)(-\frac{1}{2},\frac{1}{2})$ & $(1,1)(-1,\frac{1}{2})$ \\ \hline
SO(1,5) chirality & $-$ & $+$ & $+$ & $-$ & $+$ \\ \hline 
Boundary condition & $-$ & $+$ & $+$ & $-$ & $+$ \\ \hline
\end{tabular} \vspace{-1ex}
\caption{The fermion contents in the model where the representations under gauge symmetry, SO(1,5) chirality, and the boundary conditions are shown.\label{Fermions}} \vspace{-1ex}
\end{table}
%

The gauge symmetry of the model is $G=$SU(3)$\times$SU(2)$\times$U(1)$_Y \times$U(1)$_X$ defined on the six-dimensional spacetime 
with gauge coupling constants $g_{6a}$ in six dimensions where index $a=\{X,1,2,3 \}$ distinguishes the gauge symmetries U(1)$_X$, U(1)$_Y$, SU(2) and SU(3).
The extra U(1)$_X$ is introduced, which is associated with a background gauge field $A^B_\phi$ given by~\cite{Manton:1979kb,RandjbarDaemi:1982hi,background} 
\begin{equation}
\label{BGAphi}
A^B_\phi=\hat{Q}_X \cos \theta, 
\end{equation}
where $\hat{Q}_X$ is U(1)$_X$ charge operator, in order to obtain massless chiral fermions in four dimensions. 
We then assign the U(1)$_X$ charge $\hat{Q}_X=\frac{1}{2}$ to fermions as the simplest case in which massless SM fermions appear in four dimensions.
Fermions in six dimensions are thus introduced as in Table~\ref{Fermions} whose zero modes are corresponding to SM fermions.
Then the action of our model in six dimensions is written as 
\begin{align}
\label{6Daction}
S_{6D} = & \int dx^4 R^2 \sin \theta d \theta d \phi \large[ ( \bar{Q}, \bar{U}, \bar{D}, \bar{L}, \bar{E} ) i \Gamma^M D_M (Q, U, D, L, E)^T \nonumber  \\
& - g^{MN} g^{KL} \sum_a \frac{1}{4 (g_{6a})^2} Tr[F_{a MK} F_{a NL}] \nonumber \\
&- \frac{R^2 }{2 \xi (g_{6a} )^2 } \biggl[ (\partial_\mu A^\mu)^2 + \frac{\xi^2}{R^4 \sin^2 \theta} \left( \partial_\theta (\sin \theta A_\theta) + \frac{1}{\sin \theta} \partial_\phi A_\phi \right) \nonumber \\
& - \frac{2 \xi }{R^2 \sin \theta} (\partial_\mu A^\mu )\left(   \partial_\theta (\sin \theta A_\theta) + \frac{1}{\sin \theta} \partial_\phi A_\phi \right)    \biggr] \nonumber \\
& +\bar{c} \left( \partial^\mu D_\mu + \frac{\xi}{R^2 \sin \theta} \partial_\theta (\sin \theta D_\theta) + \frac{\xi}{R^2 \sin^2 \theta} \partial_\phi (D_\phi) \right) c \nonumber \\
&  +(D^M H)^\dagger (D_M H)- \mu^2 H^\dagger H + \frac{\lambda_6}{4} (H^\dagger H)^2 \nonumber \\
& + [ Y_u Q \bar{U} H^* + Y_d Q \bar{D} H + Y_e L \bar{E} H + {\rm h. c.} ] \large]
\end{align}
where 
the $\Gamma^M$ in the first line in the RHS are gamma matrices in six-dimensions defined as $\Gamma^M = \{\gamma^\mu \otimes I_2, \gamma_5 \otimes i \sigma_1, \gamma_5 \otimes i \sigma_2 \}$ 
with the four-dimensional Gamma matrices $\gamma^\mu$,
the third to the fourth lines are gauge fixing terms with gauge fixing parameter $\xi$, 
the fifth line corresponds to a ghost term for non-Abelian gauge group,
the sixth line denotes a Lagrangian for Higgs field,
and the last line denotes a Yukawa interactions with Yukawa couplings $Y_{u,d,e}$  in six dimensions.
%

We here note that the U(1)$_X$ gauge symmetry should be broken to avoid massless gauge boson. 
In \cite{Maru:2009wu}, we discussed that the U(1)$_X$ symmetry is anomalous and is broken at the quantum level, so that its gauge boson should be heavy as UV cutoff scale. 
However, it is also discussed that if U(1)$_X$ is broken at high scale classical monopole configuration is changed by the mass term of gauge boson, 
and spontaneous compactification mechanism is spoiled~\cite{Dohi:2010vc,Dohi:2014fqa}. 
Thus it would be needed to introduce some mechanism to recover monopole configulation.
As another possibility, we can consider lower breaking scale of U(1)$_X$ than compactification scale if U(1)$_X$ is not anomalous introducing suitable fermion content. 
This case will give a $Z'$ boson which would be relatively light, and it can accommodate with observed data if the gauge coupling of U(1)$_X$ is small enough. 
Also its first KK mode will be heavier than lightest KK mode of SM particles since the zero mode is not massless and quantum correction would be small due to small coupling, 
and it will not be DM candidate consistent with our analysis below.
In this paper, we will not further discuss U(1)$_X$ issue since it is beyond the scope of our discussion.

The KK masses for each particles are obtained from kinetic terms in Eq.~(\ref{6Daction}), 
by expanding fields on six dimensions using KK mode functions. 
The fermions $\Psi_\pm^{(\pm \gamma_5)}$ are expanded in terms of the eigenfunctions of the Dirac operator on $S^2$ which are given as 
\begin{align}
\label{Fmode}
\Psi_{\ell m(\neq 0)}(\theta,\phi) &= 
\begin{pmatrix} \tilde{\alpha}_{\ell m}(z,\phi) \\
\tilde{\beta}_{\ell m}(z,\phi) \end{pmatrix} 
=
\frac{e^{\ell m\phi}}{\sqrt{2 \pi}} 
\begin{pmatrix} C_{\tilde{\alpha}}^{\ell m} (1-z)^{\frac{1}{2}|m|} (1+z)^{\frac{1}{2}|m|} 
P_{\ell-|m|}^{(|m|,|m|)}(z)  \\
C_{\tilde{\beta}}^{\ell m} (1-z)^{\frac{1}{2}|m+1|} (1+z)^{\frac{1}{2}|m-1|} 
P_{\ell -|m|}^{(|m+1|,|m-1|)}(z) \end{pmatrix}, \\
\label{Fmode0}
\Psi_{\ell 0}(\theta,\phi) &= 
\begin{pmatrix} \tilde{\alpha}_{\ell 0}(z) \\
\tilde{\beta}_{\ell -10}(z) \end{pmatrix} 
=
\frac{1}{\sqrt{2 \pi}} 
\begin{pmatrix} C_{\tilde{\alpha}}^{\ell 0} P_{\ell}^{(0,0)}(z)  \\
C_{\tilde{\beta}}^{\ell-10} \sqrt{1-z^2} 
P_{\ell -1}^{(1,1)}(z) \end{pmatrix},
\end{align}
where $P_\ell^{(m,n)}(z)$ is Jacobi polynomial with $z=\cos \theta$, $C_{\tilde{\alpha}(\tilde{\beta})}^{\ell m}$ are the normalization constants
determined by $\int d \Omega \tilde{\alpha}^* \tilde{\alpha} (\tilde{\beta}^* \tilde{\beta})=1$, and 
the indices $\{ \ell, m \}$ corresponds to angular momentum quantum numbers on two-sphere specifying KK modes \cite{A.A.Abrikosov, Maru:2009wu}.
These mode functions satisfy 
\begin{align}
\label{DiracO-S2}
i \hat{D} \Psi_{\ell m} = 
- \frac{1}{R} \Bigl[ \sigma_1 \left( \partial_\theta + \frac{\cos \theta}{2} \right) + \sigma_2 \left( \frac{1}{\sin \theta} \partial_\phi + \hat{Q}_X \cot \theta \right) \Bigl]  \Psi_{\ell m} 
= M_{\ell} \Psi_{\ell m},
\end{align}
where $M_\ell = \frac{\sqrt{\ell(\ell+1)}}{R}$ corresponding to KK mass and the $i \hat{D}$ is Dirac operator on $S^2/Z_2$ with background gauge field in Eq.~(\ref{BGAphi}).
They also satisfy the completeness relation 
\begin{equation}
\label{complete}
\sum_{\ell =0}^{\infty} \sum_{m =- \ell}^{\ell} \Psi_{\ell m} (z, \phi) \Psi_{\ell m}^{\dagger} (z', \phi') = \delta(z-z') \delta(\phi-\phi'). 
\end{equation}
A fermion $\Psi_+^{(\pm \gamma_5)}(x,\theta,\phi)$ on $M^4 \times S^2/Z_2$, satisfying boundary condition Eq.~(\ref{BCF}), are expanded such as 
\begin{align}
\label{Fexpand}
\Psi_+^{(\pm \gamma_5)}(x,\theta,\phi) = \sum_{\ell m} 
\frac{1}{\sqrt{2}} \begin{pmatrix} [\tilde{\alpha}_{\ell m}(\theta, \phi) \pm \tilde{\alpha}_{\ell m}(\pi-\theta,-\phi)] P_{R} \psi_{ \ell m}(x) \\ 
[\tilde{\beta}_{\ell m}(\theta,\phi) \mp \tilde{\beta}_{\ell m}(\pi-\theta,-\phi)] P_L \psi_{ \ell m}(x) \end{pmatrix} 
\end{align}
where a $\psi_{\ell m}$ shows an SO(1,3) Dirac femion, $P_{L(R)}$ are Chiral projection operators in four-dimensions, and $\{ P_L, P_R \}$ are interchanged in RHS for $\Psi_-^{(\pm \gamma_5)}(x,\theta,\phi)$.
Then the kinetic and KK mass terms for KK modes are given in terms of  $\psi_{\ell m}$ such that
\begin{align}
\label{KM-F}
\int d \Omega \Psi_\pm^{(\pm \gamma_5)}(x,\theta,\phi) i \Gamma^M \partial_M \Psi_\pm^{(\pm \gamma_5)}(x,\theta,\phi) 
= \sum_{\ell m} \epsilon_{\ell m} \bigl[ \bar{\psi}_{ \ell m}(x) i \gamma^\mu \partial_\mu \psi_{ \ell m}(x) 
\pm  M_\ell \bar{\psi}_{ \ell m}(x) \gamma_5 \psi_{ \ell m}(x)  \bigr],
\end{align}
where $\epsilon_{\ell m}$ is $0$ for $\{ \ell = {\rm odd}, m=0 \}$ and unity for other modes.
We also need to take into account Yukawa couplings of Higgs zero mode and fermion non-zero KK modes to obtain mass spectrum of the KK particles after the electroweak symmetry breaking. 
The Yukawa coupling with Higgs zero mode $H^{00}$ is written as
\begin{equation}
L_Y = \sum_{\ell m} \frac{Y}{\sqrt{4 \pi R^2}} \bigl[ \bar{\psi}_{\ell m}^{F }(x) H^{00}(x) P_L \psi_{ \ell m}^{f } (x) + \bar{\psi}_{ \ell m}^{F }(x) H^{00}(x) P_R \psi_{ \ell m}^{f } (x)  \bigr] + {\rm h. c.},
\end{equation}
where $\psi^F = Q, L$ correspond to SU(2) doublets and $\psi^f = U, D, E$ correspond to SU(2) singlets.
After the zero mode Higgs getting vacuum expectation value (VEV), we have the mass term of the KK modes of the form 
\begin{equation}
\label{mf-matrix}
L_{\psi {\rm mass}} =
\begin{pmatrix} \bar{\psi}_{ \ell m}^F & \bar{\psi}_{ \ell m}^f \end{pmatrix}
\begin{pmatrix} M_\ell & m_{\rm SM} \\ m_{\rm SM} & -M_\ell \end{pmatrix}
\begin{pmatrix} \psi_{ \ell m}^F & \psi_{ \ell m}^f \end{pmatrix},
\end{equation}
where $m_{\rm SM}$s express the masses in the SM, and the mass spectrum is obtained by diagonalizing the mass matrix.

The four-dimensional components of gauge fields $A_\mu(x,\theta, \phi)$ are expanded in terms of linear combination of spherical harmonics, satisfying boundary condition Eq.~(\ref{BCA}), such that 
\begin{align}
\label{Aexpand}
A_{\mu}(x,\theta, \phi) = \sum_{\ell m} Y^+_{\ell m} (\theta, \phi) A_{\ell m \mu}(x), 
\end{align}
where $ Y^+_{\ell m} (\theta, \phi) = (i)^{\ell+ m}[Y_{\ell m}(\theta,\phi)+(-1)^\ell Y_{\ell -m}(\theta,\phi)]/\sqrt{2}$ for $m \neq 0$ and $Y^+_{\ell 0}(\theta)=Y_{\ell 0}(\theta)(0) $ for $m=0$ with $\ell=$even(odd), 
respectively.
Then the kinetic and KK mass terms for KK modes are obtained as
\begin{align}
\label{KM-A}
& \int d \Omega \Bigl[ -\frac{1}{4} F^{\mu \nu}(x,\theta,\phi) F_{\mu \nu}(x,\theta,\phi) -\frac{1}{2} g^{\mu \alpha} g^{\nu \beta} F_{\mu \alpha}(x,\theta,\phi) F_{\mu \beta}(x,\theta,\phi) \nonumber \\
& \supset  \int d \Omega \Bigl[ -\frac{1}{4} F^{\mu \nu}(x,\theta,\phi) F_{\mu \nu}(x,\theta,\phi) + \frac{1}{2} A^{\mu}(x,\theta,\phi) \hat{L}^2 A_{\mu}(x,\theta,\phi)  \Bigr]  \nonumber \\
&= \sum_{\ell m} \epsilon_{\ell m} \Bigl[ -\frac{1}{4} F^{\mu \nu}_{\ell m}(x) F_{\ell m \mu \nu}(x) + \frac{1}{2} M_\ell^2 A^\mu_{\ell m }(x) A_{\ell m \mu}(x) \Bigr],
\end{align}
where $\hat{L}^2 = -(1/\sin \theta) \partial_\theta (\sin \theta \partial_\theta) - (1/\sin^2 \theta) \partial_\phi^2$ is the square of angular momentum operator 
and $\epsilon_{\ell m}$ is the same as in Eq.~(\ref{KM-F}).
After electroweak symmetry breaking, KK modes of $W^\pm$ and $Z$ boson also obtain the contribution of SM mass $m_W^2 W_\mu^{+ \ell m} W^{\mu - \ell m}$ and $m_Z^2 Z_\mu^{\ell m} Z^{\mu \ell m}$
respectively.
%

The Lagrangian quadratic in $\{A_\theta, A_\phi \}$ is given by 
\begin{align}
L_{A_\theta, A_\phi \ {\rm quadratic}} =& \frac{1}{2 g^2} \sin \theta \Bigl[ (\partial_\mu A_\theta)(\partial^\mu A_\theta) + (\partial_\mu \tilde{A}_\phi)(\partial^\mu \tilde{A}_\phi)
- \frac{1}{R^2 \sin^2 \theta} ((\partial_\theta \sin \theta \tilde{A}_\phi)-(\partial_\phi A_\theta))^2 \nonumber \\
& - \frac{\xi}{ R^2 \sin^2 \theta} ( (\partial_\theta \sin \theta A_\theta)+ \partial_\phi \tilde{A}_\phi )^2 \Bigr],
\end{align}
where $\tilde{A}_\phi = A_\phi/\sin \theta$ and the second line in the RHS corresponds to a gauge fixing term.
This Lagrangian is not diagonal for $\{A_\theta, A_\phi \}$ so that we carry out the substitution 
\begin{align} 
\label{substitution1}
A_\theta(x,\theta,\phi) = \partial_\theta \phi_2 (x,\theta,\phi) - \frac{1}{\sin \theta} \partial_\phi \phi_1(x,\theta,\phi), \\
\label{substitution2}
\frac{A_\phi(x,\theta,\phi)}{\sin \theta} = \partial_\theta \phi_1 (x,\theta,\phi) + \frac{1}{\sin \theta} \partial_\phi \phi_2(x,\theta,\phi), 
\end{align}
to diagonalize the quadratic terms.
Then the quadratic terms become 
\begin{align}
\label{quadra-ex}
L_{A_\theta, A_\phi \ {\rm quadratic}} =& \frac{1}{g^2} \sin \theta \Bigl[ \partial_\mu \phi_1 \partial^\mu(\hat{L}^2 \phi_1) +\partial_\mu \phi_2 \partial^\mu(\hat{L}^2 \phi_2) 
-\frac{1}{ R^2} (\hat{L}^2 \phi_1)^2 - \frac{\xi}{R^2} (\hat{L}^2 \phi_1)^2 \Bigr].
\end{align}
Thus the mode functions for extra dimensional components gauge filed $\phi_{i}$ is given as 
\begin{equation}
\tilde{Y}_{\ell m}(\theta, \phi) = \frac{1}{\sqrt{\ell(\ell+1)}} Y_{\ell m}(\theta, \phi)
\end{equation} 
where factor of $1/\sqrt{\ell (\ell+1)}$ is required for normalization due to extra $\hat{L}^2$ factor in Eq.~(\ref{quadra-ex}).
Then $\phi_{i}(\theta, \phi, x)$ is expanded as 
\begin{align}
\label{phiexpand}
\phi_i(x,\theta, \phi) = \sum_{\ell m} \tilde{Y}^+_{\ell m} (\theta, \phi) \phi_{i\ell m}(x), 
\end{align}
taking into account the boundary condition Eq.~(\ref{BCex}).
Therefore KK mass terms for $\phi_i$ are obtained as 
\begin{align}
\label{KM-phi}
& \frac{1}{2} \int d\Omega \Bigl[  \partial_\mu \phi_1(x,\theta,\phi) \partial^\mu (\hat{L}^2 \phi_1(x,\theta,\phi)) +  \partial_\mu \phi_2(x,\theta,\phi) \partial^\mu (\hat{L}^2 \phi_2(x,\theta,\phi) ) 
-\frac{1}{R^2} (\hat{L}^2 \phi_1)^2-\frac{\xi^2}{R^2} (\hat{L}^2 \phi_2)^2  \Bigr]  \nonumber \\
&= \sum_{\ell m} \epsilon_{\ell m} \Bigl[ \partial_\mu \phi_{1 \ell m} \partial^\mu \phi_{1 \ell m}+\partial_\mu \phi_{2 \ell m} \partial^\mu \phi_{2 \ell m } - M_\ell^2 \phi_{1 \ell m} \phi_{1 \ell m}- \xi M_\ell^2 \phi_{2 \ell m} \phi_{2 \ell m} \Bigr]
\end{align}
where $\epsilon_{\ell m}$ is $0$ for $\{ \ell = {\rm odd}, m=0 \}$ and unity for other modes,
the gauge fixing parameter $\xi$ is taken as $1$ in our analysis applying Feynman-t'Hooft gauge,  and the KK modes of $\phi_2$ are interpreted as Nambu-Goldstone (NG) bosons.
These NG bosons will be eaten by KK modes of four-dimensional components of gauge field giving their longitudinal component.
After electroweak symmetry breaking, KK modes corresponding to extra components for $W^\pm$ and $Z$ fields 
obtain the contribution of SM mass $m_W^2 W_{1,2}^{+ \ell m} W_{1,2}^{ - \ell m}$ and $m_Z^2 Z_{1,2}^{\ell m} Z_{1,2}^{ \ell m}$ respectively.

The mode function for the scalar field is also given by the Spherical Harmonics $Y_{\ell m}(\theta, \phi)$, and $H(x, \theta, \phi)$ on $M^4 \times S^2/Z_2$ is expanded as
\begin{align}
\label{Hexpand}
H(x,\theta, \phi) = \sum_{\ell m} Y^+_{\ell m} (\theta, \phi) H_{\ell m}(x), 
\end{align}
taking into account the boundary condition Eq.~(\ref{BCH}).
Thus the kinetic and KK mass terms are obtained as
\begin{equation}
\label{KM-H}
\int d \Omega \Bigl[ \partial^\mu H^\dagger (x,\theta,\phi) \partial_\mu  H(x,\theta, \phi) -\hat{L}^2 H^\dagger(x,\theta,\phi) H(x,\theta, \phi) \Bigr] =
\sum_{\ell m} \epsilon_{\ell m} \Bigl[ \partial^\mu H_{ \ell m}^\dagger \partial_\mu H_{ \ell m} - M_\ell^2 H_{ \ell m}^\dagger H_{ \ell m} \Bigr],
\end{equation} 
and we also have SM Higgs mass contribution $m_{H}^2 (H^{\ell m})^\dagger H^{\ell m} $.
%

%
Therefore the KK masses are given in general, at tree level, as 
\begin{equation}
m^2_\ell = \frac{\ell(\ell+1)}{R^2} + m_{SM}^2,
\end{equation}
which is characterized by angular momentum number $\ell$ but independent of $m$.
Also the KK parity is defined as $(-1)^m$ for each KK particles due to discrete $Z_2'$ symmetry of  $(\theta,\phi) \rightarrow (\theta,\phi+\pi)$ 
which is understood as the symmetry under the exchange of two fixed points on $S^2/Z_2$ orbifold $(\frac{\pi}{2},0)$ and $(\frac{\pi}{2},\pi)$. 
Thus we can confirm the stability of the lightest KK particle with odd parity.

\subsection{Propagators on the six-dimensions with $S^2/Z_2$ extra space \label{propagator}}
Here we discuss the Feynman rules for propagators and vertices on the $M^4 \times S^2/Z_2$ for a fermion and a gauge boson.
We first consider propagators on $M^4 \times S^2$ and then derive those of after orbifolding taking into account the boundary conditions of fields on $S^2/Z_2$.
The propagator of fermion $S_F (x-x',z-z',\phi-\phi') = \langle \Psi^{}(x,\theta,\phi) \bar{\Psi}^{}(x,\theta,\phi) \rangle$ on $M^4 \times S^2$, here $\Psi$ being Dirac fermion,  is defined as an 
inverse matrix of the Dirac operator on six dimensions
\begin{align}
i \Gamma^M \partial_M = i \Gamma^\mu \partial_\mu + i \Gamma^\alpha \partial_\alpha,
\end{align} 
where the $i \Gamma^\alpha \partial_\alpha = \gamma_5 \otimes i \hat{D} $ is the extra-dimensional components of the Dirac operator in six dimensions 
with $i \hat{D}$ given in Eq.~(\ref{DiracO-S2}).
Then the propagator should satisfy
\begin{equation}
\label{propagator-condition}
i \Gamma^M \partial_M S_F(x-x',z-z',\phi-\phi') = \frac{i}{R^2}\delta^{(4)} (x-x') \delta (z-z') \delta (\phi-\phi'),
\end{equation}
where $R^2$ factor on the RHS is appeared to compensate mass dimensions.
Then the $S_F(x-x',z-z',\phi-\phi')$ can be written, using mode function in Eq.~(\ref{Fmode}) and (\ref{Fmode0}), as 
\begin{align}
\label{S2propaF}
S_F(x-x',z-z',\phi-\phi') = \sum_{\ell =0}^{\infty} \sum_{m=-\ell}^{\ell} \int \frac{d^4 p}{(2\pi)^4} 
\frac{i}{\Gamma^{\mu} p_{\mu} + i \Upsilon_5 M_\ell} \frac{\Psi_{\ell m}(z,\phi)}{R} \frac{\Psi^{\dagger}_{\ell m}(z',\phi')}{R} e^{-ip \cdot (x-x')},
\end{align}
where we can confirm that it satisfies Eq.~(\ref{propagator-condition}) applying Eq.~(\ref{DiracO-S2}) and completeness condition Eq.~(\ref{complete}) for mode functions. 
The propagator on $M^4 \times S^2/Z_2$ is obtained in terms of the linear combination $\Psi^{(\pm \gamma_5)}(x,z,\phi)= (\Psi(x, z,\phi)\pm \Upsilon_5 \Psi(x,-z,-\phi) )/2$ 
satisfying the boundary condition Eq.~(\ref{BCF}),  
such that 
\begin{align}
\tilde{S}_F^{(\pm \gamma_5)} (x-x',z-z',\phi-\phi') =& \langle \Psi^{(\pm \gamma_5)}(x,\theta,\phi) \bar{\Psi}^{(\pm \gamma_5)}(x,\theta,\phi) \rangle \nonumber \\
=& \frac{1}{4} \Big[ \langle \Psi(x, z, \phi) \bar{\Psi}(x', z', \phi') \rangle \mp \langle \Psi(x, z, \phi) \bar{\Psi}(x', -z', -\phi') \rangle \Upsilon_5  \nonumber \\
& \pm \Upsilon_5  \langle \Psi(x, -z, -\phi) \bar{\Psi}(x', z', \phi') \rangle - \Upsilon_5 \langle \Psi(x, -z, -\phi) \bar{\Psi}(x', -z', -\phi') \rangle \Upsilon_5 \Big],
\end{align}
where each $\langle \Psi \bar{\Psi} \rangle$ are given by Eq.~(\ref{S2propaF}).
Then it can be simplified as
\begin{align}
\label{propa-F}
& \tilde{S}_F^{(\pm \gamma_5)}(x-x',z-z',\phi-\phi')  \nonumber \\
& =\frac{1}{2} \sum_{\ell =0}^{\infty} \sum_{m,m'=-\ell}^{\ell}  \int \frac{d^4 p}{(2\pi)^4} 
\biggl[ \frac{i\delta_{mm'}}{\Gamma^{\mu} p_{\mu} + i \Upsilon_5 M_l} 
\mp  \frac{(-1)^{\ell+m} i\delta_{-mm'}}{\Gamma^{\mu} p_{\mu} + i \Upsilon_5 M_{\ell} } \Upsilon_5 \biggr] 
\frac{\Psi_{\ell m}(z,\phi)}{R} \frac{\Psi^{\dagger}_{\ell m'}(z',\phi')}{R} e^{-ip \cdot (x-x')},
\end{align}
where the sign factor $(-1)^{\ell+ m}$ is derived from the relation $\Psi_{\ell m}(-z, -\phi)=(-1)^{\ell +m} \Psi_{\ell -m}(z, \phi)$.
This propagator has similar structure as the minimal UED case except for the sign factor~\cite{Georgi:2000ks,Cheng:2002iz}.

The propagator of gauge boson, $D_F^{\mu \nu}(x-x',z-z',\phi-\phi')$, on $M^4 \times S^2$ is defined to satisfy 
\begin{equation}
\label{GpropaCond}
\biggl[ \partial_{\mu} \partial^{\mu} + \frac{1}{R^2}\hat{L}^2 \biggr] D_F^{\mu \nu}(x-x',z-z',\phi-\phi') 
= \frac{i g^{\mu \nu}}{R^2} \delta^{(4)} (x-x') \delta (z-z') \delta (\phi-\phi'),
\end{equation} 
in Feynman-t'Hooft gauge, where the operator in LHS comes from the kinetic term shown in Eq.~(\ref{KM-A}).
Then the propagator is expressed by Spherical Harmonics as
\begin{equation}
D_F^{\mu \nu}(x-x',z-z',\phi-\phi') = \sum_{\ell =0}^{\infty} \sum_{m=-\ell}^{\ell} \int \frac{d^4 p}{(2\pi)^4}
\frac{-i g^{\mu \nu}}{p^2-M_\ell^2} \frac{Y_{\ell m}(z,\phi)}{R} \frac{Y^*_{\ell m}(z',\phi')}{R} e^{-ip \cdot (x-x')} 
\end{equation}
which satisfies Eq.~(\ref{GpropaCond}) with completeness relation of Spherical Harmonics.
The propagator on $M^4 \times S^2/Z_2$ is obtained by taking into account the boundary condition Eq.~(\ref{BCA}) as the case of fermion propagator. 
Substituting $Y_{\ell m}(\theta, \phi)$ to the linear combination $(Y_{\ell m}(\theta, \phi)+ Y_{\ell m}(\pi-\theta,-\phi) )/2$, satisfying the boundary condition , we obtain
\begin{align}
\label{propa-G}
&\tilde{D}_F^{\mu \nu}(x-x',z-z',\phi-\phi') \nonumber \\
& = \frac{1}{2} \sum_{\ell=0}^{\infty} \sum_{m,m'=-\ell}^{\ell}  \int \frac{d^4 p}{(2\pi)^4}
\frac{-i g^{\mu \nu}}{p^2-M_{\ell}^2} \bigl[ \delta_{mm'}+(-1)^\ell\delta_{-mm'} \bigr] 
\frac{Y_{\ell m}(z,\phi)}{R} \frac{Y^*_{\ell m'}(z',\phi')}{R} e^{-ip \cdot (x-x')},
\end{align}
where the factor of $(-1)^\ell$ is derived from the relation $Y_{\ell m}(\pi-\theta,-\phi)=(-1)^\ell Y_{\ell -m}(\theta, \phi)$.

The propagators for $\phi_{1,2}$, the linear combination of extra dimensional component of gauge field $A_{\theta, \phi}$, are similarly given as four dimensional component which satisfy 
\begin{equation}
\label{prop-cond}
\biggl[ \partial_{\mu} \partial^{\mu}  + \frac{1}{R^2} \hat{L}^2 \biggr] \hat{L}^2 D_F^{ex}(x-x',z-z',\phi-\phi') 
= -\frac{ i }{R^2} \delta^{(4)} (x-x') \delta (z-z') \delta (\phi-\phi'),
\end{equation} 
where the minus sign in the RHS comes from the minus sign in $\theta \theta$ and $\phi \phi$ components of the metric $g_{MN}$, 
and extra $\hat{L}^2$ factor comes from LHS of Eq.~(\ref{KM-phi}) due to the substitutions of Eqs.~(\ref{substitution1}) and (\ref{substitution2}).
Thus the propagators before and after orbifolding are obtained as 
\begin{align}
& {D}^{ex}_F(x-x',z-z',\phi-\phi') = \sum_{\ell=0}^{\infty} \sum_{m,m'=-\ell}^{\ell}  \int \frac{d^4 p}{(2\pi)^4} \frac{i}{p^2-M_\ell^2}
\frac{\tilde{Y}_{\ell m}(z,\phi)}{R} \frac{\tilde{Y}^*_{\ell m'}(z',\phi')}{R} e^{-ip \cdot (x-x')}, \\
\label{propa-Gex}
& \tilde{D}^{ex}_F(x-x',z-z',\phi-\phi') \nonumber \\
&  = \frac{1}{2} \sum_{\ell=0}^{\infty} \sum_{m,m'=-\ell}^{\ell}  \int \frac{d^4 p}{(2\pi)^4}
\frac{i}{p^2-M_\ell^2}[\delta_{mm'} + (-1)^\ell\delta_{-mm'}] 
\frac{\tilde{Y}_{\ell m}(z,\phi)}{R} \frac{\tilde{Y}^*_{\ell m'}(z',\phi')}{R} e^{-ip \cdot (x-x')}, 
\end{align}
where $\tilde{Y}_{\ell m} = Y_{\ell m}/\sqrt{\ell (\ell+1)}$,
for both $\phi_1$ and $\phi_2$ in the Feynman-t'Hooft gauge, and the propagator on $M^4 \times S^2/Z_2$ is obtained in the same way as four dimensional components 
using $(\tilde{Y}_{\ell m}(\theta, \phi)+ \tilde{Y}_{\ell m}(\pi-\theta,-\phi) )/2$ for mode functions.

The propagator for scalar boson is obtained similar to the gauge boson case as
\begin{equation}
D_F(x-x',z-z',\phi-\phi') = \sum_{\ell =0}^{\infty} \sum_{m=-\ell}^{\ell} \int \frac{d^4 p}{(2\pi)^4}
\frac{i}{p^2-M_\ell^2} \frac{Y_{\ell m}(z,\phi)}{R} \frac{Y^*_{\ell m}(z',\phi')}{R} e^{-ip \cdot (x-x')}. 
\end{equation}
This propagator satisfies the condition 
\begin{equation}
\label{prop-condH}
\biggl[ \partial_{\mu} \partial^{\mu}  + \frac{1}{R^2} \hat{L}^2 \biggr]  D_F(x-x',z-z',\phi-\phi') 
= -\frac{ i }{R^2} \delta^{(4)} (x-x') \delta (z-z') \delta (\phi-\phi').
\end{equation} 
Then after orbifolding, we obtain 
\begin{align}
\label{propa-S}
& \tilde{D}_F(x-x',z-z',\phi-\phi') \nonumber \\
&   = \frac{1}{2} \sum_{\ell =0}^{\infty} \sum_{m=-\ell}^{\ell} \sum_{m'=-\ell}^{\ell} \int \frac{d^4 p}{(2\pi)^4}
\frac{i}{p^2-M_\ell^2}[\delta_{mm'} + (-1)^\ell \delta_{-mm'}] 
\frac{Y_{\ell m}(z,\phi)}{R} \frac{Y^*_{\ell m'}(z',\phi')}{R} e^{-ip \cdot (x-x')}. 
\end{align}


\subsection{Vertices on the six-dimensions with $S^2/Z_2$ extra space \label{vertices}}

The Feynman rules for the vertices in the model, along with the propagators, are obtained in terms of mode functions.
As an example, we consider a gauge interaction of a chiral fermion $\Psi_{\pm}^{(\pm \gamma_5)}$ written by
\begin{equation}
\int d^4x d \Omega g_{6a} \bar{\Psi}_\pm^{(\pm \gamma_5)} \Gamma^\mu T^a_i  A_\mu^i \Psi_\pm^{(\pm \gamma_5)} 
= \int d^4x d \Omega g_{6a}  \bar{\Psi}^{(\pm \gamma_5)} \Gamma^\mu T^a_i A_\mu^i \frac{1\pm \Gamma_7}{2} \Psi^{(\pm \gamma_5)},
\end{equation}
where $T^a$s are generators of a gauge symmetry.
When the fields $\bar{\Psi}^{(\pm \gamma_5)}$, $\Psi^{(\pm \gamma_5)}$ and $A_\mu$ on the interaction make contractions, 
applying the propagators of $\Psi$ and $A_\mu$ given in Eq.~(\ref{propa-F}) and Eq.~(\ref{propa-G}), we get the propagators on momentum space, 
\begin{align}
& \frac{1}{2} \biggl[ \frac{i\delta_{mm'}}{\Gamma^{\mu} p_{\mu} + i \Upsilon_5 M_l} \mp  \frac{(-1)^{\ell+m} i\delta_{-mm'}}{\Gamma^{\mu} p_{\mu} + i \Upsilon_5 M_{\ell} } \Upsilon_5 \biggr],  \\
& \frac{1}{2}  \frac{-i g^{\mu \nu}}{k^2-M_{\ell}^2} \bigl[ \delta_{mm'}+(-1)^\ell\delta_{-mm'} \bigr],
\end{align}
and the mode functions 
$\Psi_{\ell m}(z,\phi)  (\Psi^\dagger_{\ell m}(z,\phi))$ and $Y_{\ell m}(z,\phi)$ are left as a vertex factor.
Thus the vertex factor of $\psi_{\ell_1 m_1} A_{\ell_2 m_2}^i \psi_{\ell_3 m_3}$ coupling for each KK modes combinaton is derived as
\begin{align}
\label{vertex-ex}
& i \frac{g_{6a}}{R} T^a_i \int d \Omega \Psi^\dagger_{\ell_1 m_1}(z, \phi) Y_{\ell_2 m_2}(z,\phi) \Gamma^\mu \frac{1\pm \Gamma_7}{2} \Psi_{\ell_3 m_3}(z,\phi) \nonumber \\
& = i \frac{g_{6a}}{R} T^a_i \left[ \left( \int d \Omega \tilde{\alpha}^*_{\ell_1 m_1} Y_{\ell_2 m_2} \tilde{\alpha}_{\ell_3 m_3} \right) \gamma^\mu P_{R(L)} 
+ \left( \int d \Omega \tilde{\beta}^*_{\ell_1 m_1} Y_{\ell_2 m_2} \tilde{\beta}_{\ell_3 m_3} \right) \gamma^\mu P_{L(R)} \right] \nonumber \\ 
& \equiv i \frac{g_{6a}}{R} T^a_i \gamma^\mu \left[ I^\alpha_{\ell_1 m_1; \ell_2 m_2; \ell_3 m_3} P_{R(L)} + I^\beta_{\ell_1 m_1; \ell_2 m_2; \ell_3 m_3} P_{L(R)} \right]
\end{align}
where the integration in RHS is over $S^2/Z_2$, the mode functions for fermion are given in Eq.~(\ref{Fmode}) and (\ref{Fmode0}), 
and indices of $T_a$ are understood to be contracted with KK fermions $\psi_{\ell_1 m_1}$ and $\psi_{\ell_3 m_3}$.
Other vertex factors are derived in the same way and the full list of the vertex factors in the model is given in Appendix.~\ref{vertex-list}.

\section{One loop quantum correction of KK masses \label{Sec:OneLoop}}
In this section, we derive one-loop corrections to KK mass $M_\ell$ which does not include SM mass $m_{SM}$, being purely obtained from extra-dimensional kinetic term, 
by calculating relevant one-loop diagrams.
Readers who are interested in our results might skip this section discussing technical details.
%
\begin{figure}[t] 
\begin{minipage}{0.45 \hsize}
\begin{center}
\includegraphics[width=70mm]{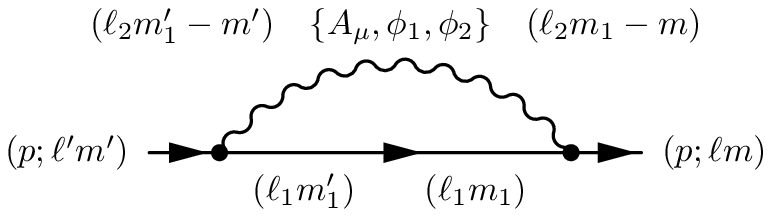}
\caption{One loop diagram for correction to KK masses of fermion with virtual gauge bosons including extra dimensional components.
\label{Loop1}}
\end{center}
\end{minipage}
\hspace{8mm}
\begin{minipage}{0.45 \hsize}
\begin{center}
\includegraphics[width=70mm]{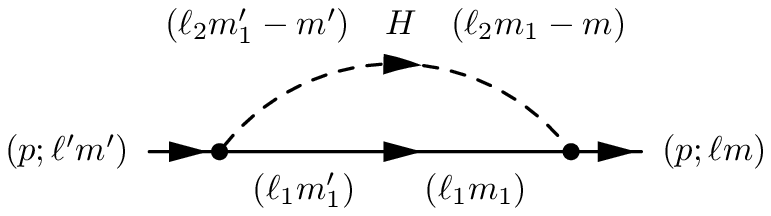} 
\caption{One loop diagram for correction to KK masses of fermion with virtual Higgs bosons.
\label{Loop2}}
\end{center}
\end{minipage} 
\end{figure}
%
\subsection{Correction to KK masses of fermion}
The one loop diagrams relevant to correction to KK masses of fermion are given as Fig.~\ref{Loop1} and \ref{Loop2} where corresponding quantum numbers $\{ \ell, m \}$ 
are shown for each external and internal lines.
The Fig.~\ref{Loop1} shows the virtual gauge boson contributions including both four-dimensional components and extra-dimensional components, 
and the Fig.~\ref{Loop2} shows the virtual Higgs boson contribution.
Here we discuss the one loop diagram Fig.~\ref{Loop1} with virtual $A_\mu$ for external $\Psi_+^{(\pm \gamma_5)}$ case to show the structure of the quantum correction.
We can calculate the diagram, by applying the propagators Eq.~(\ref{propa-F}) and (\ref{propa-G}) and vertex Eq.~(\ref{vertex-ex}), such that
\begin{align}
\label{LoopFig1}
-i \Sigma^{\rm Fig.1}(p; \ell m; \ell' m') = & \frac{1}{4R^2}
\sum_{\ell_1=0}^{\ell_{max}} \sum_{\ell_2=0}^{\ell_{max}} \sum_{m_1=-\ell_1}^{\ell_1} \sum_{m_1'=-\ell_1}^{\ell_1} \int_\Lambda \frac{d^4 k}{(2 \pi)^4} \nonumber \\
& \times \biggl[ \frac{-i}{(p-k)^2-M^2_{\ell_2}} (\delta_{m_1-m, m_1'-m'}+(-1)^{\ell_2} \delta_{-(m_1-m),m_1'-m'}) \nonumber \\
& \quad \times (i g_{6a} T^a \gamma^\mu) [I^\alpha_{\ell_1 m_1'; \ell_2 m_1'-m'; \ell' m'} P_R +I^\beta_{\ell_1 m_1'; \ell_2 m_1'-m'; \ell' m' } P_L ] \nonumber \\
& \quad \times \frac{i}{\sla{k}+i \gamma_5 M_{\ell_1}} (\delta_{m_1, m_1'} \mp (-1)^{\ell_1+m_1} \delta_{-m_1, m_1'} \gamma_5) \nonumber \\
& \quad \times (i g_{6a} T^a \gamma_\mu) [I^\alpha_{\ell m; \underline{\ell_2 m_1-m}; \ell_1 m_1 } P_R+I^\beta_{\ell m; \underline{\ell_2 m_1-m}; \ell_1 m_1} P_L] \biggr],
\end{align}
where $P_L$ and $P_R$ on the RHS will be interchanged for $\Psi_-^{(\pm \gamma_5)}$ 
and the underline for the KK numbers in the vertex factor $I^{\alpha(\beta)}_{\ell m; \underline{\ell_2 m_1-m}; \ell_1 m_1 } $ shows the corresponding spherical harmonics is conjugate one $Y^*_{\ell_2 m_1-m}$.
The quantum numbers $\ell_1$, $\ell_2$, $m_1$ and $m_1'$ are associated with the KK modes in the loop which are summed over.
In this paper we employ, for all one-loop diagram calculations, 
the cut-off $\Lambda$ for four momentum integration and $\ell_{max}$ for angular momentum sum, in order to regularize the loop diagram.
Summing over $m_1'$, the products of Kronecker delta and vertex factors $I^x$ ($x= \alpha$ or $\beta$) are arranged as
\begin{align}
\label{Cdelta}
& \sum_{m_1'}  (\delta_{m_1-m, m_1'-m'}+(-1)^{\ell_2} \delta_{-(m_1-m),m_1'-m'})  (\delta_{m_1, m_1'} \pm (-1)^{\ell_1+m_1} \delta_{-m_1, m_1'} \gamma_5) \nonumber \\
& \qquad \times I^{x}_{\ell_1 m_1'; \ell_2 m_1'-m'; \ell' m'} I^{x'}_{\ell m; \underline{\ell_2 m_1-m}; \ell_1 m_1 }  \nonumber \\
& = (\delta_{m, m'} \pm (-1)^{\ell'+m} \delta_{-m, m'} \gamma_5) I^{x}_{\ell_1 m_1;\ell_2 m_1-m;\ell' m} I^{x'}_{\ell m;\ell_2 m-m_1;\ell_1 m_1} \nonumber \\
& \quad  + (-1)^{\ell_2}(\delta_{2 m_1, m+m'} \pm (-1)^{\ell'+m} \delta_{2m_1, m-m'} \gamma_5)  I^{x}_{\ell_1 m_1;\ell_2 -m_1+m;\ell' 2m_1-m} I^{x'}_{\ell m;\ell_2 m-m_1;\ell_1 m_1},
\end{align}
where we used the relation $I^{x}_{\ell_1 -m_1;\ell_2-m_2;\ell_3 -m_3} = (-1)^{\ell_1+\ell_2+\ell_3+m_1+m_3} I^{x}_{\ell_1 m_1;\ell_2 m_2;\ell_3 m_3}$ obtained from 
the definition of the vertex factor given in Eq.~(\ref{vertex-ex}).
The first term in RHS of Eq.~(\ref{Cdelta}) corresponds to bulk contribution conserving KK-number $m$, and the second term corresponds to the boundary contribution violating 
the conservation of the KK-number $m$.
Thus the one-loop diagram contribution can be separated to the bulk  and boundary contributions such that
\begin{align}
- i \Sigma^{\rm Fig.1} (p;\mu; \ell m ; \ell' m') =& - i \Sigma^{\rm Fig.1}_{\rm bulk}  (p; \ell m ; \ell' m')  - i \Sigma^{\rm Fig.1}_{\rm bound}  (\mu; \ell m ; \ell' m') 
\end{align}
where bulk contribution is proportional to $m$ conserving Kronecker deltas $\{ \delta_{m,m'}, \delta_{-m,m'} \}$, 
and the boundary contribution is proportional to $m$ violating Kronecker deltas $\{\delta_{2m_1,m+m'}, \delta_{2m_1,m-m'} \}$.
As usual we combine denominator in the momentum integral using Feynman parameter $\alpha$, and carry out Wick rotation.
The momentum integral becomes
\begin{align}
\int_\Lambda \frac{d^4 k}{(2 \pi)^4} \frac{1}{[(p-k)^2-M^2_{\ell_2}][k^2-M_{\ell_1}^2] } = \frac{i}{(4\pi)^2} \int_0^1 d \alpha \int_0^{\Lambda^2} 
\frac{k_E^2 d k_E^2}{[k_E^2 - \alpha (1-\alpha) p^2 + M_{\ell_1}^2 (1-\alpha) + M_{\ell_2}^2 \alpha ]^2} 
\end{align}
where we performed the substitution $k-\alpha p \to k$, $k_E$ is a Euclidean 4-momentum, and the integration over $k_E^2$ is regularized by cut-off $\Lambda$.
We estimate the momentum integration numerically taking cut-off $\Lambda$ as parameter for bulk contribution, while we only left leading log-divergent part 
with renormalization scale $\mu$ for boundary contribution such that 
\begin{equation}
\label{logdiv}
\int_0^{\Lambda^2} 
\frac{k_E^2 d k_E^2}{[k_E^2 - \alpha (1-\alpha) p^2 + M_{\ell_1}^2 (1-\alpha) + M_{\ell_2}^2 \alpha ]^2}  \rightarrow \log \Bigl( \frac{\Lambda^2}{\mu^2} \Bigr),
\end{equation}
assuming vanishing contribution at the cut-off scale $\Lambda$.
Then after arranging the terms, Eq.~(\ref{LoopFig1}) is organized as
\begin{align}
\label{LoopFbulk}
-i \Sigma_{\rm bulk}^{\rm Fig.1}(p; \ell m; \ell' m')  = & \Bigl[ \sla{p} (i \Sigma_{\rm bulk}^{{\rm Fig.1}, L}(p; \ell m; \ell' m') P_{L} +i \Sigma_{\rm bulk}^{{\rm Fig.1}, R}(p; \ell m; \ell' m') P_{R}) \nonumber \\
& + i \gamma_5 (i \tilde{\Sigma}_{\rm bulk}^{{\rm Fig.1}, L}(p; \ell m; \ell' m') P_{L} + i\tilde{\Sigma}_{\rm bulk}^{{\rm Fig.1}, R}(p;\ell m; \ell' m') P_{R}) \Bigr]  \\
\label{LoopFbound}
-i \Sigma_{\rm bound}^{\rm Fig.1}(p; \ell m; \ell' m') 
= & \Bigl[ \sla{p} (i \Sigma_{\rm bound}^{{\rm Fig.1}, L}(\mu; \ell m; \ell' m') P_{L} + i \Sigma_{\rm bound}^{{\rm Fig.1}, R}(\mu; \ell m;\ell' m') P_{R}) \nonumber \\
&  + i \gamma_5 (i \tilde{\Sigma}_{\rm bound}^{{\rm Fig.1}, L}(\mu;\ell m; \ell' m') P_{L} + i \tilde{\Sigma}_{\rm bound}^{{\rm Fig.1},R}(\mu; \ell m; \ell'm') P_{R}) \Bigr],
\end{align}
where the coefficients $\Sigma^{L(R)}$ and $\tilde{\Sigma}^{L(R)}$ correspond to the terms proportional to $\sla{p}$ and $i \gamma_5$ respectively, associated with $P_{L(R)}$.
These coefficients are explicitly given by
\begin{align}
\label{SigmaFig1a}
i\Sigma_{\rm bulk}^{{\rm Fig.1}, L(R)}(p;\ell m; \ell' m')  
& = i \frac{g_{6a}^2}{64 \pi^2 R^2} (T^a)^2  \sum_{\ell_1=0}^{\ell_{max}} \sum_{\ell_2=0}^{\ell_{max}} \sum_{m_1=-\ell_1}^{\ell_1}  \int_0^1 d \alpha \int^{\Lambda^2}_{0} 
\frac{x dx}{[x + \Delta ]^2} (-1)^{m_1-m} \nonumber \\
&  \quad \times 2 \alpha I^{\beta(\alpha)}_{\ell_1 m_1;\ell_2 m_1-m;\ell' m} I^{\beta(\alpha)}_{\ell m;\ell_2 m-m_1;\ell_1 m_1} \bigl( \delta_{m,m'} \pm (-1)^{\ell'+m} \delta_{-m,m'} \gamma_5 \bigr) ,  \\
\label{SigmaFig1b}
i \tilde{\Sigma}_{\rm bulk}^{{\rm Fig.1}, L(R)}(p; \ell m;\ell' m')  
& = i \frac{g_{6a}^2}{64 \pi^2 R^2} (T^a)^2  \sum_{\ell_1=0}^{\ell_{max}} \sum_{\ell_2=0}^{\ell_{max}} \sum_{m_1=-\ell_1}^{\ell_1}  \int_0^1 d \alpha \int^{\Lambda^2}_{0} 
\frac{x dx}{[x + \Delta ]^2} (-1)^{m_1-m} \nonumber \\
&  \quad \times 4 M_{\ell_1} I^{\alpha(\beta)}_{\ell_1 m_1;\ell_2 m_1-m;\ell' m} I^{\beta(\alpha)}_{\ell m;\ell_2 m-m_1;\ell_1 m_1} \bigl( \delta_{m,m'} \pm (-1)^{\ell'+m} \delta_{-m,m'} \gamma_5 \bigr) , \\
\label{SigmaFig1c}
i\Sigma_{\rm bound}^{{\rm Fig.1}, L(R)}(\mu;\ell m; \ell' m') 
&= i \frac{g_{6a}^2}{64 \pi^2 R^2} (T^a)^2  \sum_{\ell_1=0}^{\ell_{max}} \sum_{\ell_2=0}^{\ell_{max}} \sum_{m_1=-\ell_1}^{\ell_1} 
(-1)^{m_1-m+\ell_2} \log \Bigl( \frac{\Lambda^2}{\mu^2} \Bigr) \nonumber \\
& \quad \times I^{\beta(\alpha)}_{\ell_1 m_1;\ell_2 -m_1+m;\ell' 2m_1-m} I^{\beta(\alpha)}_{\ell m;\ell_2 m-m_1;\ell_1 m_1} \ \nonumber \\ 
& \quad \times \bigl( \delta_{2m_1,m+m'} \pm (-1)^{\ell'+m} \delta_{2m_1,m-m'} \gamma_5 \bigr),   \\
\label{SigmaFig1d}
i \tilde{\Sigma}_{\rm bound}^{{\rm Fig.1}, L(R)}(\mu; \ell m; \ell' m') 
&= i \frac{g_{6a}^2}{64 \pi^2 R^2} (T^a)^2  \sum_{\ell_1=0}^{\ell_{max}} \sum_{\ell_2=0}^{\ell_{max}} \sum_{m_1=-\ell_1}^{\ell_1} 
(-1)^{m_1-m+\ell_2} \log \Bigl( \frac{\Lambda^2}{\mu^2} \Bigr) \nonumber \\
& \quad \times 4 M_{\ell_1} I^{\alpha(\beta)}_{\ell_1 m_1;\ell_2 -m_1+m;\ell' 2m_1-m} I^{\beta(\alpha)}_{\ell m;\ell_2 m-m_1;\ell_1 m_1} \nonumber \\
& \quad \times \bigl( \delta_{2m_1,m+m'} \pm (-1)^{\ell'+m} \delta_{2m_1,m-m'} \gamma_5 \bigr),
\end{align}
where $\Delta = -\alpha (1-\alpha)p^2 + (1-\alpha) M_{\ell_1}^2 + \alpha M_{\ell_2}^2$.
We also note that there appear sign factors of $(-1)^{\{ \ell, m \}}$ from propagators in $S^2/Z_2$ orbifold and vertices, 
which lead different sign contributions from each KK modes in the loop. 
Then the contributions depend on $\ell_{max}$ differently for even or odd $\ell_{max}$ as we will see below.
Also we can numerically check that external $\ell$ and $m$ are conserved so that $\ell = \ell'$ and $m=m'$ cases give non-zero contribution, 
which is satisfied for other diagrams.
The other diagrams are also calculated in the same manner, and
we list the contributions from each diagrams in Table~\ref{OneLoopF}. 
Therefore we obtain one-loop level correction to the Lagrangian such that
\begin{align}
\label{1Loop-F}
\delta L_{\rm 1-loop} =&  
\Sigma_{\rm bulk}^L(p^2;\ell m;\ell m) \bar{\psi}_{\ell m L}  i \gamma^{\mu} \partial_{\mu} \psi_{\ell m L}  
+\Sigma_{\rm bulk}^R(p^2;\ell m;\ell m)  \bar{\psi}_{\ell m R}  i \gamma^{\mu} \partial_{\mu} \psi_{\ell m R} 
\nonumber \\ 
& + \frac{1}{2} \Bigl[ \tilde{\Sigma}_{bulk}^L(p^2;\ell m;\ell m)+\tilde{\Sigma}_{bulk}^R(p^2;\ell m;\ell m)] \bar{\psi}_{\ell m} i \gamma_5 \psi_{\ell m} \Bigr] \nonumber \\
& +\Sigma_{\rm bound}^L(\mu;\ell m;\ell m) \bar{\psi}_{\ell m L}  i \gamma^{\mu} \partial_{\mu} \psi_{\ell m L}  
+\Sigma_{\rm bound}^R(\mu;\ell m;\ell m)  \bar{\psi}_{\ell m R}  i \gamma^{\mu} \partial_{\mu} \psi_{\ell m R} 
\nonumber \\ 
& + \frac{1}{2} \Bigl[ \tilde{\Sigma}_{\rm bound}^L(\mu;\ell m;\ell m)+\tilde{\Sigma}_{\rm bound}^R(\mu;\ell m;\ell m)] \bar{\psi}_{\ell m} i \gamma_5 \psi_{\ell m} \Bigr],
\end{align}
where each coefficients $\Sigma^{L(R)}_{\rm bulk}$,  $\Sigma^{L(R)}_{\rm bound}$, $\tilde{\Sigma}^{L(R)}_{\rm bulk}$ and $\tilde{\Sigma}^{L(R)}_{\rm bound}$ 
are understood as the sum of corresponding contributions from all diagrams.

To obtain one-loop corrections to KK mass, we take into account a renormalization condition for each KK modes.
The kinetic and KK mass terms of each KK modes of fermion are given in Eq.~(\ref{KM-F}).
Defining as a renormalization of fields, $\psi_{R(L)} \rightarrow \sqrt{Z_{R(L)}} \psi_{R(L)} = \sqrt{1+\delta_{R(L)}} \psi_{R(L)}$, the terms in Eq.~(\ref{KM-F}) become  
\begin{align}
& \bar{\psi}_{L \ell m} i \gamma^\mu \partial_\mu \psi_{L \ell m} + \bar{\psi}_{R \ell m} i \gamma^\mu \partial_\mu \psi_{R \ell m}
+ i M_\ell \bar{\psi}_{L \ell m} \gamma_5 \psi_{R \ell m} + i M_\ell \bar{\psi}_{R \ell m} \gamma_5 \psi_{L \ell m} \nonumber \\
& \delta_L  \bar{\psi}_{L \ell m} i \gamma^\mu \partial_\mu \psi_{L \ell m} + \delta_R \bar{\psi}_{R \ell m} i \gamma^\mu \partial_\mu \psi_{R \ell m} 
+ \frac{1}{2}(\delta_L+\delta_R)( i M_\ell \bar{\psi}_{L \ell m} \gamma_5 \psi_{R \ell m} + i M_\ell \bar{\psi}_{R \ell m} \gamma_5 \psi_{L \ell m} ),
\end{align}
where we have taken up to the first order in $\delta_{R(L)}$, which corresponds to the counter terms.
Thus the contribution from these counter terms is shown by
\begin{equation}
{\rm conter terms}(p;\ell m) = i \sla{p} [\delta_L P_L+\delta_R P_R] + i (i \gamma_5 M_\ell) (P_L + P_R) \frac{1}{2} [\delta_L+ \delta_R].
\end{equation}
Then we employ the renormalization conditions at a cut off scale $\Lambda$ as
\begin{align}
\bigl( i \delta_L + i \Sigma_{bulk}^L(p^2;\ell m;\ell m) )\mid_{ p^2 =-\Lambda^2 , M_{\ell max}^2=\Lambda^2} = 0, \\
\bigl( i \delta_R + i \Sigma_{bulk}^R(p^2;\ell m;\ell m) )\mid_{p^2 =-\Lambda^2 , M_{\ell max}^2=\Lambda^2} = 0,
\end{align}
which corresponds to the requirement that the kinetic term has canonical form at the cut-off scale, and the counter terms $\delta_{R(L)}$ are determined from these conditions.
Combining contributions from the one-loop diagrams and the counter terms, we obtain the corrected kinetic and KK mass terms such that 
\begin{align}
\delta L =&  \delta_L \bar{\psi}_{L \ell m} i \gamma^{\mu} \partial_{\mu} \psi_{L \ell m}  + \delta_{R} \bar{\psi}_{R \ell m} i \gamma^{\mu} \partial_{\mu} \psi_{R \ell m} \nonumber \\
& + \frac{1}{2} (\delta_L+\delta_R)( \bar{\psi}_{L \ell m} i M_\ell \gamma_5 \psi_{R \ell m} +\bar{\psi}_{R \ell m} i M_\ell \gamma_5 \psi_{L \ell m})  
+ \delta L_{\rm 1-loop}.
\end{align}
Therefore, normalizing the kinetic terms by 
\begin{equation}
\label{normalize-f}
\psi_{R(L) \ell m} \rightarrow \left(1 + \Sigma_{\rm bulk}^{R(L)}(p^2;\ell m;\ell m)+\Sigma_{\rm bound}^{R(L)}(\mu;\ell m;\ell m) + \delta_{R(L)}  \right)^{-\frac{1}{2}} \psi_{R(L) \ell m},
\end{equation}
we obtain the corrected KK mass of 
\begin{align}
\label{mass-F}
\tilde{M}_{\ell m} \simeq & M_{\ell} + \frac{1}{2} \left[   (\tilde{\Sigma}^L_{\rm bulk}+ \tilde{\Sigma}^R_{\rm bulk})(p^2;\ell m;\ell m) +(\tilde{\Sigma}^L_{\rm bound}+\tilde{\Sigma}^R_{\rm bound})(\mu;\ell m;\ell m) \right] 
\nonumber \\
& - \frac{1}{2}  M_{\ell  } \Bigl[ (\Sigma^L_{\rm bulk}+ \Sigma^R_{\rm bulk})(p^2;\ell m;\ell m)+ (\Sigma^L_{\rm bound} +\Sigma^R_{\rm bound})(\mu;\ell m;\ell m)  \Bigr]  \nonumber \\
\equiv & M_{\ell} +(\delta M)^\psi_{\ell m}
%
%
\end{align}
for each KK modes of fermions, where we have taken the first order of $\Sigma$ and $\delta$ in RHS.


\begin{figure}[tb] 
\begin{minipage}{0.45 \hsize}
\begin{center}
\includegraphics[width=70mm]{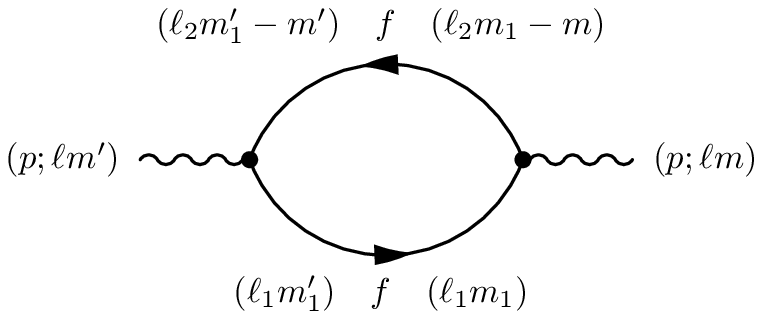}
\caption{One loop diagram for correction to KK masses of gauge bosons $\{ A_\mu, \phi_1, \phi_2 \}$ with virtual fermions.
\label{Loop3}}
\end{center}
\end{minipage}
\hspace{8mm}
\begin{minipage}{0.45 \hsize}
\begin{center}
\includegraphics[width=70mm]{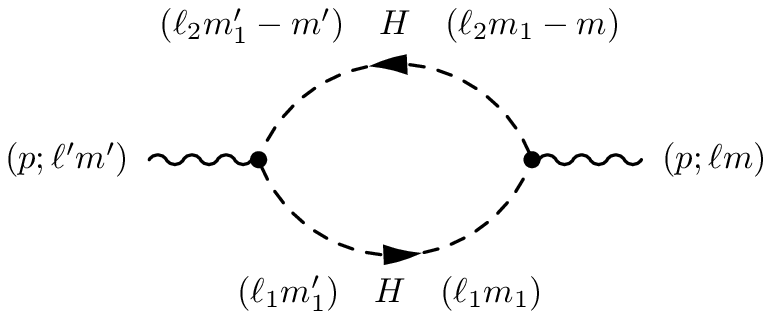} 
\caption{One loop diagram for correction to KK masses of gauge bosons $\{ A_\mu, \phi_1, \phi_2 \}$ with virtual Higgs bosons.
\label{Loop4}}
\end{center}
\end{minipage} 
\end{figure}
\begin{figure}[tb] 
\begin{minipage}{0.45 \hsize}
\begin{center}
\includegraphics[width=70mm]{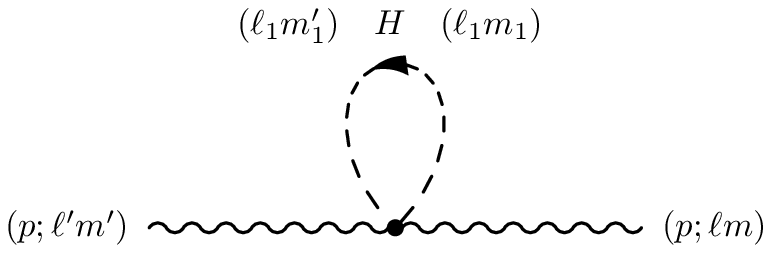}
\caption{One loop diagram for correction to KK masses of gauge bosons $\{ A_\mu, \phi_1, \phi_2 \}$ with virtual Higgs bosons.
\label{Loop5}}
\end{center}
\end{minipage}
\hspace{8mm}
\begin{minipage}{0.45 \hsize}
\begin{center}
\includegraphics[width=70mm]{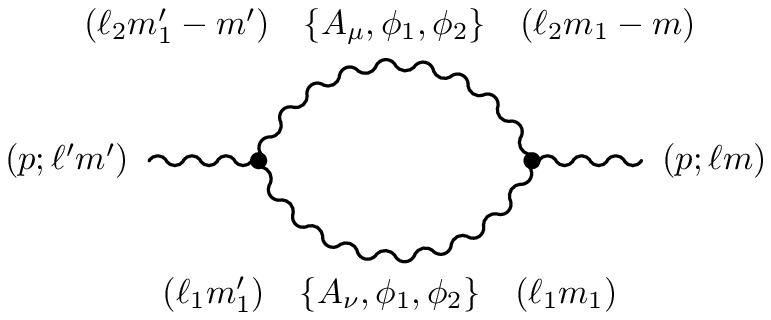} 
\caption{One loop diagram for correction to KK masses of gauge bosons $\{ A_\mu, \phi_1, \phi_2 \}$ with virtual gauge bosons including extra components.
\label{Loop6}}
\end{center}
\end{minipage} 
\end{figure}

\subsection{Correction to KK masses of gauge boson}

\begin{figure}[tb] 
\begin{center}
\includegraphics[width=70mm]{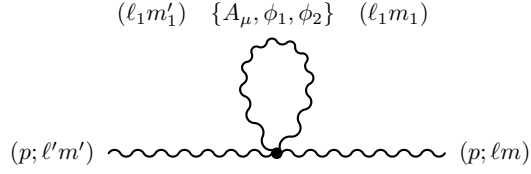}
\caption{One loop diagram for correction to KK masses of gauge bosons $\{ A_\mu, \phi_1, \phi_2 \}$ with virtual gauge bosons including extra components.
\label{Loop7}}
\end{center}
%
\end{figure}

The one loop diagrams corresponding to correction to KK masses of gauge bosons are given in Fig.~\ref{Loop3}-\ref{Loop7} for both 
four-dimensional components  and extra-dimensional components in external line, where corresponding quantum numbers $\{ \ell, m \}$ 
are shown for each external and internal lines.
The Fig.~\ref{Loop3} shows the fermion loop, the Fig.~\ref{Loop4} and \ref{Loop5} show Higgs boson loops from 3-point and 4-point gauge interactions respectively, 
and the Fig.~\ref{Loop6} and \ref{Loop7} show the gauge boson loops from 3-point and 4-point self-interactions respectively.
These diagrams are calculated in the same manner as the case of KK mass of fermions, and we summarize the results in Appendix \ref{LoopCalculation}.
Separating bulk and boundary contributions, we can organize the one loop diagram contributions generally as  
\begin{align}
\label{OneLoopGSeparate}
& i \Pi_{\mu \nu}(p^2;\mu; \ell m; \ell m) = i \Pi_{\mu \nu}^{\rm bulk}(p^2; \ell m; \ell m) + i \Pi_{\mu \nu}^{\rm bound}(p^2;\mu; \ell m; \ell m), \\ 
\label{OneLoopGBulkBound}
& i \Pi_{\mu \nu}^{\rm bulk (bound)}(p^2(p^2;\mu); \ell m; \ell m) = (p^2 g_{\mu \nu} - p_\mu p_\nu) i \Pi_{\rm bulk(bound)}(p^2(p^2;\mu); \ell m; \ell m)  \nonumber \\
& \hspace{55mm} + g_{\mu \nu} i \tilde{\Pi}_{\rm bulk(bound)}(p^2(p^2;\mu); \ell m; \ell m),
\end{align}
where we also separated the terms proportional to $(p^2 g_{\mu \nu} - p_\mu p_\nu)$ contributing the correction of kinetic term of $A_\mu$ and the terms proportional to $g_{\mu \nu}$
in RHS of second line.
The contributions to the coefficients $\Pi(\tilde{\Pi})$s from each diagrams are summarized in Table~\ref{OneLoopG} and \ref{OneLoopG-NA}.
Thus the correction to the Lagrangian is 
\begin{align}
\label{1Loop-A}
\delta L_{\rm 1 loop} =& \frac{1}{4}  \Pi_{\rm bulk}(p^2;\ell m;\ell m)  F^{\mu \nu}_{lm} F_{\ell m \mu \nu}
 + \frac{1}{2} \tilde{\Pi}_{\rm bulk}(p^2;\ell m;\ell m) A^{\mu}_{\ell m}  A_{\ell m \mu} \nonumber \\ 
 &+ \frac{1}{4}  \Pi_{\rm bound}(p^2;\mu;\ell m;\ell m)  F^{\mu \nu}_{lm} F_{\ell m \mu \nu}
 + \frac{1}{2} \tilde{\Pi}_{\rm bound}(p^2;\mu;\ell m;\ell m) A^{\mu}_{\ell m}  A_{\ell m \mu} 
\end{align}
where $\Pi(\tilde{\Pi})$s in the RHS are understood as the sum of corresponding contributions from all diagrams.
%

%
As in the fermion case, we consider the renormalization condition to obtain one-loop corrections to KK masses for KK gauge bosons.
The kinetic and KK mass terms of each KK modes of gauge boson are given in Eq.~(\ref{KM-A}).
Defining as a renormalization of gauge field, $A_\mu \rightarrow \sqrt{Z_3} A_\mu =\sqrt{1+\delta_3} A_\mu$, these terms become 
\begin{equation}
-\frac{1}{4} F^{\mu \nu}_{\ell m} F_{\ell m \mu \nu} + \frac{1}{2} M_\ell^2 A^\mu_{\ell m } A_{\ell m \mu} 
-\frac{1}{4} \delta_3 F^{\mu \nu}_{\ell m} F_{\ell m \mu \nu} + \frac{1}{2} \delta_3 M_\ell^2 A^\mu_{\ell m } A_{\ell m \mu},
\end{equation}
where we have taken up to the first order in $\delta_3$ corresponding to counter terms.
Thus the contribution from these counter terms is 
\begin{equation}
{\rm counter term}(p^2;\ell m) = -i(p^2g_{\mu \nu} -p_{\mu} p_{\nu}) \delta_3 +i M_{\ell}^2 \delta_3.
\end{equation}
Then we employ the renormalization condition at a cut-off scale $\Lambda$ as
\begin{equation}
(\Pi_{bulk}(p^2;\ell m;\ell m)-\delta_3) \mid_{p^2=-\Lambda^2,M_{\ell max}^2=\Lambda^2} = 0
\end{equation}
which requires the canonical form of kinetic term at the cut-off scale as in the fermion case, 
and the counter term $\delta_3$ is determined by the condition.
Combining contributions from one-loop diagrams and counter terms, we obtain the one-loop corrections to kinetic and KK mass terms such that
\begin{align}
\delta L = -\frac{1}{4} \delta_3 F^{\mu \nu}_{\ell m }F_{\ell m \mu \nu} + \frac{1}{2} \delta_3 A^{\mu}_{\ell m} \hat{L}^2 A_{\ell m \mu} + \delta L_{\rm 1-loop}.
\end{align}
Therefore, normalizing the kinetic terms by 
\begin{equation}
\label{NormalizeA}
A^{\mu}_{\ell m} \rightarrow \left( 1-\Pi_{\rm bulk}(p^2;\ell m;\ell m)- \Pi_{bound}(p^2;\mu;\ell m;\ell m)+\delta_3 \right)^{-\frac{1}{2}} A^{\mu}_{\ell m},
\end{equation}
we obtain one-loop corrections to KK mass
\begin{align}
\label{mass-G}
\tilde{M}_{\ell}^2 = &  M_\ell^2 + \tilde{\Pi}_{\rm bulk}(p^2;\ell m;\ell m)+\tilde{\Pi}_{\rm bound}(p^2;\mu;\ell m;\ell m)  + M_{\ell}^2 \bigl[ \Pi_{\rm bulk}(p^2;\ell m;\ell m)+ \Pi_{\rm bound}(p^2;\mu;\ell m;\ell m) \bigr], \nonumber \\
\equiv & M_\ell^2 + (\delta M^2)^{A_\mu}_{\ell m},
\end{align}
for each KK modes of gauge bosons, where we have taken the first order in $\Pi$ and $\delta_3$ in RHS.

The extra dimensional components of gauge fields $A_{\theta, \phi}$ are considered as scalar fields in four dimensions, and the KK mass eigenstates are given by $\phi_i$
which are defined as Eq.~(\ref{substitution1}) and (\ref{substitution2}).
Then the kinetic and KK mass term of $\phi_i$ are the same form as that of scalar field.
For $\phi_{i}$,  we write the contributions from one-loop diagrams to corrections of diagonal mass terms $\phi_i \phi_i$ such that 
\begin{align}
\label{OneLoopPhiSeparate}
& i \Theta^{(i)}(p^2;\mu; \ell m; \ell m') = i \Theta^{(i)}_{\rm bulk}(p^2;\mu; \ell m; \ell m') + i \Theta^{(i)}_{\rm bound}(p^2;\mu; \ell m; \ell m'), \\ 
\label{OneLoopPhiBulkBound}
& i \Theta^{(i)}_{\rm bulk (bound)}(p^2(p^2;\mu); \ell m; \ell m') = p^2  i \Theta^{(i)}_{\rm bulk(bound)}(p^2=0(p^2=0;\mu); \ell m; \ell m')  \nonumber \\
& \hspace{60mm} - i \tilde{\Theta}^{(i)}_{\rm bulk(bound)}(p^2=0(p^2=0;\mu); \ell m; \ell m'),
\end{align} 
where the bulk and the boundary contributions are separated, and each numerical factors can be calculated in the same manner as the above cases.
The RHS of Eq.~(\ref{OneLoopPhiBulkBound}) corresponds to the expansion in terms of $p^2$, and we omit $p^2=0$ in $\Theta(\tilde{\Theta})$ hereafter. 
We note that there are contributions to off-diagonal $\phi_1 \phi_2$ terms. 
However this contributions are negligibly small compared to diagonal terms, and we just ignore these contributions in this paper. 
The contributions to these coefficients $\Theta(\tilde{\Theta})$s from each diagrams are summarized in Table~\ref{OneLoopGex}, \ref{OneLoopGexNA11},  \ref{OneLoopGexNA22} and  \ref{OneLoopGexNA12}.
Thus the correction to the Lagrangian is 
\begin{align}
\label{1Loop-phi}
\delta L_{\rm 1 loop} =&  \Theta_{\rm bulk}^{(i)}(\ell m;\ell m) \partial_\mu \phi_{i \ell m} \partial^\mu \phi_{i \ell m} - \tilde{\Theta}_{\rm bulk}^{(i)}(\ell m;\ell m)  \phi_{i \ell m}  \phi_{i \ell m} 
\nonumber \\
& +  \Theta_{\rm bound}^{(i)}(\mu;\ell m;\ell m) \partial_\mu \phi_{i \ell m} \partial^\mu \phi_{i \ell m} - \tilde{\Theta}_{\rm bound}^{(i)}(\mu;\ell m;\ell m)  \phi_{i \ell m}  \phi_{i \ell m} 
\end{align}
where $\Theta$s in the RHS are understood as the sum of corresponding contributions from all diagrams.

Then we consider the renormalization condition to obtain the one-loop corrections to KK masses for $\phi_{1,2}$.
Defining as a renormalization of fields $\phi_i \rightarrow \sqrt{Z_3^{(i)}} \phi_i = \sqrt{1+\delta_3^{(i)}} \phi_i$, these terms become 
\begin{align}
& \partial_\mu \phi_{1 \ell m} \partial^\mu \phi_{1 \ell m}+\partial_\mu \phi_{2 \ell m} \partial^\mu \phi_{2 \ell m } - M_\ell^2 \phi_{1 \ell m} \phi_{1 \ell m}-  M_\ell^2 \phi_{2 \ell m} \phi_{2 \ell m} \nonumber \\
& + \delta_3^{(1)} \partial_\mu \phi_{1 \ell m} \partial^\mu \phi_{1 \ell m}+ \delta_3^{(2)} \partial_\mu \phi_{2 \ell m} \partial^\mu \phi_{2 \ell m } 
- \delta_3^{(1)} M_\ell^2 \phi_{1 \ell m} \phi_{1 \ell m}- \delta_3^{(2)}  M_\ell^2 \phi_{2 \ell m} \phi_{2 \ell m}
\end{align}
where we have taken up to the first order of $\delta_3^{(i)}$ corresponding to counter terms.
Thus the contribution from the counter terms is 
\begin{equation}
{\rm counter term}(p^2;\ell m) = i p^2 \delta_3^{(i)} -i M_{\ell}^2 \delta_3^{(i)}.
\end{equation}
Then we employ the renormalization condition at a cut-off scale $\Lambda$ as
\begin{equation}
(\Theta_{\rm bulk}(\ell m;\ell m) + \delta_3^{(i)}) \mid_{M_{\ell max}^2=\Lambda^2} = 0
\end{equation}
in the same way as four dimensional component case, and the counter term $\delta_3^{(i)}$ is determined by the condition.
Combining contributions from the one-loop diagrams and counter terms, we obtain the one-loop corrections to kinetic and KK mass terms as 
\begin{align}
\delta L = \delta_3^{(i)} \partial_\mu \phi_{i \ell m} \partial^\mu \phi_{i \ell m} - M_{\ell}^2 \delta_3^{(i)} \phi_{i \ell m}  \phi_{i \ell m} + \delta L_{\rm 1-loop}.
\end{align}
Therefore, normalizing the kinetic terms by 
\begin{equation}
\label{NormalizePhi}
\phi_{i \ell m} \rightarrow \left( 1 + \Theta_{\rm bulk}(\ell m;\ell m) + \Theta_{\rm bound}(\mu;\ell m;\ell m)+\delta_3^{(i)} \right)^{-\frac{1}{2}} \phi_{i \ell m},
\end{equation}
we obtain the one-loop corrections to KK mass 
\begin{align}
\label{mass-Gex}
\tilde{M}_{\ell m}^2 
=& M_\ell^2 + \tilde{\Theta}_{\rm bulk}(\ell m;\ell m)+\tilde{\Theta}_{\rm bound}(\mu;\ell m;\ell m)
- M_{\ell}^2 \big[ \Theta_{\rm bulk}(\ell m;\ell m)+\Theta_{\rm bound}(\mu;\ell m;\ell m) \big] \nonumber \\
\equiv & M_{\ell}^2 + (\delta M^2)^{\phi_i}_{\ell m},
\end{align}
for each KK modes of $\phi_i$, where we took first order of $\Theta$ and $\delta$ in RHS.

\subsection{Correction to KK masses of scalar boson}
\begin{figure}[t] 
\begin{minipage}{0.45 \hsize}
\begin{center}
\includegraphics[width=70mm]{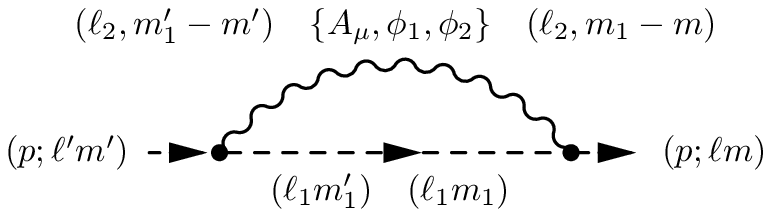} 
\caption{One loop diagram for correction to KK masses of Higgs bosons with virtual gauge bosons including extra components
\label{Loop11}}
\end{center}
\end{minipage}
\hspace{8mm}
\begin{minipage}{0.45 \hsize}
\begin{center}
\includegraphics[width=70mm]{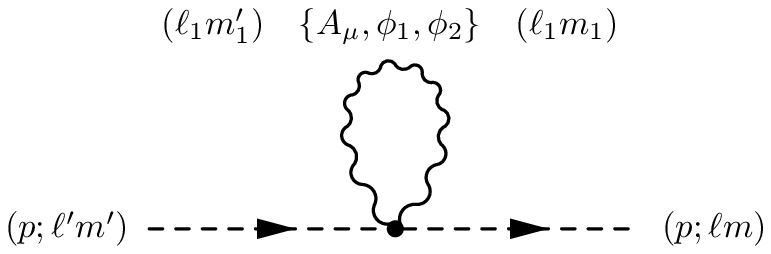}
\caption{One loop diagram for correction to KK masses of Higgs bosons with virtual gauge bosons including extra components.
\label{Loop10}}
\end{center}
\end{minipage} 
\end{figure}

\begin{figure}[t] 
\begin{minipage}{0.45 \hsize}
\begin{center}
\includegraphics[width=70mm]{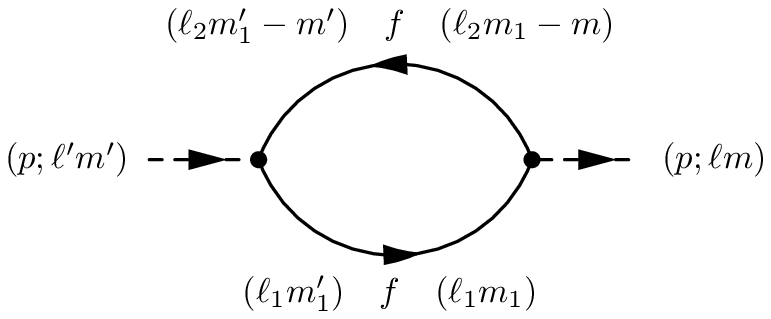} 
\caption{One loop diagram for correction to KK masses of Higgs bosons with virtual fermions.
\label{Loop9}}
\end{center}
\end{minipage}
\hspace{8mm}
\begin{minipage}{0.45 \hsize}
\begin{center}
\includegraphics[width=70mm]{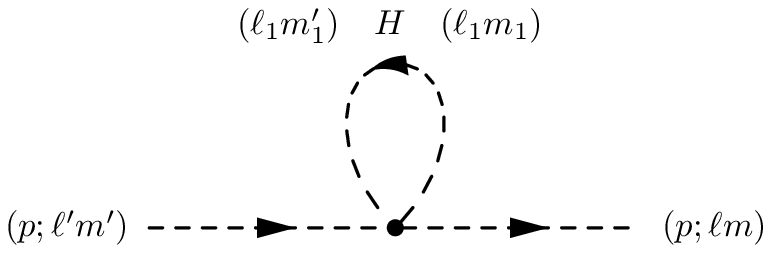}
\caption{One loop diagram for correction to KK masses of Higgs bosons with virtual Higgs bosons.
\label{Loop8}}
\end{center}
\end{minipage} 
\end{figure}

The one loop diagrams corresponding to correction to KK masses of Higgs boson are given as Figs.~\ref{Loop8}-\ref{Loop11} where corresponding quantum numbers $\{ \ell, m \}$ 
are shown for each external and internal lines. 
The Figs.~\ref{Loop11} and \ref{Loop10}
show the gauge boson loops including both four-dimensional components and extra-dimensional components from 
4-point and 3-point gauge interactions. 
The Fig.~\ref{Loop9} and \ref{Loop8} show the fermion and the Higgs boson loop contributions, respectively.
Quantum correction to the KK masses of scalar boson can be organized in
the
same way as that of $\phi_i$ such that 
\begin{align}
\label{OneLoopHSeparate}
& i \Xi(p^2;\mu; \ell m; \ell m') = i \Xi_{\rm bulk}(p^2; \ell m; \ell m') + i \Xi_{\rm bound}(p^2;\mu; \ell m; \ell m'), \\ 
\label{OneLoopHBulkBound}
& i \Xi_{\rm bulk (bound)}(p^2(p^2;\mu); \ell m; \ell m') = p^2  i \Xi_{\rm bulk(bound)}(p^2=0(p^2=0;\mu); \ell m; \ell m') \nonumber \\
& \hspace{60mm} - i \tilde{\Xi}_{\rm bulk(bound)}(p^2=0(p^2=0;\mu); \ell m; \ell m'),
\end{align}
where we list the contribution from each diagrams in Table~\ref{OneLoopS} and \ref{OneLoopS-F}.
The RHS of Eq.~(\ref{OneLoopHBulkBound}) corresponds to the expansion in terms of $p^2$ as in the previous case, and we omit $p^2=0$ in $\Xi(\tilde{\Xi})$ hereafter. 
Thus the correction to the Lagrangian is 
\begin{align}
\label{1Loop-H}
\delta L_{\rm 1 loop} =&  \Xi_{\rm bulk}^{(i)}(\ell m;\ell m) \partial_\mu H_{ \ell m}^\dagger \partial^\mu H_{\ell m} - \tilde{\Xi}_{\rm bulk}^{(i)}(\ell m;\ell m)  H_{ \ell m}^\dagger  H_{ \ell m} 
\nonumber \\
& +  \Xi_{\rm bound}^{(i)}(\mu;\ell m;\ell m) \partial_\mu H_{ \ell m}^\dagger \partial^\mu H_{i \ell m} - \tilde{\Xi}_{\rm bound}^{(i)}(\mu;\ell m;\ell m)  H_{ \ell m}^\dagger  H_{ \ell m} 
\end{align}
where $\Xi(\tilde{\Xi})$s in the RHS are understood as the sum of corresponding contributions from all diagrams.

The renormalization condition and one-loop corrections to KK mass are also discussed as in the case of $\phi_{i}$.
Repeating the same procedures as the case of $\phi_i$, we obtain counter terms
\begin{equation}
\delta_H \partial_\mu H_{ \ell m}^\dagger \partial^\mu H_{ \ell m} - M_{\ell}^2 \delta_H H_{ \ell m}^\dagger  H_{ \ell m},
\end{equation}
by the renormalization of field $H \rightarrow \sqrt{Z_H} H =\sqrt{1+\delta_H}$.
We then apply the renormalization condition 
\begin{equation}
(\Xi_{\rm bulk}(\ell m;\ell m)-\delta_H) \mid_{M_{\ell max}^2=\Lambda^2} = 0,
\end{equation}
at a cut-off scale $\Lambda$, requiring the canonical form of kinetic term at the cut-off scale.
Then we carry out the normalization of fields 
\begin{equation}
\label{NormalizeH}
H_{ \ell m} \rightarrow \left( 1 + \Xi_{\rm bulk}(\ell m;\ell m) + \Xi_{\rm bound}(\mu;\ell m;\ell m)+\delta_H \right)^{-\frac{1}{2}} H_{ \ell m},
\end{equation}
same as above cases.
Therefore we obtain the one-loop corrections to KK mass 
\begin{align}
\label{mass-H}
\tilde{M}_{\ell m}^2 =& M_\ell^2 + \tilde{\Xi}_{\rm bulk}(\ell m;\ell m)+\tilde{\Xi}_{\rm bound}(\mu;\ell m;\ell m) - M_\ell^2 \bigl[ \Xi_{\rm bulk}(\ell m;\ell m)+\Xi_{\rm bound}(\mu;\ell m;\ell m) \bigr] \nonumber \\
\equiv & M_\ell^2 + (\delta M^2)^H_{\ell m},
\end{align}
for each KK mode of scalar boson, where we have taken the first order in $\Xi$ and $\delta$ in RHS.

\section{KK mass spectrum and LKP \label{Sec:correctedMasses}}

In this section, we estimate the one-loop corrections to KK masses for the first KK modes of SM  fermions shown in Table~\ref{Fermions}, SM gauge bosons, 
and Higgs boson.
In estimating these corrections, we take the renormalization scale $\mu$ and the external momentum $p^2$ 
to be the first KK mass $\mu^2 = 2/R^2$ and $p^2 = -2/R^2$ respectively, and four-momentum cut-off scale $\Lambda$ as a parameter. 
Then angular momentum cut-off $\ell_{max}$ is taken to be the maximum integer satisfying $\sqrt{\ell_{max}(\ell_{max}+1)}/R \leq \Lambda$.
We specify the lightest KK particle which is expected to be the dark matter candidate.

To estimate the one-loop corrections to KK masses for SM fermions, we need to calculate coefficients $\Sigma(\tilde{\Sigma})$s in Eq.~(\ref{1Loop-F}) from one-loop diagrams.
For each fermions in Table~{\ref{Fermions}},  the coefficients $\Sigma(\tilde{\Sigma})$s obtain contributions from each diagrams such that
\begin{align}
&\Sigma (Q) = \Sigma^{QG_\mu}_{\rm Fig.\ref{Loop1}} + \Sigma^{QW_\mu}_{\rm Fig.\ref{Loop1}} + \Sigma^{ QB_\mu}_{\rm Fig.\ref{Loop1}} 
+ \sum_{i=1,2} \Bigl[ \Sigma^{QG_{i}}_{\rm Fig.\ref{Loop1}} + \Sigma^{QW_{i}}_{\rm Fig.\ref{Loop1}} + \Sigma^{ QB_{i}}_{\rm Fig.\ref{Loop1}} \Bigr] + \Sigma^{QH}_{\rm Fig.\ref{Loop2}} \\
&\Sigma (U) = \Sigma^{UG_\mu}_{\rm Fig.\ref{Loop1}} + \Sigma^{UB_\mu}_{\rm Fig.\ref{Loop1}} 
+ \sum_{i=1,2}  \Bigl[ \Sigma^{UG_{i}}_{\rm Fig.\ref{Loop1}} + \Sigma^{UB_{i}}_{\rm Fig.\ref{Loop1}} \Bigr] + \Sigma^{UH}_{\rm Fig.\ref{Loop2}} \\
&\Sigma (D) = \Sigma^{DG_\mu}_{\rm Fig.\ref{Loop1}} + \Sigma^{D B_\mu}_{\rm Fig.\ref{Loop1}} 
+ \sum_{i=1,2} \Bigl[ \Sigma^{DG_{i}}_{\rm Fig.\ref{Loop1}} + \Sigma^{D B_{i}}_{\rm Fig.\ref{Loop1}} \Bigr] + \Sigma^{DH}_{\rm Fig.\ref{Loop2}} \\
&\Sigma (L) =  \Sigma^{LW_\mu}_{\rm Fig.\ref{Loop1}} + \Sigma^{L B_\mu}_{\rm Fig.\ref{Loop1}} 
+ \sum_{i=1,2} \Bigl[ \Sigma^{LW_{i}}_{\rm Fig.\ref{Loop1}} + \Sigma^{L B_{i}}_{\rm Fig.\ref{Loop1}} \Bigr]  + \Sigma^{LH}_{\rm Fig.\ref{Loop2}}  \\
&\Sigma (E) =  \Sigma^{EB_\mu}_{\rm Fig.\ref{Loop1}} + \sum_{i=1,2} \Bigl[ \Sigma^{EB_{i}}_{\rm Fig.\ref{Loop1}} \Bigr] + \Sigma^{EH}_{\rm Fig.\ref{Loop2}}, 
\end{align}
where $\tilde{\Sigma}$s are given in the same way, and each terms in the RHS are the contributions from different loop diagrams with particles running in a loop shown as superscripts;
the $G_\mu$, $W_\mu$ and $B_\mu$ denote the SU(3), SU(2) and U(1)$_Y$ gauge fields respectively, 
the $G_i$, $W_i$ and $B_i$ denote corresponding extra dimensional components $\phi_i$, and the H denotes the Higgs boson.
Here we take into account contribution from Higgs-loop only for fermions
in the 3rd generations 
since the Yukawa couplings for the 1st and the 2nd generations are small. 
The contributions to each coefficients are obtained from Table~\ref{OneLoopF} applying corresponding gauge couplings $g_{6a}$ and $(T^a)^2$ factor 
regarding the particles inside a loop; $g_{6a=1,2,3}$ for U(1)$_Y$, SU(2) and SU(3) gauge coupling, and $(T^a)^2  = \{Y^2, C_2(N), C_2(N)  \}$ where $C_2(N)=(N^2-1)/2N$ for fundamental representation.
We also note that each gauge couplings, Yukawa couplings and the Higgs self coupling in six-dimensions $\{ g_{6a}, Y_{u,d,e}, \lambda_6 \}$ 
are related to that in four-dimensions as $g_a(y_{u,e,d}) = g_{6a}(Y_{u,d,e})/\sqrt{4 \pi R^2}$ and $\lambda = \lambda_6/(4 \pi R^2)$ 
since the factor $4\pi R^2$ comes from the surface area of $S^2$ and $1/\sqrt{4 \pi}$ factor comes from mode functions for zero mode.
%

\begin{figure}[tb] 
\includegraphics[width=80mm]{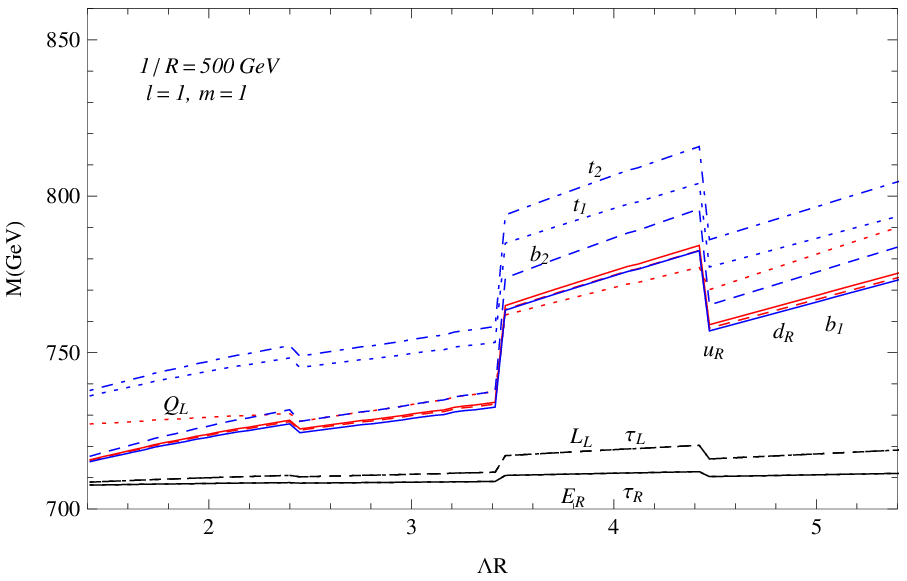} 
\includegraphics[width=80mm]{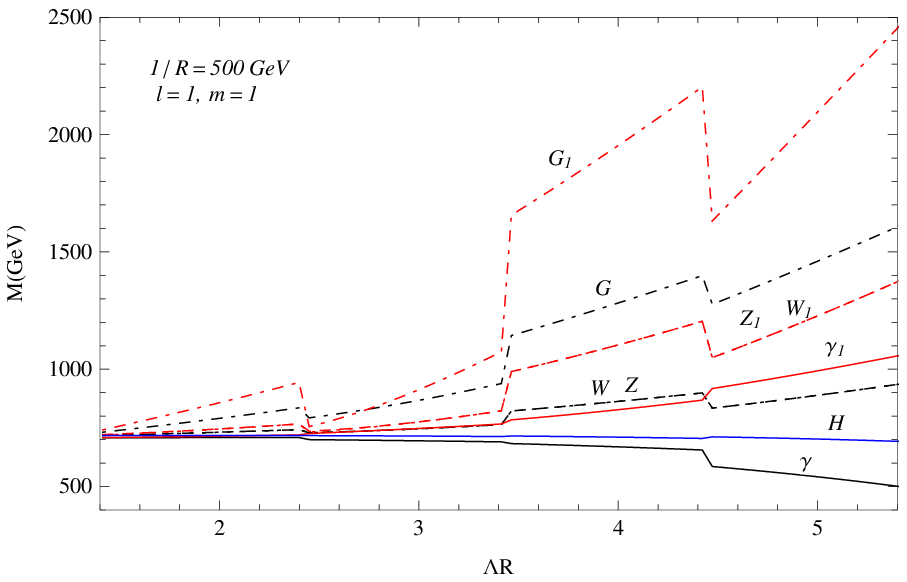} 
\caption{The KK masses for the first KK modes at one-loop level. 
The left panel shows the KK masses for fermions and the right panel shows that of gauge bosons as the function of $\Lambda R$.
The angular momentum cut-off $\ell_{max}$ is taken as the maximum integer satisfying $\sqrt{\ell_{max}(\ell_{max}+1)}/R  \leq \Lambda$.
\label{KKmass}} 
\end{figure}

The mass matrix for fermion Eq.~(\ref{mf-matrix}) is corrected by quantum correction as 
\begin{align}
L_{m_f} = 
\begin{pmatrix} \bar{\psi}_{\ell m }^F \\ \bar{\psi}_{\ell m }^f \end{pmatrix}^T 
\begin{pmatrix}
M_\ell + (\delta M)_{\ell m}^F  & m_{\rm SM}  -\frac{1}{2}  m_{\rm SM} \delta N_{\ell m; \Lambda} \\ m_{\rm SM} -\frac{1}{2} m_{\rm SM} \delta N_{\ell m; \Lambda} & - (M_\ell + (\delta M)_{\ell m}^f)
\end{pmatrix}
\begin{pmatrix} \psi_{\ell m }^F \\ \psi_{\ell m }^f \end{pmatrix}
\end{align}
where $ (\delta M)^{f(F)}_{\ell m}$ is given by Eq.~(\ref{mass-F}) and
indices $F(f)$ show corresponding fermion comes from SU(2) doublet(singlet).
Here we write one-loop correction to KK mass $(\delta M)^{f(F)}_{\ell m}$ as a product of $1/R$ and dimensionless numerical factor $\Delta_{\ell m; \Lambda }^{f(F)}$ 
which depends on $\{ \ell, m \}$ and $\Lambda$, such as 
\begin{equation}
\label{deltaM}
(\delta M)^{f(F)}_{\ell m} = \Delta_{\ell m; \Lambda }^{f(F)} \frac{1}{R},
\end{equation}
and we list the numerical factor $\Delta_{\ell m; \Lambda }^{f(F)}$ in Table~\ref{NumericalCoefficients} for the first KK mode.
The correction to off-diagonal components are induced by normalization of field Eq.~(\ref{normalize-f}) such that
\begin{align}
\label{deltaNF}
\delta N_{\ell m; \Lambda} = (\Sigma^L_{\rm bulk}+\Sigma^L_{\rm bound})^F(p^2;\mu;\ell m;\ell m) + \delta_L^F  + (\Sigma^R_{\rm bulk}+\Sigma^R_{\rm bound})^f(p^2;\mu;\ell m;\ell m) + \delta_R^f  ,
\end{align}
where we have taken the first order of $\{ \Sigma, \delta \}$, and the values of $\delta N_{\ell m;\Lambda}$ is shown in Table~\ref{DeltaN} for the first KK mode.
We thus obtain the KK mass eigenvalues 
\begin{align}
\tilde{m}_{\ell m}^{f_1} =&  \frac{1}{2} \Bigl[  (\delta M)^f_{\ell m} -(\delta M)^F_{\ell m} 
+ 2 \sqrt{M_{\ell}^2 + m_{SM}^2} \Bigl[ 1+ \frac{M_{\ell}((\delta M)^f_{\ell m} +(\delta M)^F_{\ell m}) -m_{SM} \delta N_{\ell m;\Lambda} }{M_{\ell}^2 + m_{SM}^2} \Bigr] \Bigr], \\
\tilde{m}_{\ell m}^{f_2} =& - \frac{1}{2} \Bigl[  (\delta M)^f_{\ell m} -(\delta M)^F_{\ell m} 
- 2 \sqrt{M_{\ell}^2 + m_{SM}^2} \Bigl[ 1+ \frac{M_{\ell}((\delta M)^f_{\ell m} +(\delta M)^F_{\ell m}) - m_{SM} \delta N_{\ell m;\Lambda} }{M_{\ell}^2 + m_{SM}^2} \Bigr] \Bigr],
\end{align}
for each KK modes of fermions, where we have taken the first order of $\{ \delta M, \delta N \}$ in the RHS. 
The one-loop corrections to masses of  the first KK modes for each fermion are shown in Table~\ref{correctedMasses} and left-panel of Fig.~\ref{KKmass}.

\renewcommand{\arraystretch}{1.3}
\begin{table}[t] 
\begin{center}
\begin{tabular}{|c||c|c|c|c|} \hline
$\ell_{max} $ & $2$ & $3$ & $4$ & $5$   \\ \hline \hline
$\Delta_{11,\Lambda}^{E_R}$ & 0.00232 & 0.00726 & 0.00650 & 0.0167 \\ \hline
$\Delta_{11,\Lambda}^{L_L}$ & 0.00635 & 0.0199 & 0.0178 & 0.0455 \\ \hline
$\Delta_{11,\Lambda}^{Q_L}$ & 0.0418 & 0.110 & 0.126 & 0.242 \\ \hline
$\Delta_{11,\Lambda}^{U_R}$ & 0.0370 & 0.116 & 0.104 & 0.265 \\ \hline
$\Delta_{11,\Lambda}^{D_R}$ & 0.0362 & 0.113 & 0.101 & 0.259 \\ \hline
$\Delta_{11,\Lambda}^{b_L}$ & 0.0418 & 0.133 & 0.116 & 0.304 \\ \hline
$\Delta_{11,\Lambda}^{b_R}$ & 0.0345 & 0.113 & 0.100 & 0.261 \\ \hline
$\Delta_{11,\Lambda}^{t_L}$ & 0.0434 & 0.136 & 0.120 & 0.310 \\ \hline
$\Delta_{11,\Lambda}^{t_R}$ & 0.0362 & 0.118 & 0.103 & 0.272 \\ \hline
\end{tabular} \vspace{-1ex}
\caption{The numerical coefficients $\Delta_{\ell m }^{f(F)}$ in Eq.~(\ref{deltaM}) for $\ell=1, |m|=1$ as a function of $\ell_{max}$, where $b_{L(R)}$ and $t_{L(R)}$ are 
KK b quark and KK top quark from the SU(2) doublet(singlet). \label{NumericalCoefficients}} \vspace{-1ex}
\end{center}
\end{table}
\renewcommand{\arraystretch}{1.3}
\begin{table}[t] 
\begin{center}
\begin{tabular}{|c||c|c|c|c|} \hline
$\ell_{max} $ & $2$ & $3$ & $4$ & $5$   \\ \hline \hline
$\Delta_{11,\Lambda}^{B_\mu}$ & $-0.0467$ & $-0.141$ & $-0.637$ & $-1.128$ \\ \hline
$\Delta_{11,\Lambda}^{W_\mu}$ & 0.100 & 0.677 & 0.758 & 2.57 \\ \hline
$\Delta_{11,\Lambda}^{G_\mu}$ & 0.513 & 3.23 & 4.52 & 13.6 \\ \hline
$\Delta_{11,\Lambda}^{B_1}$ & $0.120$ & $0.453$ & $1.36$ & $2.86$ \\ \hline
$\Delta_{11,\Lambda}^{W_1}$ & 0.118 & 1.89 & 2.38 & 10.2 \\ \hline
$\Delta_{11,\Lambda}^{G_1}$ & 0.283 & 8.98 & 8.65 & 46.6 \\ \hline
$\Delta_{11,\Lambda}^{H}$ & $-0.0064$ & $-0.0167$ & $-0.0362$ & $-0.0269$ \\ \hline
\end{tabular} \vspace{-1ex}
\caption{The numerical coefficients $\Delta_{\ell m }^{A_\mu,\phi_1}$ in Eq.~(\ref{DeltaMG}) and (\ref{deltaMPhi}) for $\ell=1, |m|=1$ as a function of $\ell_{max}$. 
\label{NumericalCoefficientsG}} \vspace{-1ex}
\end{center}
\end{table}
\renewcommand{\arraystretch}{1.3}
\begin{table}[t] 
\begin{center}
\begin{tabular}{|c||c|c|c|c|} \hline
$\ell_{max} $ & $2$ & $3$ & $4$ & $5$   \\ \hline \hline
$\delta N_{11; \Lambda}^{\tau}$ & 0.00237 & 0.00223 & 0.00400 & 0.00274 \\ \hline
$\delta N_{11; \Lambda}^{b}$ & 0.0229 & 0.0276 & 0.0397 & 0.0356 \\ \hline
$\delta N_{11;\Lambda}^{t}$ & 0.0244 & 0.0301 & 0.0432 & 0.0404 \\ \hline
$\delta N_{11;\Lambda}^{B_\mu}$ & $-0.0036$ & $-0.0098$ & $-0.0154$ & $-0.0250$ \\ \hline
$\delta N_{11;\Lambda}^{W_\mu}$ & $-0.0014$ & $-0.0067$ & $-0.0129$ & $-0.0240$ \\ \hline
$\delta N_{11;\Lambda}^{B_1}$ & $-0.00267$ & $-0.000777$ &$ -0.00622$ & $-0.000748$ \\ \hline
$\delta N_{11;\Lambda}^{W_1}$ & $-0.00532$ & $-0.00152$ & $-0.0124$ & $-0.00149$ \\ \hline
$\delta N_{11;\Lambda}^{H}$ & $-0.00364$ & $-0.00662$ & $-0.0151$ & $-0.0232$ \\ \hline
\end{tabular} \vspace{-1ex}
\caption{The numerical coefficients $\delta N_{\ell m; \Lambda }^{\psi,A,\phi_1,H}$ in Eq.~(\ref{deltaNF}), (\ref{deltaNA}), (\ref{deltaNPhi}) and (\ref{deltaNH}) 
for $\ell=1, |m|=1$ as a function of $\ell_{max}$. Here the coefficients of only third generation of fermions are shown for fermion. \label{DeltaN}} \vspace{-1ex}
\end{center}
\end{table}
To estimate the one-loop corrections to KK masses for four-dimensional gauge bosons, we need to calculate coefficients $\Pi$s in Eq.~(\ref{1Loop-A}) from one-loop diagrams.
For each gauge bosons, the coefficients $\Pi(\tilde{\Pi})$s obtain contributions from each diagrams such that
\begin{align}
\label{Correction-G}
\Pi (G_\mu) = & \sum_f \Pi^{ff}_{\rm Fig.\ref{Loop3}} +  \Pi^{HH}_{\rm Fig.\ref{Loop4}} +  \Pi^{H}_{\rm Fig.\ref{Loop5}} + \Pi^{G_\mu G_\mu}_{\rm Fig.\ref{Loop6}}+ \Pi^{ G_\mu}_{\rm Fig.\ref{Loop7}} \nonumber \\
&+ \sum_{i=1,2} \Pi^{G_\mu G_i}_{\rm Fig.\ref{Loop6}} + \sum_{i=1,2} \Pi^{ G_i}_{\rm Fig.\ref{Loop7}}+ \sum_{i,j=1,2} \Pi^{G_i G_j}_{\rm Fig.\ref{Loop6}} + \Pi^{c_G c_G}_{\rm Fig.\ref{Loop4}}, \\
\label{Correction-W}
\Pi (W_\mu) =& \sum_f \Pi^{ff}_{\rm Fig.\ref{Loop3}} +  \Pi^{HH}_{\rm Fig.\ref{Loop4}}+  \Pi^{H}_{\rm Fig.\ref{Loop5}} + \Pi^{W_\mu W_\mu}_{\rm Fig.\ref{Loop6}}+ \Pi^{ W_\mu}_{\rm Fig.\ref{Loop7}} \nonumber \\
&+ \sum_{i=1,2} \Pi^{W_\mu W_i}_{\rm Fig.\ref{Loop6}}+ \sum_{i=1,2} \Pi^{ W_i}_{\rm Fig.\ref{Loop7}}+ \sum_{i,j=1,2} \Pi^{W_i W_j}_{\rm Fig.\ref{Loop6}} + \Pi^{c_W c_W}_{\rm Fig.\ref{Loop4}}, \\
\label{Correction-B}
\Pi (B_\mu) =& \sum_f \Pi^{ff}_{\rm Fig.\ref{Loop3}} +  \Pi^{HH}_{\rm Fig.\ref{Loop4}}+  \Pi^{H}_{\rm Fig.\ref{Loop5}}, 
\end{align}
where $\tilde{\Pi}$s are given in the same way, each terms in the RHS are the contributions from different loop diagrams with particles running in a loop shown as superscripts,
and $c_{G,W}$ are ghost fields corresponding to $G_\mu$ and $W_\mu$ respectively.
The contributions to each coefficients are obtained from Table~\ref{OneLoopG} and \ref{OneLoopG-NA}. 
%
For a U(1)$_Y$ gauge field $B_\mu$ and a SU(2) gauge field $W^3_\mu$, the mass matrix  at one-loop level is given as 
\begin{equation}
\begin{pmatrix} B_{\ell m \mu} \\ W^3_{\ell m \mu} \end{pmatrix}^T
\begin{pmatrix}
M_\ell^2 + (\delta M^2 )_{\ell m}^B  + \frac{1}{4} g_1^2 v^2(1 + \delta N^B) & \frac{1}{4} g_1 g_2 v^2 \Bigl( 1 +  \frac{1}{2} \delta N^W + \frac{1}{2} \delta N^B \Bigr) 
\\ \frac{1}{4} g_1 g_2 v^2 \Bigl( 1 +  \frac{1}{2} \delta N^W + \frac{1}{2} \delta N^B \Bigr) & M_\ell^2 + (\delta M^2 )_{\ell m}^W + \frac{1}{4} g_1^2 v^2(1 + \delta N^W)
\end{pmatrix}
\begin{pmatrix} B^{\mu}_{\ell m} \\ W^{3 \mu}_{\ell m} \end{pmatrix}
\end{equation}
where $(\delta M^2 )_{\ell m}^B$ and $(\delta M^2 )_{\ell m}^W$ are given by Eq.~(\ref{mass-G}) with coefficients $\Pi(\tilde{\Pi})$s of Eqs.~(\ref{Correction-W}) and  (\ref{Correction-B}) for $B_\mu$ and $W_\mu^i$ respectively.
Here we write $(\delta M^2)^{A_\mu}_{\ell m}$ as a product of $1/R^2$ and numerical factor as in the fermion case, where $A_\mu = \{B_\mu, W_\mu, G_\mu \}$ , such as 
\begin{equation}
\label{DeltaMG}
(\delta M^2)^{A_\mu}_{\ell m} = \Delta_{\ell m; \Lambda }^{A_\mu} \frac{1}{R^2},
\end{equation}
and we list the numerical factors $\Delta_{\ell m; \Lambda }^{A_\mu}$ in
Table~\ref{NumericalCoefficients} for the first KK mode.
The correction to non-diagonal elements is induced by  renormalization of field Eq.~(\ref{NormalizeA}) such that
\begin{equation}
\label{deltaNA}
\delta N^{W_\mu,B_\mu}_{\ell m; \Lambda}=\Pi_{\rm bulk}(W,B)(p^2;\ell m;\ell m)+ \Pi_{\rm bound}(W,B)(p^2;\mu;\ell m;\ell m)-\delta_3(W,B)
\end{equation}
where we have taken up to the first order in $\Pi$s, and the values of
$\delta N^{W,B}$ are shown in Table~\ref{DeltaN} for the first KK mode. 
Diagonalizing the mass matrix, we obtain the KK masses of KK-photon and KK-Z boson at one-loop level such that
\begin{align}
\tilde{m}_{\gamma^\ell_\mu}^2 \simeq & M_\ell^2 +\frac{1}{2} m_Z^2 +\frac{1}{2} [ (\delta M^2)^B_{\ell m}+(\delta M^2)^W_{\ell m}  -(m_Z^2 -m_W^2) \delta N^{B^\mu}_{\ell m; \Lambda} - m_W^2 \delta N^{W^\mu}_{\ell m; \Lambda} ]  \nonumber \\ 
&- \frac{1}{2} \sqrt{((\delta M^2)^B_{\ell m}- (\delta M^2)^W_{\ell m} )^2 + m_Z^4}, \\
\tilde{m}_{Z^\ell_\mu}^2 \simeq & M_\ell^2 +\frac{1}{2} m_Z^2 +\frac{1}{2} [ (\delta M^2)^B_{\ell m}+(\delta M^2)^W_{\ell m}  -(m_Z^2 -m_W^2) \delta N^{B^\mu}_{\ell m; \Lambda} - m_W^2 \delta N^{W^\mu}_{\ell m; \Lambda} ]  \nonumber \\ 
&+ \frac{1}{2} \sqrt{((\delta M^2)^B_{\ell m}- (\delta M^2)^W_{\ell m} )^2 + m_Z^4}, 
\end{align} 
where we ignored terms proportional to $\delta N_G^{B,W}$ in the square root in the second line of RHS for both equation due to the smallness of the factor.
For KK modes of $W^{\pm}$ bosons and Gluon, the one-loop corrections to KK masses are 
\begin{align}
&(\tilde{m}_\ell^{W^\pm_\mu})^2= M_{\ell}^2 + m_W^2 + (\delta M^2)^{W}_{\ell m} + m_W^2 \delta N^{W^\mu}_{\ell m; \Lambda},  \\
&(\tilde{m}_\ell^{G})^2 =M_\ell^2 + (\delta M^2)^G_{\ell m}, 
\end{align}
where $( \delta M^2)^G_{\ell m}$ is given by Eq.~(\ref{mass-G}) with coefficients Eq.~(\ref{Correction-G}).
The corrected masses of  the first KK modes for each gauge bosons are shown in the Table~\ref{correctedMasses} and the right-panel of Fig.~\ref{KKmass}.

\begin{table}[t] 
\begin{center}
\begin{tabular}{|c||c|c|c|c|} \hline
$\ell_{max} $ & $2$ & $3$ & $4$ & $5$  \\ \hline \hline
$\tilde{m}_{11}(E_R)$ (GeV) & 708.3 & 710.7 & 710.4   & 715.4 \\ \hline 
$\tilde{m}_{11}(L_L)$ (GeV) & 710.3 & 717.1 & 716.0 & 729.9 \\ \hline
 $\tilde{m}_{11}(\tau_1)$ (GeV) & 708.3 & 710.7 & 710.4 & 715.4 \\ \hline  
  $\tilde{m}_{11}(\tau_2)$ (GeV) & 710.3 & 717.1 & 716.0 & 729.9 \\ \hline 
  $\tilde{m}_{11}(d_R)$ (GeV) & 725.2 & 763.8.4 & 757.8 & 836.8 \\ \hline 
  $\tilde{m}_{11}(u_R)$ (GeV) & 725.6 & 765.0 & 758.9 & 839.6 \\ \hline 
  $\tilde{m}_{11}(Q_L)$ (GeV) & 728.0 & 762.1 & 770.1 & 827.9 \\ \hline 
  $\tilde{m}_{11}(b_1)$ (GeV) & 724.4 & 763.6 & 756.9 & 837.7 \\ \hline 
 $\tilde{m}_{11}(b_2)$ (GeV) & 728.0 & 773.9 & 765.2 & 859.3 \\ \hline 
  $\tilde{m}_{11}(t_1)$ (GeV) & 745.2 & 785.0 & 777.4 & 860.1 \\ \hline
  $\tilde{m}_{11}(t_2)$ (GeV) & 748.9 & 794.0 & 786.1 & 879.2  \\ \hline  
   $\tilde{m}_{11}(\gamma)$ (GeV) & 701.4 & 684.7 & 587.2 & 428.4 \\ \hline 
      $\tilde{m}_{11}(Z)$ (GeV) & 727.7 & 820.7 & 833.0 & 1071. \\ \hline 
   $\tilde{m}_{11}(W)$ (GeV) & 729.0 & 822.1 & 834.2 & 1072. \\ \hline 
   $\tilde{m}_{11}(G)$ (GeV) & 792.8 & 1144. & 1276. & 1973. \\ \hline 
   $\tilde{m}_{11}(\gamma_1)$ (GeV) & 727.9 & 785.7 & 919.0 & 1105.  \\ \hline
  $\tilde{m}_{11}(Z_1)$ (GeV) & 733.6 & 988.9 & 1049. & 1746.  \\ \hline
   $\tilde{m}_{11}(W_1)$ (GeV) & 732.3 & 990.0 & 1050. & 1746. \\ \hline
    $\tilde{m}_{11}(G_1)$ (GeV) & 755.4 & 1656. & 1632. & 3485. \\ \hline 
   $\tilde{m}_{11}(H)$ (GeV) & 717.1 & 715.3 & 711.7 & 713.3 \\ \hline
\end{tabular} \vspace{-1ex}
\caption{The one-loop correction the masses of the first KK modes ($\ell=1,|m|=1$) for each cut off scale where $\ell_{max} =2,3,4,5$ are taken as reference values. 
Here we have taken the inverse of $S^2/Z_2$ radius as $1/R=500$ GeV, where the tree-level first KK mass corresponds to $\sqrt{2}/R \simeq 707$ GeV. \label{correctedMasses}} \vspace{-1ex}
\end{center}
\end{table}

To estimate the one-loop corrections to KK masses for physical extra-dimensional components of gauge fields $\phi_1$, we need to calculate coefficients $\Theta(\tilde{\Theta})$s in Eq.~(\ref{1Loop-phi}) from one-loop diagrams.
For each gauge bosons, the coefficients $\Theta(\tilde{\Theta})$s obtain contributions from each diagrams such that
\begin{align}
\label{Correction-Gi}
\Theta (G_{1}) = & \sum_f \Theta^{(1)ff}_{\rm Fig.\ref{Loop3}} +  \Theta^{(1)HH}_{\rm Fig.\ref{Loop4}}+  \Theta^{(1)H}_{\rm Fig.\ref{Loop5}} + \Theta^{(1)G_\mu G_\mu}_{\rm Fig.\ref{Loop6}}
+ \Theta^{(1) G_\mu}_{\rm Fig.\ref{Loop7}} \nonumber \\
&+ \sum_{a=1,2} \Theta^{(1) G_\mu G_a}_{\rm Fig.\ref{Loop6}}+ \sum_{a=1,2} \Theta^{(1) G_a}_{\rm Fig.\ref{Loop7}}+ \sum_{a,b=1,2} \Theta^{(1) G_a G_b}_{\rm Fig.\ref{Loop6}} + \Theta^{(1) c_G c_G}_{\rm Fig.\ref{Loop4}}, \\
\label{Correction-Wi}
\Theta (W_{1}) =& \sum_f \Theta^{(1)ff}_{\rm Fig.\ref{Loop3}} +  \Theta^{(1)HH}_{\rm Fig.\ref{Loop4}}+  \Theta^{(1)H}_{\rm Fig.\ref{Loop5}} 
+ \Theta^{(1)W_\mu W_\mu}_{\rm Fig.\ref{Loop6}}+ \Theta^{(1) W_\mu}_{\rm Fig.\ref{Loop7}} \nonumber \\
&+ \sum_{a=1,2} \Theta^{(1)W_\mu W_a}_{\rm Fig.\ref{Loop6}}+ \sum_{a=1,2} \Theta^{(1)W_a}_{\rm Fig.\ref{Loop7}}+ \sum_{a,b=1,2} \Theta^{(1)W_aW_b}_{\rm Fig.\ref{Loop6}}+\Theta^{(1) c_Wc_W}_{\rm Fig.\ref{Loop4}}, \\
\label{Correction-Bi}
\Theta (B_{1}) = &\sum_f \Theta^{(1)ff}_{\rm Fig.\ref{Loop3}} +  \Theta^{(1)HH}_{\rm Fig.\ref{Loop4}}+  \Theta^{(1)H}_{\rm Fig.\ref{Loop5}}, 
\end{align}
where $\tilde{\Theta}$s are given in the same way, and each terms in the RHS are the contributions from different loop diagrams with particles running in a loop shown as superscripts.
The contributions to each coefficients are obtained from Table~\ref{OneLoopGex} and \ref{OneLoopGexNA11}. 
%
The one-loop corrections to KK masses for $\phi_1$ are similarly obtained as four dimensional case.
For $B_{1}$ and $W^3_{1}$, the corrected mass matrix is given as
\begin{equation}
\begin{pmatrix} B_{1 \ell m} \\ W^3_{1 \ell m } \end{pmatrix}^T
\begin{pmatrix}
M_\ell^2 + (\delta M^2 )_{\ell m}^{B_1}  + \frac{1}{4} g_1^2 v^2(1 + \delta N^{B_1}_{\ell m; \Lambda}) & \frac{1}{4} g_1 g_2 v^2 \Bigl( 1 +  \frac{1}{2} \delta N^{W_1}_{\ell m;\Lambda} + \frac{1}{2} \delta N^{B_1}_{\ell m;\Lambda} \Bigr) 
\\ \frac{1}{4} g_1 g_2 v^2 \Bigl( 1 +  \frac{1}{2} \delta N^{W_1}_{\ell m;\Lambda} + \frac{1}{2} \delta N^{B_1}_{\ell m;\Lambda} \Bigr) & M_\ell^2 + (\delta M^2 )_{\ell m}^{W_1}  + \frac{1}{4} g_1^2 v^2(1 + \delta N^{W_1}_{\ell m;\Lambda})
\end{pmatrix}
\begin{pmatrix} B_{1 \ell m} \\ W^{3}_{1 \ell m} \end{pmatrix}
\end{equation}
where $\tilde{M}_{B_{1 \ell m}}^2$ and $\tilde{M}_{W_{ 1 \ell m}}^2$ are given by Eq.~(\ref{mass-Gex}) with coefficients in Eqs.~(\ref{Correction-Wi}) and (\ref{Correction-Bi}) for $B_{i}$ and $W_i$ respectively.
Here we write $(\delta M^2)^{\phi_1}_{\ell m}$ in the same form as Eq.~(\ref{DeltaMG}), where $\phi_1=\{B_1, W_1, G_1 \}$, as 
\begin{equation}
\label{deltaMPhi}
(\delta M^2)^{\phi_1}_{\ell m} = \Delta_{\ell m }^{\phi_1} \frac{1}{R^2},
\end{equation}
and we list the numerical factor $\Delta_{\ell m }^{\phi_1}$ in Table~\ref{NumericalCoefficients} for the first KK mode.
The correction to off-diagonal elements is induced by the renormalization of field Eq.~(\ref{NormalizePhi}) such that
\begin{equation}
\label{deltaNPhi}
\delta N_{\ell m;\Lambda}^{W_1,B_1} = -\Theta_{\rm bulk}(W,B)(\ell m;\ell m)- \Theta_{\rm bound} (W,B)(\mu;\ell m;\ell m) - \delta_3^{(1)}(W,B)
\end{equation}
where we have taken up to the first order of $\Theta(\tilde{\Theta})$, and the values of $\delta N^{W_1,B_1}_{\ell m;\Lambda}$ are shown in Table~\ref{DeltaN} for the first KK mode.
Diagonalizing the mass matrix, we obtain the one-loop corrections to KK masses of extra dimensional components KK-photon and KK-Z boson such that
\begin{align}
\tilde{m}_{\gamma^\ell_{1}}^2 \simeq & M_\ell^2 +\frac{1}{2} m_Z^2 +\frac{1}{2} [ (\delta M^2)^{B_1}_{\ell m}+(\delta M^2)^{W_1}_{\ell m}  -(m_Z^2 -m_W^2) \delta N^{B_1}_{\ell m;\Lambda} - m_W^2 \delta N^{W_1}_{\ell m;\Lambda} ] \nonumber \\ 
&- \frac{1}{2} \sqrt{((\delta M^2)^{B_1}_{\ell m}- (\delta M^2)^W_{\ell m} )^2 + m_Z^4}, \\
\tilde{m}_{Z^\ell_{1}}^2 \simeq & M_\ell^2 +\frac{1}{2} m_Z^2 +\frac{1}{2} [ (\delta M^2)^{B_1}_{\ell m}+(\delta M^2)^{W_1}_{\ell m}  -(m_Z^2 -m_W^2) \delta N^{B_1}_{\ell m;\Lambda} - m_W^2 \delta N^{W_1}_{\ell m; \Lambda} ]  \nonumber \\ 
&+ \frac{1}{2} \sqrt{((\delta M^2)^{B_1}_{\ell m}- (\delta M^2)^{W_1}_{\ell m} )^2 + m_Z^4},
\end{align} 
where we ignored the terms proportional to $\delta N^{B,W}_{\phi}$ in square root in the second line of RHS for both equation due to the smallness of the factor.
For KK modes of $W^{\pm}$ bosons and Gluon, the one-loop corrections to KK masses are 
\begin{align}
&(\tilde{m}_\ell^{W^\pm_{1}})^2= M_{\ell}^2 + m_W^2 + (\delta M^2)^{W_{1}}_{\ell m} + m_W^2 \delta N^{W_1}_{\ell m; \Lambda},  \\
&(\tilde{m}_\ell^{G_{1}})^2 =M_\ell^2 + (\delta M^2 )^{G_{1}}_{\ell m}, 
\end{align}
where $(\delta M^2)^{G_1}_{ \ell m} $ is given by Eq.~(\ref{mass-Gex}) with coefficients in Eq.~(\ref{Correction-Gi}).
The one-loop corrections to masses of the first KK modes for each $\phi_1$ are shown in Table~\ref{correctedMasses} and the right-panel of Fig.~\ref{KKmass}.

To estimate the one-loop corrections to KK masses for SM Higgs boson, we need to calculate coefficients $\Xi(\tilde{\Xi})$ in Eq.~(\ref{1Loop-H}) from one-loop diagrams.
The coefficients $\Xi(\tilde{\Xi})$ obtain contributions from each diagrams such that
\begin{align}
\Xi(H)  =&  \sum_f \Xi^{ff}_{\rm Fig.\ref{Loop9}} +  \Xi^{H}_{\rm Fig.\ref{Loop8}} + \Xi^{W_\mu H}_{\rm Fig.\ref{Loop11}}+ \Xi^{B_\mu H}_{\rm Fig.\ref{Loop11}}
+ \Xi^{W_\mu}_{\rm Fig.\ref{Loop10}}+ \Xi^{B_\mu}_{\rm Fig.\ref{Loop10}} \nonumber \\
&+ \sum_{i} \Xi^{ W_i}_{\rm Fig.\ref{Loop10}}+ \sum_{i} \Xi^{ B_i}_{\rm Fig.\ref{Loop10}}+ \sum_{i} \Xi^{ W_i H}_{\rm Fig.\ref{Loop11}} + \sum_i \Xi^{ B_i H}_{\rm Fig.\ref{Loop11}}  
\end{align}
where $\tilde{\Xi}$ is given in the same way, and each terms in the RHS are the contributions from different loop diagrams with particles propagating inside a loop shown as  superscripts.
The one-loop corrections to KK masses for the KK Higgs boson are also obtained in a similar way such as 
\begin{equation}
(\tilde{m}_\ell^{H} )^2= M_{\ell}^2+ m_h^2 + (\delta M^2)^H_{\ell m} +  m_h^2  \delta N^H_{\ell m;\Lambda}
\end{equation}
where the fourth term in RHS is induced by normalization of field Eq.~(\ref{NormalizeH}) such that 
\begin{equation}
\label{deltaNH}
\delta N^H_{\ell m;\Lambda} = -\Xi_{\rm bulk}(\ell m;\ell m)-\Xi_{\rm bound}(\mu;\ell m;\ell m) - \delta_3^H
\end{equation}
taking up to the first order of $\Xi$s, and the values of $\delta N^{H}$ are shown in Table~\ref{DeltaN}.
The $(\delta M^2)^{H}_{\ell m}$ is also written as Eq.~(\ref{DeltaMG}) such that 
\begin{equation}
(\delta M^2)^{H}_{\ell m} = \Delta_{\ell m; \Lambda }^{H} \frac{1}{R^2},
\end{equation}
and we list the numerical factor $\Delta_{\ell m; \Lambda }^{H}$ in Table~\ref{NumericalCoefficients}.
Also the one-loop corrections to masses of the first KK modes for Higgs bosons are shown in Table~\ref{correctedMasses} and the right-panel of Fig.~\ref{KKmass}.

The Table~\ref{correctedMasses} shows the one-loop corrections to KK mass of the first KK modes $\ell=1, |m|=1$ for each SM particles where we take cut off $\ell_{max}=2,3,4,5$ as reference values.
From the Table we can see that the first KK photon has the lightest KK mass for any cut-off scale.
The first KK photon thus is a promising candidate of the dark matter in the model.
The smallest correction to the KK photon mass is due to the negative contributions from fermion loops while the non-Abelian gauge bosons obtain large contribution from the loops associated with 
self-interactions.
We also find that negative contribution from fermion loops to KK masses
of $\phi_1$ tends to be less effective compared with the case of four-dimensional components, 
and the one-loop corrections to the KK masses become larger than that of four-dimensional components.
We show the cut-off scale dependence of the first KK masses in the Fig.~\ref{KKmass}, 
where we ignored the SM masses of $e, \mu, u, d, s, c$ and write as $E_R, L, u_R, d_R, Q_L$ for SU(2) singlet charged lepton, doublet lepton, singlet quarks and doublet quarks respectively.
%
%
We also note that zig-zag shaped lines appear in Fig.~\ref{KKmass}  
where the discontinuity of the plot corresponds to increase of $\ell_{max}$ at $\Lambda R = \sqrt{\ell_{max} (\ell_{max}+1) }$ and 
the corrected masses tend to be small for even $\ell_{\rm max}$ compared to odd one.
This tendency is due to the sign factor of $(-1)^{ \{ \ell, m \} }$ which appears from propagator and vertices on $S^2/Z_2$ and 
gives different sign contribution from each KK modes in the loops as mentioned below Eq.~(\ref{SigmaFig1d}).
Since number of negative and positive sign contributions is different for even or odd $\ell_{\rm max}$ case,
the corrected mass depend on $\ell_{\rm max}$ differently for even or odd case.
%
%
The quantum corrections tend to be large for the large cut-off scale since the six dimensional loop expansion parameter $\epsilon=\frac{g_a^2}{32 \pi^2} (R \Lambda)^2$ becomes large. 
This indicates that the perturbation is not reliable and we cannot take a large cut-off scale, which is a generic feature of non-renormalizable theories.

\section{Summary and discussions \label{summary}}
We have investigated the quantum correction of Kaluza-Klein masses in the $S^2/Z_2$ universal extra dimensional model in order to determine the lightest KK particle 
which would be the dark matter candidate.

We first derived the Feynman rules for propagators and vertices on the six dimensional spacetime $M^4 \times S^2/Z^2$.
%
%
%
%
We then calculated the one-loop diagrams relevant to one-loop corrections to the KK masses with the Feynman rules.
The contributions from the one-loop diagrams can be separated to the bulk contribution conserving KK number $m$ and the boundary contribution violating the KK number conservation, 
which is the similar structure as in the minimal UED case.
In calculating these diagrams, we introduced the cut-off for both 4-momentum integration and angular momentum sum inside a loop by defining 
the 4-momentum cut-off scale $\Lambda$ and the angular momentum number cut-off  $\ell_{max}$.
Thus the bulk contribution is estimated with cut-off scale $\Lambda$ as an upper bound of momentum integration while 
the boundary contribution is estimated by taking leading log-divergent part assuming vanishing contribution at the cut-off scale.
The contributions from the one loop diagrams are then organized as the form of Eq.~(\ref{1Loop-F}), (\ref{1Loop-A}), (\ref{1Loop-phi}) and (\ref{1Loop-H}) for 
fermions, gauge bosons, extra dimensional components gauge bosons and scalar bosons respectively, and the explicit form of the contributions are summarized 
in the Table~\ref{OneLoopF}, \ref{OneLoopG}, \ref{OneLoopG-NA}, \ref{OneLoopGex}, \ref{OneLoopGexNA11}, \ref{OneLoopGexNA12}, \ref{OneLoopS} and \ref{OneLoopS-F}.
Taking into account the renormalization conditions, we finally obtained the formula to derive the corrected KK masses as
Eq.~(\ref{mass-F}), (\ref{mass-G}), (\ref{mass-Gex}) and (\ref{mass-H}) for fermions, gauge bosons, extra dimensional components gauge bosons and scalar bosons respectively.

At last we estimated the corrected KK masses of first KK mode ($\ell=1,|m|=1$) for each SM particles applying the formulas in Section~\ref{Sec:OneLoop}.
The resulting corrected masses, taking $1/R=500$GeV as a reference value,  are summarized in Table~\ref{correctedMasses} for each cut-off.
We then find that the lightest KK particle is the 1st KK photon $\gamma^{11}_\mu$ as in the minimal UED case, and it can be identified as the promising DM candidate.
We have also shown the numerical plot of the one-loop correction to the first KK modes as a function of cut-off scale $\Lambda$ in the Fig.~\ref{KKmass}. 
We have found that the corrected masses tend to be small for even $\ell_{max}$ compared to odd one due to the sign factor $(-1)^{\{ \ell, m\}}$ which appears in loop diagram calculation.
The quantum correction tend to be large for higher cut-off since the
number of modes inside a loop becomes very large, which indicates that we can not take cut-off scale too large.

It would be very interesting to study the relic density of the dark matter candidate, $\gamma^{11}_\mu$, with the corrected mass spectrum, 
which will give constraints for the $S^2/Z_2$ radius $R$ and the cut-off scale $\Lambda$.
Furthermore the one-loop corrections to the mass spectrum would lead a prediction of the experimental signature of our model and the analysis is left for future work.

\acknowledgments
The work of N.M. was supported in part by the Grant-in-Aid for the Ministry of Education, Culture, Sports, Science, and Technology, Government of Japan (No. 24540283).
The work of T.N. was supported in part by NSC Grant No. 102-2811-M-006-035.
The work of J.S. was supported in part by the Grant-in-Aid for the Ministry of Education, Culture, Sports, Science, and Technology, Government of Japan (Nos. 24340044 and 25105009).
%

\appendix

%




\section{Summary lists of vertices in our model \label{vertex-list}}
Here we summarize the Feynman rules for vertices in our model.  
The vertices including fermions are derived from gauge interactions and Yukawa interactions
\begin{align}
& \int dx^4 d \Omega g_{6a}  \bar{\Psi}_{\pm} \Gamma^M T^a_i A_M^i \Psi_\pm, \\
& \int dx^4 d \Omega Y_f \bar{\Psi}_\pm H \Psi_\mp
\end{align}
where $A_{\theta, \phi}$ in Eqs.~(\ref{substitution1}) and (\ref{substitution2}) are substituted.
The Feynman rules for these vertices are listed in Table~\ref{vertices-F} where the non-trivial factors in these vertices are given by 
integrations of mode function as Eq.~(\ref{vertex-ex}) for $I^{\alpha(\beta)}$, and 
\begin{align}
& C^\alpha_{\ell_1 m_1; \ell_2 m_2; \ell_3 m_3} = \int d \Omega \tilde{\alpha}^*_{\ell_1 m_1} \left( \partial_\theta - \frac{i}{\sin \theta} \partial_\phi \right) \tilde{Y}_{\ell_2 m_2} \tilde{\beta}_{\ell_3 m_3}, \\
& C^\beta_{\ell_1 m_1; \ell_2 m_2; \ell_3 m_3} = - \int d \Omega \tilde{\beta}^*_{\ell_1 m_1} \left( \partial_\theta + \frac{i}{\sin \theta} \partial_\phi \right) \tilde{Y}_{\ell_2 m_2} \tilde{\alpha}_{\ell_3 m_3}. 
\end{align}
%
\renewcommand{\arraystretch}{1.3}
%
\begin{table}[bt] \vspace{1ex}
\begin{tabular}{c||c} 
{\scriptsize Interaction} & {\scriptsize vertex factor} \\ \hline
{\scriptsize $\Psi_{\pm \ell_1 m_1} A_{\mu \ell_2 m_2} \Psi_{\pm \ell_3 m_3}$ }
& \scriptsize{ $i \frac{g_{6a}}{R} T^a \gamma^\mu \left[ I^\alpha_{\ell_1 m_1; \ell_2 m_2; \ell_3 m_3} P_{R(L)} + I^\beta_{\ell_1 m_1; \ell_2 m_2; \ell_3 m_3} P_{L(R)} \right]$ }\\
{\scriptsize $\Psi_{\pm \ell_1 m_1} \phi_{1 \ell_2 m_2} \Psi_{\pm \ell_3 m_3}$  }
& \scriptsize{ $i \frac{g_{6a}}{R} T^a \gamma_5 \left[ C^\alpha_{\ell_1 m_1; \ell_2 m_2; \ell_3 m_3 } P_{L(R)} + C^\beta_{\ell_1 m_1; \ell_2 m_2; \ell_3 m_3} P_{R(L)} \right]$} \\
{\scriptsize $\Psi_{\pm \ell_1 m_1} \phi_{2 \ell_2 m_2} \Psi_{\pm \ell_3 m_3}$  }
& {\scriptsize $i \frac{g_{6a}}{R} T^a\gamma_5 \left[ i C^\alpha_{\ell_1 m_1; \ell_2 m_2; \ell_3 m_3} P_{L(R)} - i C^\beta_{ \ell_1 m_1; \ell_2 m_2; \ell_3 m_3} P_{R(L)} \right]$ } \\
{\scriptsize $\Psi_{\mp \ell_1 m_1} H_{ \ell_2 m_2} \Psi_{\pm \ell_3 m_3}$  }
& {\scriptsize $ i \frac{Y_{f}}{R} \left[ I^\alpha_{\ell_1 m_1; \ell_2 m_2; \ell_3 m_3} P_{R(L)} + I^\beta_{\ell_1 m_1; \ell_2 m_2; \ell_3 m_3} P_{L(R)} \right] $ } \\ \hline
\end{tabular} \vspace{-1ex}
\caption{The vertex factors for interactions including fermions where $Y_f = Y_{u,d,e}$.\label{vertices-F}} \vspace{-1ex}
\end{table}
%
\begin{table}[t] \vspace{1ex}
\begin{tabular}{c||c} 
{\scriptsize Interaction} & {\scriptsize vertex factor} \\ \hline
{\scriptsize $ H_{\ell_1 m_1}^\dagger(p_1) A_{\mu \ell_2 m_2} H_{\ell_3 m_3}(p_2)$ }
& {\scriptsize $i \frac{g_{6a}}{R} T^a (p_1^\mu + p_2^\mu) J^1_{\underline{\ell_1 m_1}; \ell_2 m_2; \ell_3 m_3}$ } \\
{\scriptsize $ H_{\ell_1 m_1}^\dagger(p_1) \phi_{1 \ell_2 m_2} H_{\ell_3 m_3}(p_2)$ }
& {\scriptsize $- \frac{g_{6a}}{R} T^a \big[ J^2_{\underline{\ell_1 m_1}; \ell_2 m_2; \ell_3 m_3}-J^2_{ \ell_3 m_3; \ell_2 m_2;\underline{\ell_1 m_1} } \big]$ }\\
{\scriptsize $ H_{\ell_1 m_1}^\dagger(p_1) \phi_{2 \ell_2 m_2} H_{\ell_3 m_3}(p_2)$ }
&{\scriptsize $- \frac{g_{6a}}{R} T^a \big[ J^3_{ \underline{\ell_1 m_1}; \ell_2 m_2; \ell_3 m_3 } - J^3_{ \ell_3 m_3; \ell_2 m_2;\underline{\ell_1 m_1} } \big] $ } \\
{\scriptsize $A_{\ell_1 m_1}^\mu A_{\ell_2 m_2}^\nu H^\dagger_{\ell_3 m_3} H_{\ell_4 m_4}$ }
& {\scriptsize $2 i \frac{g_{6a}^2}{R^2} T^a_i T^a_j g_{\mu \nu} K^1_{ \ell_1 m_1; \ell_2 m_2; \underline{\ell_3 m_3}; \ell_4 m_4 }$ }\\
{\scriptsize  $\phi_{1(2) \ell_1 m_1} \phi_{1(2) \ell_2 m_2} H^\dagger_{\ell_3 m_3} H_{\ell_4 m_4}$ }
 &{\scriptsize $-2 i \frac{g_{6a}^2}{R^2} T^a_i T^a_j K^2_{ \ell_1 m_1; \ell_2 m_2; \underline{\ell_3 m_3}; \ell_4 m_4} $ }\\
{\scriptsize  $H_{\ell_1 m_1} H_{\ell_2 m_2} H_{\ell_3 m_3} H_{\ell_4 m_4}$ }
 & {\scriptsize $-  i \frac{\lambda_6}{R^2}  g_{\mu \nu} K^1_{ \ell_1 m_1; \ell_2 m_2; \ell_3 m_3; \ell_4 m_4 }$ } \\ \hline
\end{tabular} \vspace{-1ex}
\caption{The vertex factors for interactions including Higgs bosons.\label{vertices-H}} \vspace{-1ex}
\end{table}
%
%
The vertices for self-interactions of gauge boson are derived from the terms in the Lagrangian
\begin{align}
\label{Self-interaction}
\int dx^4 d \Omega \left[ -\frac{1}{4} F^i_{\mu \nu} F^{i \mu \nu} + \frac{1}{2 R^2} F_{\theta \mu}^i F_\theta^{i \mu} 
+ \frac{1}{2 R^2 \sin^2 \theta} F_{\phi \mu}^i F_\phi^{i \mu} - \frac{1}{2 R^4 \sin^2 \theta} F^i_{\theta \phi} F^i_{\theta \phi} \right]
\end{align}
where $i$ is the index of gauge group.
The first term of Eq.~(\ref{Self-interaction}) in the bracket is expanded as 
\begin{align}
\label{4dgauge}
-\frac{1}{4} F_{\mu \nu}^i F^{i \mu \nu} =&  - \frac{1}{4} (\partial_{\mu} A^i_{\nu} - \partial_{\nu} A^i_{\mu} )(\partial^{\mu} A^{i\nu} - \partial^{\nu} A^{i \mu} ) \nonumber \\
& - g f^{ijk} \partial_{\mu} A_{\nu}^i A^{j \mu} A^{k \nu} - \frac{1}{4} g^2 f^{ijm} f^{klm} A^{i}_{\mu} A^j_{\nu} A^{k \mu} A^{l \nu},
\end{align}
where the second line of RHS gives three- and four-point self-interaction of gauge boson summarized in Table~\ref{vertices-A}.
The second and the third term of Eq.~(\ref{Self-interaction}) give interactions between $A_\mu$ and $A_{\theta, \phi}$.
By substituting $A_{\theta, \phi}$ of Eqs.~(\ref{substitution1}) and (\ref{substitution2}), we obtain interaction terms
\begin{align}
\label{Aphi}
&\frac{1}{2 R^2} F_{\theta \mu}^i F_\theta^{i \mu} + \frac{1}{2 R^2 \sin^2 \theta} F_{\phi \mu}^i F_\phi^{i \mu} \nonumber \\
& \supset  \frac{g_{6a}}{ R^2} f^{ijk} \biggl[ 
\frac{1}{\sin \theta} ( \partial_{\phi} A^i_{\mu} \partial_{\theta} \phi^j_1 A^{k \mu} -\partial_{\theta} A^i_{\mu} \partial_{\phi} \phi^j_1 A^{k \mu} )
+ \partial_{\theta} A^i_{\mu} \partial_{\theta} \phi^j_2 A^{k \mu} + \frac{1}{\sin^2 \theta} \partial_{\phi} A^i_{\mu} \partial_{\phi} \phi^j_2 A^{k \mu} \nonumber \\
& \qquad \qquad 
- \partial_{\theta} \partial_{\mu} \phi_1^i \partial_{\theta} \phi^j_1 A^{k \mu}-\frac{1}{\sin^2 \theta}  \partial_{\phi} \partial_{\mu} \phi_1^i \partial_{\phi} \phi^j_1 A^{k \mu}
- \partial_{\theta} \partial_{\mu} \phi_2^i \partial_{\theta} \phi^j_2 A^{k \mu}-\frac{1}{\sin^2 \theta}  \partial_{\phi} \partial_{\mu} \phi_2^i \partial_{\phi} \phi^j_2 A^{k \mu}
\nonumber \\ & \qquad \qquad
-\frac{1}{\sin \theta}( \partial_{\theta} \partial_{\mu} \phi_1^i \partial_{\phi} \phi^j_2 A^{k \mu}+ \partial_{\phi} \partial_{\mu} \phi_2^i \partial_{\theta} \phi^j_1 A^{k \mu}
- \partial_{\theta} \partial_{\mu} \phi_2^i \partial_{\phi} \phi^j_1 A^{k \mu}-  \partial_{\phi} \partial_{\mu} \phi_1^i \partial_{\theta} \phi^j_2 A^{k \mu} ) \biggr]
\nonumber \\ & + \frac{g_{6a}^2}{2R^2} f^{ijm} f^{klm} \biggl[
\partial_{\theta} \phi_1^i \partial_{\theta} \phi^k_1 A^{j}_{\mu} A^{l \mu}+\frac{1}{\sin^2 \theta}  \partial_{\phi} \phi_1^i \partial_{\phi} \phi^k_1 A_{\mu}^j A^{l \mu} 
+ \partial_{\theta} \phi_2^i \partial_{\theta} \phi^k_2 A^{j}_{\mu} A^{l \mu}+\frac{1}{\sin^2 \theta}  \partial_{\phi} \phi_2^i \partial_{\phi} \phi^k_2 A_{\mu}^j A^{l \mu} 
\nonumber \\ & \qquad \qquad \qquad + \frac{1}{\sin \theta}
(  \partial_{\theta} \phi_1^i \partial_{\phi} \phi^k_2 A^{j}_{\mu} A^{l \mu}+  \partial_{\phi} \phi_2^i \partial_{\theta} \phi^k_1 A_{\mu}^j A^{l \mu} 
-  \partial_{\theta} \phi_2^i \partial_{\phi} \phi^k_1 A^{j}_{\mu} A^{l \mu}-  \partial_{\phi} \phi_1^i \partial_{\theta} \phi^k_2 A_{\mu}^j A^{l \mu} ) \biggr] 
\end{align}
providing vertex factors from the third line to the seventh line in Table~\ref{vertices-A}.
The last term of Eq.~(\ref{Self-interaction}) gives interactions between $\phi_i$s after substituting $A_{\theta, \phi}$ of Eqs.~(\ref{substitution1}) and (\ref{substitution2}) such that
\begin{align}
& - \frac{1}{2 R^4 \sin^2 \theta} F^i_{\theta \phi} F^i_{\theta \phi} \nonumber \\
\supset &
\frac{g_{6a}}{R^4} f^{ijk} \biggl[ 
(\hat{L}^2 \phi_1^i) \biggl( \partial_{\theta} \phi^j_2 \partial_{\theta} \phi_1^k - \frac{1}{\sin^2 \theta}  \partial_{\phi} \phi^j_1 \partial_{\phi} \phi_2^k \biggr) 
+ \frac{1}{\sin \theta} (\hat{L}^2 \phi^i_1) \partial_{\theta} \phi^j_2 \partial_{\phi} \phi^k_2  
-\frac{1}{\sin \theta} (\hat{L}^2 \phi^i_1) \partial_{\phi} \phi^j_1 \partial_{\theta} \phi^k_1 \biggr] \nonumber \\
& - \frac{g_{6a}^2}{2R^4} f^{ijm} f^{klm} \biggl[ 
\frac{1}{\sin^2 \theta} \partial_{\phi} \phi^i_1 \partial_{\theta} \phi^j_1  \partial_{\phi} \phi^k_1 \partial_{\theta} \phi^l_1 
+\frac{1}{\sin^2 \theta} \partial_{\theta} \phi^i_2 \partial_{\phi} \phi^j_2  \partial_{\theta} \phi^k_2 \partial_{\phi} \phi^l_2 
- \frac{1}{\sin \theta} \partial_{\theta} \phi^i_2 \partial_{\theta} \phi^j_1  \partial_{\phi} \phi^k_1 \partial_{\theta} \phi^l_1 \nonumber \\
& \qquad \qquad \qquad
-\frac{1}{\sin \theta} \partial_{\phi} \phi^i_1 \partial_{\theta} \phi^j_1  \partial_{\theta} \phi^k_2 \partial_{\theta} \phi^l_1   
+\frac{1}{\sin^3 \theta} \partial_{\phi} \phi^i_1 \partial_{\theta} \phi^j_1  \partial_{\phi} \phi^k_1 \partial_{\phi} \phi^l_2 
+\frac{1}{\sin^3 \theta} \partial_{\phi} \phi^i_1 \partial_{\phi} \phi^j_2  \partial_{\phi} \phi^k_1 \partial_{\theta} \phi^l_1  \nonumber \\
& \qquad \qquad \qquad
+ \frac{1}{\sin \theta} \partial_{\theta} \phi^i_2 \partial_{\theta} \phi^j_1  \partial_{\theta} \phi^k_2 \partial_{\phi} \phi^l_2
+ \frac{1}{\sin \theta} \partial_{\theta} \phi^i_2 \partial_{\phi} \phi^j_2  \partial_{\theta} \phi^k_2 \partial_{\theta} \phi^l_1   
-\frac{1}{\sin^3 \theta} \partial_{\theta} \phi^i_2 \partial_{\phi} \phi^j_2  \partial_{\phi} \phi^k_1 \partial_{\phi} \phi^l_2 \nonumber \\
& \qquad \qquad \qquad 
-\frac{1}{\sin^3 \theta} \partial_{\phi} \phi^i_1 \partial_{\phi} \phi^j_2  \partial_{\theta} \phi^k_2 \partial_{\phi} \phi^l_2  
+ \partial_{\theta} \phi^i_2 \partial_{\theta} \phi^j_1  \partial_{\theta} \phi^k_2 \partial_{\theta} \phi^l_1 
+\frac{1}{\sin^4 \theta} \partial_{\phi} \phi^i_1 \partial_{\phi} \phi^j_2  \partial_{\phi} \phi^k_1 \partial_{\phi} \phi^l_2 \nonumber \\
& \qquad \qquad \qquad
-\frac{1}{\sin^2 \theta} \partial_{\theta} \phi^i_2 \partial_{\theta} \phi^j_1  \partial_{\phi} \phi^k_1 \partial_{\phi} \phi^l_2 
-\frac{1}{\sin^2 \theta} \partial_{\theta} \phi^i_2 \partial_{\phi} \phi^j_2  \partial_{\phi} \phi^k_1 \partial_{\theta} \phi^l_1 \nonumber \\
& \qquad \qquad \qquad 
-\frac{1}{\sin^2 \theta} \partial_{\phi} \phi^i_1 \partial_{\theta} \phi^j_1  \partial_{\theta} \phi^k_2 \partial_{\phi} \phi^l_2 
-\frac{1}{\sin^2 \theta} \partial_{\phi} \phi^i_1 \partial_{\phi} \phi^j_2  \partial_{\theta} \phi^k_2 \partial_{\theta} \phi^l_1 \biggr],    
\end{align}
providing vertex factors from the eighth line to the thirteenth line in Table~\ref{vertices-A}.
The vertex factor associated with ghost fields $c^i$ is also obtained from the guage-ghost interaction in Eq.~(\ref{6Daction}), and it is summarized in both
the fourteenth and the fifteenth vertex factors in Table~\ref{vertices-A}.
The non-trivial coefficients appearing in the vertex factors in Table~\ref{vertices-A} are given by integration of spherical harmonics and its derivative with respect to extra spatial coordinates. 
%
\begin{table}[t] \vspace{1ex}
\begin{tabular}{c||c} 
{\scriptsize Interaction} & {\scriptsize vertex factor } \\ \hline
{\scriptsize $A^{i \mu}_{\ell_1 m_1}(k) A^{j \nu}_{\ell_2 m_2}(p) A^{k \rho}_{\ell_3 m_3}(q)$ }
& {\scriptsize $\frac{g_{6a}}{R} f_a^{ijk} \left[ g^{\mu \nu} (k-p)^\rho + g^{\nu \rho} (p-q)^\mu + g^{\rho \mu} (q-k)^\nu \right] J^1_{ \ell_1m_1;\ell_2m_2;\ell_3m_3 }$ } \\
{\scriptsize $A^{i \mu}_{\ell_1 m_1} A^{j \nu}_{\ell_2 m_2} A^{k \rho}_{\ell_3 m_3} A^{l \sigma}_{\ell_4 m_4}$  }
& {\scriptsize $-i \frac{g_{6a}^2}{R^2} [ f_a^{ijm} f^{klm} (g^{\mu \rho}g^{\nu \sigma}-g^{\mu \sigma}g^{\nu \rho}) + f^{ikm} f^{jlm} (g^{\mu \nu}g^{\rho \sigma}-g^{\mu \sigma}g^{\nu \rho})$ } \\
& {\scriptsize $ \hspace{10mm} +f^{ilm} f^{jkm} (g^{\mu \nu}g^{\rho \sigma}-g^{\mu \rho}g^{\nu \sigma})] K^1_{ \ell_1m_1;\ell_2m_2;\ell_3m_3;\ell_4 m_4 }$ } \\
{\scriptsize $A^{i \mu}_{\ell_1 m_1} A^{j \nu}_{\ell_2 m_2} \phi^{k }_{1 \ell_3 m_3}$ }
& {\scriptsize $-i \frac{g_{6a}}{R^2} f^{ijk} g^{\mu \nu}  J^4_{ \ell_1m_1;\ell_2m_2;\ell_3m_3 }$ } \\
{\scriptsize $A^{i \mu}_{\ell_1 m_1} A^{j \nu}_{\ell_2 m_2} \phi^{k }_{2 \ell_3 m_3}$ }
& {\scriptsize $-i \frac{g_{6a}}{R^2} f^{ijk} g^{\mu \nu}  J^5_{ \ell_1m_1;\ell_2m_2;\ell_3m_3 }$ } \\
{\scriptsize $ \phi^{i }_{1 (2) \ell_1 m_1}(p) \phi^{j }_{1(2) \ell_2 m_2}(q) A^{k \mu}_{\ell_3 m_3}$ }
& {\scriptsize $\frac{g_{6a}}{R} f^{ijk} (q-p)^\mu  J^6_{ \ell_1m_1;\ell_2m_2;\ell_3m_3 } $ } \\
{\scriptsize $ \phi^{i }_{1 \ell_1 m_1}(p) \phi^{j }_{2 \ell_2 m_2}(q) A^{k \mu}_{\ell_3 m_3}$ }
& {\scriptsize $\frac{g_{6a}}{R} f^{ijk} (q-p)^\mu  J^7_{ \ell_1m_1;\ell_2m_2;\ell_3m_3 }$ } \\
{\scriptsize $ \phi^{i }_{1  \ell_1 m_1}(p) \phi^{j }_{2 \ell_2 m_2}(q) A^{k \mu}_{\ell_3 m_3} A^{l \nu}_{\ell_4 m_4}$ }
& {\scriptsize $i \frac{g_{6a}^2}{R^2} g^{\mu \nu} (f^{ikm} f^{jlm} + f^{ilm} f^{jkm})  K^3_{ \ell_1m_1;\ell_2m_2;\ell_3m_3; \ell_4 m_4 }$ }\\
{\scriptsize  $ \phi^{i }_{1  \ell_1 m_1}(p) \phi^{j }_{1 \ell_2 m_2}(q) \phi^{k}_{2 \ell_3 m_3}$ }
 & {\scriptsize $-i \frac{g_{6a}}{R^2} f^{ijk}  J^8_{ \ell_1m_1;\ell_2m_2;\ell_3m_3 }$  }\\ 
{\scriptsize  $ \phi^{i }_{2  \ell_1 m_1}(p) \phi^{j }_{2 \ell_2 m_2}(q) \phi^{k}_{1 \ell_3 m_3}$ }
 &{\scriptsize $i \frac{g_{6a}}{R^2} f^{ijk}  J^9_{ \ell_1m_1;\ell_2m_2;\ell_3m_3 }$  }\\
{\scriptsize  $ \phi^{i }_{1  \ell_1 m_1} \phi^{j }_{1 \ell_2 m_2} \phi^{k}_{1 \ell_3 m_3}$ }
 & {\scriptsize $ -i \frac{g_{6a}}{R^2} f^{ijk}  J^{10}_{ \ell_1m_1;\ell_2m_2;\ell_3m_3 }$ }\\
  {\scriptsize $ \phi^{i }_{1(2)  \ell_1 m_1} \phi^{j }_{1(2) \ell_2 m_2} \phi^{k}_{1(2) \ell_3 m_3} \phi^{l}_{1(2) \ell_4 m_4}$ }
 &{\scriptsize $ -i \frac{g_{6a}^2}{R^2} \Bigl[ (f^{ijm}f^{klm}+f^{kjm}f^{ilm}) K^{4}_{ \ell_1m_1;\ell_2m_2;\ell_3m_3; \ell_4 m_4 }$ } \\
  &{\scriptsize $- (f^{ijm}f^{klm}+f^{ikm}f^{jlm}) K^{4}_{ \ell_2m_2;\ell_1m_1;\ell_3m_3; \ell_4 m_4 }$ } \\
 &{\scriptsize $+ (f^{ikm}f^{jlm}-f^{kjm}f^{ilm}) K^{4}_{ \ell_1m_1;\ell_3m_3;\ell_2m_2; \ell_4 m_4 } \Bigr]$ }\\
 {\scriptsize $ \phi^{i }_{1(2)  \ell_1 m_1} \phi^{j }_{2(1) \ell_2 m_2} \phi^{k}_{2(1) \ell_3 m_3} \phi^{l}_{2(1) \ell_4 m_4}$ }
 &{\scriptsize $\pm i \frac{g_{6a}^2}{R^2} \Bigl[ (f^{ijm}f^{klm}+f^{ikm}f^{jlm}) K^{5}_{ \ell_1m_1;\ell_2m_2;\ell_3m_3; \ell_4 m_4 }$ }\\
 &{\scriptsize $- (f^{ijm}f^{klm} - f^{ilm}f^{jkm}) K^{5}_{ \ell_1m_1;\ell_2m_2;\ell_4m_4; \ell_3 m_3 }$ }\\
 &{\scriptsize $- (f^{ilm}f^{jkm}+f^{ikm}f^{jlm}) K^{5}_{ \ell_1m_1;\ell_4m_4;\ell_3m_3; \ell_2 m_2 } \Bigr]$ } \\
{\scriptsize  $ \phi^{i }_{1  \ell_1 m_1} \phi^{j }_{1 \ell_2 m_2} \phi^{k}_{2 \ell_3 m_3} \phi^{l}_{2 \ell_4 m_4}$ }
 & {\scriptsize $ i \frac{g_{6a}^2}{R^2} \Bigl[ (f^{ikm}f^{jlm}+f^{jkm}f^{ilm}) K^{6}_{ \ell_1m_1;\ell_2m_2;\ell_3m_3; \ell_4 m_4 } $ } \\
 &{\scriptsize $ + (f^{ijm}f^{klm} - f^{ilm}f^{jkm}) K^{4}_{ \ell_2m_2;\ell_1m_1;\ell_3m_3; \ell_4 m_4 } $ } \\
 &{\scriptsize $ - (f^{ijm}f^{klm}+f^{ikm}f^{jlm}) K^{4}_{ \ell_1m_1;\ell_2m_2;\ell_3m_3; \ell_4 m_4 } \Bigr]$ }\\
{\scriptsize   $c^{i }_{\ell_1 m_1}(p) A^{j \mu}_{\ell_2 m_2} c^{ k }_{ \ell_3 m_3}$ }
  &{\scriptsize $- \frac{g_{6a}}{R} f^{ijk} p^\mu   J^1_{  \underline{\ell_1m_1} ;\ell_2m_2;\ell_3m_3 } $ } \\
{\scriptsize   $c^{i }_{\ell_1 m_1}(p) \phi^{j}_{2 \ell_2 m_2} c^{ k }_{ \ell_3 m_3}$ }
  &{\scriptsize $\frac{g_{6a}}{R^2} \xi f^{ijk}   J^{11}_{ \underline{\ell_1m_1} ;\ell_2m_2;\ell_3m_3 }$ } \\ \hline
\end{tabular} \vspace{-1ex}
\caption{The vertex factors for self-interactions of gauge bosons including extra-dimensional components.\label{vertices-A}} \vspace{-1ex}
\end{table}
%
%
For three-point vertices, these coefficients are given by
\begin{align}
J^1_{\ell_1 m_1; \ell_2 m_2; \ell_3 m_3} =& \int d \Omega Y_{\ell_1 m_1} Y_{\ell_2 m_2} Y_{\ell_3 m_3}, \\
J^2_{\ell_1 m_1; \ell_2 m_2; \ell_3 m_3} =&  \int d \Omega \frac{1}{\sin \theta} \left[ -\partial_\theta Y_{\ell_1 m_1} \partial_\phi \tilde{Y}_{\ell_2 m_2} Y_{\ell_3 m_3}
+\partial_\phi Y_{\ell_1 m_1} \partial_\theta \tilde{Y}_{\ell_2 m_2} Y_{\ell_3 m_3} \right], \\
J^3_{\ell_1 m_1; \ell_2 m_2; \ell_3 m_3} =& \int d \Omega  \left[ \partial_\theta Y_{\ell_1 m_1} \partial_\theta \tilde{Y}_{\ell_2 m_2} Y_{\ell_3 m_3}
+ \frac{1}{\sin^2 \theta} \partial_\phi Y_{\ell_1 m_1} \partial_\phi \tilde{Y}_{\ell_2 m_2} Y_{\ell_3 m_3} \right], \\
J^4_{\ell_1m_1;\ell_2m_2;\ell_3m_3 } =& \int d \Omega \frac{1}{\sin \theta} \bigl[ (\partial_\phi Y_{\ell_1 m_1} Y_{\ell_2 m_2} - Y_{\ell_1 m_1} \partial_\phi Y_{\ell_2 m_2} ) \partial_\theta \tilde{Y}_{\ell_3 m_3} 
\nonumber \\  & -(\partial_\theta Y_{\ell_1 m_1} Y_{\ell_2 m_2} - Y_{\ell_1 m_1} \partial_\phi Y_{\ell_2 m_2} ) \partial_\phi \tilde{Y}_{\ell_3 m_3} \bigr],  
\end{align}
\begin{align}
 J^5_{\ell_1m_1;\ell_2m_2;\ell_3m_3 } =& \int d \Omega  \bigl[ (\partial_\theta Y_{\ell_1 m_1} Y_{\ell_2 m_2} - Y_{\ell_1 m_1} \partial_\theta Y_{\ell_2 m_2} ) \partial_\theta \tilde{Y}_{\ell_3 m_3} \nonumber \\
 & + \frac{1}{\sin^2 \theta} (\partial_\phi Y_{\ell_1 m_1} Y_{\ell_2 m_2} - Y_{\ell_1 m_1} \partial_\phi Y_{\ell_2 m_2} ) \partial_\phi \tilde{Y}_{\ell_3 m_3} \bigr], \\
J^6_{\ell_1m_1;\ell_2m_2;\ell_3m_3 } =& \int d \Omega \left[ \partial_\theta \tilde{Y}_{\ell_1 m_1}\partial_\theta \tilde{Y}_{\ell_2 m_2} 
+\frac{1}{\sin^2 \theta} \partial_\phi \tilde{Y}_{\ell_1 m_1} \partial_\phi \tilde{Y}_{\ell_2 m_2} \right] Y_{\ell_3 m_3}, \\
 J^7_{ \ell_1m_1;\ell_2m_2;\ell_3m_3 } =& \int d \Omega \frac{1}{\sin \theta} \left[ \partial_\theta \tilde{Y}_{\ell_1 m_1}\partial_\phi \tilde{Y}_{\ell_2 m_2} 
- \partial_\phi \tilde{Y}_{\ell_1 m_1} \partial_\theta \tilde{Y}_{\ell_2 m_2} \right] Y_{\ell_3 m_3}, \\
  J^{8}_{\ell_1m_1;\ell_2m_2;\ell_3m_3 } =& \int d \Omega \Big[ \hat{L}^2 \tilde{Y}_{\ell_1 m_1} \left(\partial_\theta \tilde{Y}_{\ell_2 m_2} \partial_\theta \tilde{Y}_{\ell_3 m_3} 
+ \frac{1}{\sin^2 \theta} \partial_\phi \tilde{Y}_{\ell_2 m_2} \partial_\phi \tilde{Y}_{\ell_3 m_3} \right) \nonumber \\
& - \hat{L}^2 \tilde{Y}_{\ell_2 m_2} \left(\partial_\theta \tilde{Y}_{\ell_1 m_1} \partial_\theta \tilde{Y}_{\ell_3 m_3} 
+ \frac{1}{\sin^2 \theta} \partial_\phi \tilde{Y}_{\ell_1 m_1} \partial_\phi \tilde{Y}_{\ell_3 m_3} \right) \Bigr], 
\end{align}
\begin{align}
J^{9}_{\ell_1m_1;\ell_2m_2;\ell_3m_3} =& \int d \Omega \frac{1}{\sin \theta} \left[ \partial_\theta \tilde{Y}_{\ell_1 m_1} \partial_\phi \tilde{Y}_{\ell_2 m_2} 
-  \partial_\phi \tilde{Y}_{\ell_1 m_1} \partial_\theta \tilde{Y}_{\ell_2 m_2} \right] \hat{L}^2 \tilde{Y}_{\ell_3 m_3}, \\
 J^{10}_{\ell_1m_1;\ell_2m_2;\ell_3m_3 } =& \int d \Omega \frac{1}{\sin \theta} \Big[ \hat{L}^2 \tilde{Y}_{\ell_1 m_1} \left(\partial_\phi \tilde{Y}_{\ell_2 m_2} \partial_\theta \tilde{Y}_{\ell_3 m_3} 
-  \partial_\theta \tilde{Y}_{\ell_2 m_2} \partial_\phi \tilde{Y}_{\ell_3 m_3} \right) \nonumber \\
& + \hat{L}^2 \tilde{Y}_{\ell_2 m_2} \left(\partial_\theta \tilde{Y}_{\ell_1 m_1} \partial_\phi \tilde{Y}_{\ell_3 m_3} 
-  \partial_\phi \tilde{Y}_{\ell_1 m_1} \partial_\theta \tilde{Y}_{\ell_3 m_3} \right) \nonumber \\
& + \hat{L}^2 \tilde{Y}_{\ell_3 m_3} \left(\partial_\phi \tilde{Y}_{\ell_1 m_1} \partial_\theta \tilde{Y}_{\ell_2 m_2} 
-  \partial_\theta \tilde{Y}_{\ell_1 m_1} \partial_\phi \tilde{Y}_{\ell_2 m_2} \right)  \Bigr], \\
 J^{11}_{\ell_1m_1 ;\ell_2m_2;\ell_3m_3 } = & \int d \Omega Y_{\ell_1 m_1} \hat{L}^2 Y_{\ell_2 m_2} Y_{\ell_3 m_3},
\end{align}
where we adopt the convention that we put an underline for indices ${\ell m}$ of $Y^*_{\ell m}$.
The coefficients for four point vertices are also given by
\begin{align}
K^1_{\ell_1 m_1; \ell_2 m_2; \ell_3 m_3; \ell_4 m_4} =& \int d \Omega Y_{\ell_1 m_1} Y_{\ell_2 m_2} Y_{\ell_3 m_3} Y_{\ell_4 m_4}, \\
K^2_{\ell_1 m_1; \ell_2 m_2; \ell_3 m_3 ; \ell_4 m_4} =& \int d \Omega \left[ \partial_\theta \tilde{Y}_{\ell_1 m_1} \partial_\theta \tilde{Y}_{\ell_2 m_2}
+ \frac{1}{\sin^2 \theta} \partial_\phi \tilde{Y}_{\ell_1 m_1} \partial_\phi \tilde{Y}_{\ell_2 m_2}  \right] Y_{\ell_3 m_3} Y_{\ell_4 m_4}, \\
 K^3_{\ell_1m_1;\ell_2m_2;\ell_3m_3; \ell_4 m_4 } =& \int d \Omega \frac{1}{\sin \theta} \left[ \partial_\theta \tilde{Y}_{\ell_1 m_1}\partial_\phi \tilde{Y}_{\ell_2 m_2} 
-  \partial_\phi \tilde{Y}_{\ell_1 m_1} \partial_\theta \tilde{Y}_{\ell_2 m_2} \right] Y_{\ell_3 m_3} Y_{\ell_4 m_4}, \\
K^{4}_{\ell_1m_1;\ell_2m_2;\ell_3m_3; \ell_4 m_4 } =& \int d \Omega \frac{1}{\sin^2 \theta} 
\big( \partial_\phi \tilde{Y}_{\ell_1 m_1} \partial_\theta \tilde{Y}_{\ell_2 m_2} \partial_\phi \tilde{Y}_{\ell_3 m_3} \partial_\theta \tilde{Y}_{\ell_4 m_4} \nonumber \\
& + \partial_\theta \tilde{Y}_{\ell_1 m_1} \partial_\phi \tilde{Y}_{\ell_2 m_2} \partial_\theta \tilde{Y}_{\ell_3 m_3} \partial_\phi \tilde{Y}_{\ell_4 m_4} \big), \\
K^{5}_{\ell_1m_1;\ell_2m_2;\ell_3m_3; \ell_4 m_4 } =& \int d \Omega \frac{1}{\sin \theta} 
\big( \partial_\theta \tilde{Y}_{\ell_1 m_1} \partial_\theta \tilde{Y}_{\ell_2 m_2} \partial_\theta \tilde{Y}_{\ell_3 m_3} \partial_\phi \tilde{Y}_{\ell_4 m_4} \nonumber \\
& -\frac{1}{\sin^2 \theta} \partial_\phi \tilde{Y}_{\ell_1 m_1} \partial_\phi \tilde{Y}_{\ell_2 m_2} \partial_\phi \tilde{Y}_{\ell_3 m_3} \partial_\theta \tilde{Y}_{\ell_4 m_4} \big), \\
K^{6}_{\ell_1m_1;\ell_2m_2;\ell_3m_3; \ell_4 m_4 } =& - \int d \Omega  
\big( \partial_\theta \tilde{Y}_{\ell_1 m_1} \partial_\theta \tilde{Y}_{\ell_2 m_2} \partial_\theta \tilde{Y}_{\ell_3 m_3} \partial_\theta \tilde{Y}_{\ell_4 m_4} \nonumber \\
& + \frac{1}{\sin^4 \theta} \partial_\phi \tilde{Y}_{\ell_1 m_1} \partial_\phi \tilde{Y}_{\ell_2 m_2} \partial_\phi \tilde{Y}_{\ell_3 m_3} \partial_\phi \tilde{Y}_{\ell_4 m_4} \big). 
\end{align}
The vertices including Higgs field are derived from the interactions
\begin{equation}
\int dx^4 d\Omega \left[ (D^M H)^\dagger (D_M H) - \mu^2 H^\dagger H - \frac{\lambda_6}{4} (H^\dagger H)^2 \right],
\end{equation}
which are the same structure as in the SM.
The Feynman rules of these interactions are listed in Table~\ref{vertices-H},
where the non-trivial coefficients in the vertex factors have been already given above.

\renewcommand{\arraystretch}{1.35}

\begin{table}[tb] \vspace{1ex}
\begin{tabular}{c||c|l} 
{\scriptsize Diagram} & {\scriptsize Coefficients} &  \\ \hline
{\scriptsize Fig.~\ref{Loop1} }& {\scriptsize $\Sigma_{\rm bulk}^{f A_\mu, R(L)}$ }& 
{\scriptsize ${\displaystyle   \int_0^1 d \alpha \int^{\Lambda^2}_{0}  \frac{2 \alpha x dx}{[x + \Delta ]^2} (-1)^{m_1+m}
 I^{\alpha(\beta)}_{\ell_1 m_1;\ell_2 m_1-m;\ell' m} I^{\alpha(\beta)}_{\ell m;\ell_2 m-m_1;\ell_1 m_1}    }$ } \\
& {\scriptsize  $ \tilde{\Sigma}_{\rm bulk}^{f A_\mu, R(L)}$ } &
{\scriptsize ${\displaystyle   \int_0^1 d \alpha \int^{\Lambda^2}_{0} \frac{4 x dx}{[x + \Delta ]^2} (-1)^{m_1+m} 
M_{\ell_1} I^{\beta(\alpha)}_{\ell_1 m_1;\ell_2 m_1-m;\ell' m} I^{\alpha(\beta)}_{\ell m;\ell_2 m-m_1;\ell_1 m_1} }$} \\
& {\scriptsize $ \Sigma_{\rm bound}^{f A_\mu, R(L)}$} &
{\scriptsize ${\displaystyle   (-1)^{m_1 + m+\ell_2}  
I^{\alpha(\beta)}_{\ell_1 m_1;\ell_2 -m_1+m;\ell' 2m_1-m} I^{\alpha(\beta)}_{\ell m;\ell_2 m-m_1;\ell_1 m_1}}$} \\
& {\scriptsize $ \tilde{\Sigma}_{\rm bound}^{f A_\mu, R(L)}$} & 
{\scriptsize ${\displaystyle  (-1)^{m_1+ m+\ell_2}
4 M_{\ell_1} I^{\beta(\alpha)}_{\ell_1 m_1;\ell_2 -m_1+m;\ell' 2m_1-m} I^{\alpha(\beta)}_{\ell m;\ell_2 m-m_1;\ell_1 m_1} }$} \\ \hline
{\scriptsize Fig.~\ref{Loop1}} & {\scriptsize $\Sigma_{\rm bulk}^{f \phi_1, R(L)} $ } & 
{\scriptsize ${\displaystyle  -\int_0^1 d \alpha \int^{\Lambda^2}_{0} \frac{\alpha x dx}{[x + \Delta ]^2} (-1)^{m_1+m}}
C^{\alpha(\beta)}_{\ell_1 m_1;\ell_2 m_1-m;\ell' m} C^{\beta(\alpha)}_{\ell m;\ell_2 m-m_1;\ell_1 m_1} $} \\
& {\scriptsize $ \tilde{\Sigma}_{\rm bulk}^{f \phi_1 ,R(L)} $} & 
{\scriptsize ${\displaystyle    \int_0^1 d \alpha \int^{\Lambda^2}_{0} \frac{x dx}{[x + \Delta ]^2} (-1)^{m_1+m}
M_{\ell_1} C^{\beta(\alpha)}_{\ell_1 m_1;\ell_2 m_1-m;\ell' m} C^{\beta(\alpha)}_{\ell m;\ell_2 m-m_1;\ell_1 m_1} }$} \\
& {\scriptsize $  \Sigma_{\rm bound}^{f \phi_1, R(L)} $} &
{\scriptsize ${\displaystyle  -(-1)^{m_1+ m+\ell_2} 
\frac{1}{2} C^{\alpha(\beta)}_{\ell_1 m_1;\ell_2 -m_1+m;\ell' 2m_1-m} C^{\beta(\alpha)}_{\ell m;\ell_2 m-m_1;\ell_1 m_1} }$} \\
& {\scriptsize $  \tilde{\Sigma}_{\rm bound}^{f \phi_1, R(L)} $} &
{\scriptsize ${\displaystyle   (-1)^{m_1+ m+\ell_2} 
 M_{\ell_1} C^{\beta(\alpha)}_{\ell_1 m_1;\ell_2 -m_1+m;\ell' 2m_1-m} C^{\beta(\alpha)}_{\ell m;\ell_2 m-m_1;\ell_1 m_1} }$} \\ \hline
 {\scriptsize Fig.~\ref{Loop1}} & {\scriptsize $\Sigma_{\rm bulk}^{f \phi_2, R(L)} $ } & 
{\scriptsize ${\displaystyle  -\int_0^1 d \alpha \int^{\Lambda^2}_{0} \frac{\alpha x dx}{[x + \Delta ]^2} (-1)^{m_1+m}}
C^{\alpha(\beta)}_{\ell_1 m_1;\ell_2 m_1-m;\ell' m} C^{\beta(\alpha)}_{\ell m;\ell_2 m-m_1;\ell_1 m_1} $} \\
& {\scriptsize $ \tilde{\Sigma}_{\rm bulk}^{f \phi_2 ,R(L)} $} & 
{\scriptsize ${\displaystyle   - \int_0^1 d \alpha \int^{\Lambda^2}_{0} \frac{x dx}{[x + \Delta ]^2} (-1)^{m_1+m}
M_{\ell_1} C^{\beta(\alpha)}_{\ell_1 m_1;\ell_2 m_1-m;\ell' m} C^{\beta(\alpha)}_{\ell m;\ell_2 m-m_1;\ell_1 m_1} }$} \\
& {\scriptsize $  \Sigma_{\rm bound}^{f \phi_2, R(L)} $} &
{\scriptsize ${\displaystyle  -(-1)^{m_1+ m+\ell_2} 
\frac{1}{2} C^{\alpha(\beta)}_{\ell_1 m_1;\ell_2 -m_1+m;\ell' 2m_1-m} C^{\beta(\alpha)}_{\ell m;\ell_2 m-m_1;\ell_1 m_1} }$} \\
& {\scriptsize $  \tilde{\Sigma}_{\rm bound}^{f \phi_2, R(L)} $} &
{\scriptsize ${\displaystyle -  (-1)^{m_1+ m+\ell_2} 
 M_{\ell_1} C^{\beta(\alpha)}_{\ell_1 m_1;\ell_2 -m_1+m;\ell' 2m_1-m} C^{\beta(\alpha)}_{\ell m;\ell_2 m-m_1;\ell_1 m_1} }$} \\ \hline
 {\scriptsize Fig.~\ref{Loop2}} & {\scriptsize $ \Sigma_{\rm bulk}^{H,R(L)}$} & 
 {\scriptsize ${\displaystyle    \int_0^1 d \alpha \int^{\Lambda^2}_{0} \frac{\alpha x dx}{[x + \Delta ]^2} (-1)^{m_1+m} 
  I^{\beta(\alpha)}_{\ell_1 m_1;\ell_2 m_1-m;\ell' m} I^{\beta(\alpha)}_{\ell m;\ell_2 m-m_1;\ell_1 m_1} }$} \\ 
  & { \scriptsize $  \tilde{\Sigma}_{\rm bulk}^{fH,R(L)}$ }& 
{\scriptsize ${\displaystyle    \int_0^1 d \alpha \int^{\Lambda^2}_{0} \frac{x dx}{[x + \Delta ]^2} (-1)^{m_1+m} 
 M_{\ell_1} I^{\alpha(\beta) }_{\ell_1 m_1;\ell_2 m_1-m;\ell' m} I^{\beta(\alpha)}_{\ell m;\ell_2 m-m_1;\ell_1 m_1} }$} \\
 & {\scriptsize ${\displaystyle \Sigma_{\rm bound}^{fH,R(L)}}$} &
 {\scriptsize ${\displaystyle  (-1)^{m_1+m+\ell_2} 
 \frac{1}{2} I^{\beta(\alpha)}_{\ell_1 m_1;\ell_2 -m_1+m;\ell' 2m_1-m} I^{\beta(\alpha)}_{\ell m;\ell_2 m-m_1;\ell_1 m_1} } $} \\
 & {\scriptsize $  \tilde{\Sigma}_{\rm bound}^{fH,R(L)}$ } &
 {\scriptsize ${\displaystyle   (-1)^{m_1+m+\ell_2} 
 M_{\ell_1} I^{\alpha(\beta)}_{\ell_1 m_1;\ell_2 -m_1+m;\ell' 2m_1-m} I^{\beta(\alpha) }_{\ell m;\ell_2 m-m_1;\ell_1 m_1} }$} \\ \hline
\end{tabular}
\vspace{-1ex}
\caption{The one loop contributions for correction to KK masses of fermions. 
The summation symbols $ \{ \sum_{\ell_1=0}^{\ell_{max}}, \sum_{\ell_2=0}^{\ell_{max}}, \sum_{m_1=-\ell_1}^{\ell_1} \}$ and the overall factor $(T_a)^2 g_{6a}^2/64 \pi^2 R^2 ( (Y_{u,d,e})^2 /64 \pi^2 R^2)$ for fermion(Higgs) loop are omitted for all contributions in the third column.  Also $\ln (\Lambda^2/\mu^2)$ factor is omitted for boundary contributions in the third column. The subscripts $f A_\mu$ etc. show 
the propagating particles inside a loop. \label{OneLoopF}} \vspace{-1ex}
\end{table}

\begin{table}[tbh] \vspace{1ex}
\begin{tabular}{c||c|l} 
{\scriptsize Diagram} & {\scriptsize Coefficients} &  \\ \hline
{\scriptsize Fig.~\ref{Loop3}} & {\scriptsize $ \Pi_{\rm bulk}^{ff} $ } &
{\scriptsize ${\displaystyle -   \int_0^1 d \alpha  \int^{\Lambda^2}_{0} \frac{x dx}{[x + \Delta ]^2} (-1)^m  
4 \alpha (1-\alpha) \bigl( \delta_{m,m'} + (-1)^{\ell'} \delta_{-m,m'} \bigr) \times}$ } \\ 
& & {\scriptsize ${\displaystyle  [I^{\alpha}_{\ell_1 m_1;\ell' m;\ell_2 m_1-m} I^{\alpha}_{\ell_2 m_1-m;\ell -m;\ell_1 m_1}+ I^{\beta}_{\ell_1 m_1;\ell',m;\ell_2 m_1-m} I^{\beta}_{\ell_2 m_1-m;\ell -m;\ell_1 m_1} ]}$} \\ 
& {\scriptsize $  \tilde{\Pi}_{\rm bulk}^{ff}$} & 
{\scriptsize ${\displaystyle -  \int_0^1 d \alpha  \int^{\Lambda^2}_{0} \frac{x dx}{[x + \Delta ]^2} (-1)^m  
\Bigl[ (x- 2 \alpha(1-\alpha)p^2 ) \bigl( \delta_{m,m'} + (-1)^{\ell'} \delta_{-m,m'} \bigr) \times}$} \\
& & {\scriptsize ${\displaystyle  \bigl[ I^{\alpha}_{\ell_1 m_1;\ell' m;\ell_2 m_1-m} I^{\alpha}_{\ell_2 m_1-m;\ell -m;\ell_1 m_1} 
+ I^{\beta}_{\ell_1 m_1;\ell' m;\ell_2 m_1-m} I^{\beta}_{\ell_2 m_1-m;\ell -m;\ell_1 m_1} \bigr]}$} \\
& & {\scriptsize ${\displaystyle + 2  M_{\ell_1} M_{\ell_2} \bigl( \delta_{m,m'} - (-1)^{\ell'} \delta_{-m,m'} \bigr) \times}$} \\ 
& & {\scriptsize ${\displaystyle  \bigl[ I^{\alpha}_{\ell_1 m_1;\ell' m;\ell_2 m_1-m} I^{\beta}_{\ell_2 m_1-m;\ell -m;\ell_1 m_1} 
+ I^{\beta}_{\ell_1 m_1;\ell' m;\ell_2 m_1-m} I^{\alpha}_{\ell_2 m_1-m;\ell -m;\ell_1 m_1} \bigr] \Bigr]}$} \\
& {\scriptsize $   \Pi_{\rm bound}^{ff}$} &
{\scriptsize ${\displaystyle -   (-1)^{\ell_2+m_1} 
\frac{2}{3}   \bigl( \delta_{2m_1,m+m'} + (-1)^{\ell'} \delta_{2m_1,m-m'}\bigr) \bigl[ I^{\alpha}_{\ell_1 m_1;\ell' 2m_1-m;\ell_2 m-m_1} I^{\alpha}_{\ell_2 m_1-m;\ell -m;\ell_1 m_1} }$} \\ 
& & {\scriptsize ${\displaystyle  - I^{\beta}_{\ell_1 m_1;\ell' 2m_1-m;\ell_2  m-m_1} I^{\beta}_{\ell_2 m_1-m;\ell -m;\ell_1 m_1} \bigr]}$} \\
& {\scriptsize $ \tilde{\Pi}_{\rm bound}^{ff}$} & 
{\scriptsize ${\displaystyle   (-1)^{\ell_2+m_1}   \Bigl[ (M_{\ell_1}^2+M_{\ell_2}^2)   \bigl( \delta_{2m_1,m+m'} + (-1)^{\ell'} \delta_{2m_1,m-m'}\bigr) 
\bigl[ I^{\alpha}_{\ell_1m_1;\ell'2m_1-m;\ell_2m-m_1} I^{\alpha}_{\ell_2m_1-m;\ell-m;\ell_1m_1} }$ } \\ 
& & {\scriptsize ${\displaystyle  - I^{\beta}_{\ell_1m_1;\ell'2m_1-m;\ell_2m-m_1} I^{\beta}_{\ell_2m_1-m;\ell-m;\ell_1m_1} \bigr]}$} \\
& & {\scriptsize ${\displaystyle -2  M_{\ell_1} M_{\ell_2}  \bigl( \delta_{2m_1,m+m'} - (-1)^{\ell'} \delta_{2m_1,m-m'}\bigr)  \bigl[ I^{\alpha}_{\ell_1m_1;\ell' 2m_1-m;\ell_2m-m_1} I^{\beta}_{\ell_2m_1-m;\ell-m;\ell_1m_1}}$} \\ 
& & {\scriptsize ${\displaystyle - I^{\beta}_{\ell_1m_1;\ell'2m_1-m;\ell_2m-m_1} I^{\alpha}_{\ell_2m_1-m;\ell-m;\ell_1m_1} \bigr] \Bigr]}$} \\[1.5pt] \hline 
{\scriptsize Fig.~\ref{Loop4}} & {\scriptsize $ \Pi_{\rm bulk}^{H H}$} & 
{\scriptsize ${\displaystyle - \int_0^1 d \alpha \int^{\Lambda^2}_{0} \frac{x dx}{[x + \Delta ]^2} 
( 1 -4\alpha(1-\alpha) )  \bigl( \delta_{m,m'} + (-1)^{\ell'} \delta_{-m,m'} \bigr)  }$}  {\scriptsize ${\displaystyle J^1_{\ell_1 -m_1;\ell' m;\ell_2 m_1-m} J^1_{\ell_2 m-m_1;\ell -m;\ell_1 m_1} }$} \\
& {\scriptsize $ \tilde{\Pi}_{\rm bulk}^{H H}$} & 
{\scriptsize ${\displaystyle - \int_0^1 d \alpha \int^{\Lambda^2}_{0} \frac{x dx}{[x + \Delta ]^2} 
( x - (1-4\alpha (1-\alpha ))p^2 )  \bigl( \delta_{m,m'} + (-1)^{\ell'} \delta_{-m,m'} \bigr)}  $}  
 {\scriptsize ${\displaystyle J^1_{\ell_1 -m_1;\ell' m;\ell_2 m_1-m} J^1_{\ell_2 m-m_1;\ell -m;\ell_1 m_1} }$} \\
& {\scriptsize $ \Pi_{\rm bound}^{H H}$} & 
{\scriptsize ${\displaystyle -  
\frac{1}{3} (-1)^{\ell_2} \bigl( \delta_{2m_1,m+m'} + (-1)^{\ell'}\delta_{2 m_1,m-m'}  \bigr)  } $}  {\scriptsize ${\displaystyle J^1_{\ell_1 -m_1;\ell' 2m_1-m;\ell_2 -m_1+m} J^1_{\ell_2 m-m_1;\ell -m;\ell_1 m_1} }$} \\[1.5pt]
& {\scriptsize $ \tilde{\Pi}_{\rm bound}^{H H}$} & 
{\scriptsize ${\displaystyle   
(M_{\ell_1}^2+M_{\ell_2}^2) (-1)^{\ell_2} \bigl( \delta_{2m_1,m+m'} + (-1)^{\ell'}\delta_{2 m_1,m-m'}  \bigr)  }$}  
{\scriptsize ${\displaystyle J^1_{\ell_1 -m_1;\ell' 2m_1-m;\ell_2 -m_1+m } J^1_{\ell_2 m-m_1;\ell -m;\ell_1 m_1} }$}  \\ \hline
{\scriptsize Fig.~\ref{Loop5}} & {\scriptsize $\tilde{\Pi}_{\rm bulk}^{H}$} & 
{\scriptsize ${\displaystyle \int^{\Lambda^2}_0  \frac{4 x dx}{x+M_{\ell_1}^2} (-1)^{m+m_1} \delta_{m,m'}K^1_{\ell -m;\ell_1 m_1;\ell_1 -m_1;\ell' m'} }$} \\
& {\scriptsize $\bar{\Pi}_{\rm bound}^{H}$} & 
{\scriptsize ${\displaystyle - 4 M_{\ell_1}^2  (-1)^{\ell_1+m+m_1} \delta_{2m_1,m-m'} K^1_{\ell -m;\ell_1m_1;\ell_1m_1;\ell'm'} }$} 
\\ \hline
\end{tabular}
\vspace{-1ex}
\caption{The contributions from fermions and Higgs boson loop for correction to KK masses of gauge boson. 
The summation symbols and the Log divergence are omitted as in the Table~\ref{OneLoopF}. The overall factor $Tr[T^a_i T^a_j] g_{6a}^2/64 \pi^2 R^2$ is also omitted for each expression. 
A coefficient providing zero contribution is omitted. 
 \label{OneLoopG}} \vspace{-1ex}
\end{table}

\section{One loop calculations \label{LoopCalculation}}

Here we list the one-loop corrections to the KK masses for fermions, gauge bosons and a Higgs boson.

\subsection{One loop corrections to KK masses of fermion}
For fermions, the corresponding one-loop diagrams are shown in Figs.~\ref{Loop1} and \ref{Loop2}, and 
the contributions from each diagram are expressed as Eqs.~(\ref{LoopFbulk}) and (\ref{LoopFbound}).
We have calculated a diagram in Fig.~\ref{Loop1} at Sec.~\ref{Sec:OneLoop} for four dimensional components of gauge boson inside a loop and got Eqs.~(\ref{SigmaFig1a})-(\ref{SigmaFig1d}), 
which are summarized in the first part of Table~\ref{OneLoopF}. 
The contribution from $\phi_i$ to the diagram in Fig.~\ref{Loop1} is obtained by using $\bar{\Psi} \phi_i \Psi  $ vertices in Table~\ref{vertices-F} and $\phi_i$ propagator 
Eq.~(\ref{propa-Gex}). 
We then obtain the 2nd and the 3rd part of Table~\ref{OneLoopF}.

The Fig.~\ref{Loop2} can be calculated in the same way by using Yukawa coupling constants in Table~\ref{vertices-F} and scalar boson propagator Eq.~(\ref{propa-S}),
\begin{align}
\label{LoopEx}
-i \Sigma^{\rm Fig.2}(p; \ell m; \ell' m') = & \frac{1}{4R^2}
\sum_{\ell_1=0}^{\ell_{max}} \sum_{\ell_2=0}^{\ell_{max}} \sum_{m_1=-\ell_1}^{\ell_1} \sum_{m_1'=-\ell_1}^{\ell_1} \int_\Lambda \frac{d^4 k}{(2 \pi)^4} \nonumber \\
& \times \biggl[ \frac{i}{(p-k)^2-M^2_{\ell_2}} (\delta_{m_1-m, m_1'-m'}+(-1)^{\ell_2} \delta_{-(m_1-m),m_1'-m'}) \nonumber \\
& \quad \times (i Y_f ) [I^\alpha_{\ell_1 m_1'; \ell_2 m_1'-m'; \ell' m'} P_R +I^\beta_{\ell_1 m_1'; \ell_2 m_1'-m'; \ell' m' } P_L ] \nonumber \\
& \quad \times \frac{i}{\sla{k}+i \gamma_5 M_{\ell_1}} (\delta_{m_1, m_1'} + (-1)^{\ell_1+m_1} \delta_{-m_1, m_1'} \gamma_5) \nonumber \\
& \quad \times (i Y_f ) [I^\alpha_{\ell m; \underline{\ell_2 m_1-m}; \ell_1 m_1 } P_L+I^\beta_{\ell m; \underline{\ell_2 m_1-m}; \ell_1 m_1} P_R] \biggr],
\end{align}
where $Y_f$ is the corresponding Yukawa coupling constant in six-dimensions. 
The RHS is expressed as the case of Fig.~\ref{Loop1} and the results are summarized in the 4th  part of Table~\ref{OneLoopF}
%
%
\begin{table}[t] \vspace{1ex}
\begin{tabular}{c||c|l} 
{\scriptsize Diagram} & {\scriptsize Coefficients} &  \\ \hline
{\scriptsize Fig.~\ref{Loop6}} & {\scriptsize $  \Pi_{\rm bulk}^{A_\mu A_\mu}$} & 
{\scriptsize ${\displaystyle  \int^1_0 d\alpha \int_0^{\Lambda^2} dx \frac{-x}{2 [x+\Delta]^2}  [2(1-2\alpha)^2-2(1+\alpha)(2-\alpha)](\delta_{mm'}+(-1)^{\ell'}\delta_{-mm'}) J^1_{\ell'm;\underline{\ell_1m_1};\ell_2m_1-m} J^1_{\underline{\ell m}; \ell_1m_1;\underline{\ell_2m_1-m} } }$} \\ [1.5pt]
& {\scriptsize $ \tilde{\Pi}_{\rm bulk}^{A_\mu A_\mu}$} & 
{\scriptsize ${\displaystyle  \int^1_0 d\alpha \int_0^{\Lambda^2} dx \frac{x}{2[x+\Delta]^2} 
\Bigl[  [(2-\alpha)^2+3(\alpha^2-1)+ 2(1-2\alpha)^2 ]p^2 -  9x/2  \Bigr] (\delta_{mm'}+(-1)^{\ell'}\delta_{-mm'}) \times}$} \\
& & {\scriptsize ${\displaystyle  J^1_{\ell'm;\underline{\ell_1m_1}; \ell_2m_1-m}  J^1_{\underline{\ell m}; \ell_1m_1;\underline{\ell_2m_1-m} }  }$} \\[1.5pt]
& {\scriptsize $ \Pi_{\rm bound}^{A_\mu A_\mu}$} & 
{\scriptsize ${\displaystyle  (-1)^{l_2} \frac{11}{6}  (\delta_{2m_1,m+m'}+(-1)^{\ell'}\delta_{2m_1,m-m'}) }$} 
{\scriptsize ${\displaystyle J^1_{ \ell'2m_1-m;\underline{\ell_1m_1};\ell_2m-m_1 } J^1_{ \underline{\ell m};\ell_1m_1;\underline{\ell_2m_1-m} } }$} \\[1.5pt]
& {\scriptsize $ \tilde{\Pi}_{\rm bound}^{A_\mu A_\mu}$} & 
{\scriptsize ${\displaystyle  -    \biggl[ \frac{1}{4} p^2 - \frac{9}{4}(M_{l_1}^2+M_{l_2}^2) \biggr]  (-1)^{\ell_2} (\delta_{2m_1,m+m'}+(-1)^{\ell'}\delta_{2m_1,m-m'}) }$} 
{\scriptsize ${\displaystyle J^1_{\ell'2m_1-m;\underline{\ell_1m_1};\ell_2m-m_1 } J^1_{ \underline{\ell m}; \ell_1m_1;\underline{\ell_2m_1-m}}   }$} \\[1.5pt] \hline
{\scriptsize Fig.\ref{Loop7}} & {\scriptsize $ \tilde{\Pi}_{\rm bulk}^{A_\mu} $} &
{\scriptsize ${\displaystyle    \int_0^1 d \alpha \int^{\Lambda^2}_0 dx \frac{6 x}{[x+\Delta']^2}[x-(1-\alpha)^2p^2] 
\delta_{mm'}  K^1_{\ell'm;\underline{\ell m}; \ell_1m_1; \underline{\ell_1m_1}}  }$} \\ 
& {\scriptsize $ \tilde{\Pi}_{\rm bound}^{A_\mu} $} & 
{\scriptsize ${\displaystyle -  6  \ln \Bigl( \frac{\Lambda^2}{\mu^2} \Bigr) M_{\ell_1}^2(-1)^{\ell_1} \delta_{2m_1,m-m'} 
K^1_{ \ell'-2m_1+m;\underline{\ell m};\ell_1m_1; \underline{\ell_1-m_1} }  }$} \\ \hline
{\scriptsize Fig.~\ref{Loop4}} & {\scriptsize $ \Pi_{\rm bulk}^{c}$ } & 
{\scriptsize ${\displaystyle -  \int^1_0 d\alpha \int_0^{\Lambda^2} dx \frac{x}{[x+\Delta]^2} \alpha(1-\alpha) (\delta_{mm'}+(-1)^{\ell'}\delta_{-mm'})}$} 
{\scriptsize ${\displaystyle J^1_{ \underline{\ell_1m_1}; \ell'm; \ell_2m_1-m } J^1_{ \underline{\ell_2m_1-m};\underline{\ell m};l_1m_1 } }$} \\
& {\scriptsize $ \tilde{\Pi}_{\rm bulk}^{cc}$} & 
{\scriptsize ${\displaystyle \int^1_0 d\alpha \int_0^{\Lambda^2} dx \frac{x}{[x+\Delta]^2} \biggl(  \frac{x}{4}+ \alpha(1-\alpha)p^2 \biggr) (\delta_{mm'}+(-1)^{\ell'}\delta_{-mm'}) }$} 
{\scriptsize ${\displaystyle J^1_{ \underline{\ell_1m_1};\ell'm; \ell_2m_1-m } J^1_{ \underline{\ell_2m_1-m};\underline{\ell m}; \ell_1m_1 } }$} \\
& {\scriptsize $ \Pi_{\rm bound}^{cc}$} & 
{\scriptsize ${\displaystyle - (-1)^{\ell_2} \frac{1}{6}   (\delta_{2m_1,m+m'}+(-1)^{\ell'}\delta_{2m_1,m-m'}) }$} 
{\scriptsize ${\displaystyle J^1_{ \underline{\ell_1m_1};  \ell'2m_1-m; \ell_2m-m_1 } J^1_{ \underline{\ell_2m_1-m}; \underline{\ell m}; \ell_1m_1 }  }$} \\
& {\scriptsize $ \tilde{\Pi}_{\rm bound}^{cc}$} & 
{\scriptsize ${\displaystyle  (-1)^{\ell_2} \frac{1}{4}   \Bigl( p^2- M_{l_1}^2-M_{l_2}^2 \Bigr) (\delta_{2m_1,m+m'}+(-1)^{\ell'}\delta_{2m_1,m-m'})}$} 
 {\scriptsize ${\displaystyle J^1_{ \underline{\ell_1m_1}; \ell'2m_1-m;\ell_2m-m_1 } J^1_{ \underline{\ell_2m_1-m}; \underline{\ell m};\ell_1m_1 } }$} \\ \hline
{\scriptsize Fig.~\ref{Loop6}} & {\scriptsize $ \tilde{\Pi}_{\rm bulk}^{A_\mu \phi_1(\phi_2)}$} &
{\scriptsize ${\displaystyle - \frac{1}{R^2} \int^1_0 d\alpha \int_0^{\Lambda^2} dx \frac{x}{[x+\Delta]^2}  
(\delta_{mm'}+(-1)^{\ell'}\delta_{-mm'}) J^{4(5)}_{ \ell'm;\underline{\ell_1m_1}; \ell_2m_1-m } J^{4(5)}_{ \underline{\ell m}; \ell_1m_1;\underline{\ell_2m_1-m} }  }$} \\
& {\scriptsize $ \tilde{\Pi}_{\rm bound}^{A_\mu \phi_1 (\phi_2)}$} & 
{\scriptsize ${\displaystyle  -  \frac{(-1)^{\ell_2}}{R^2}   (\delta_{2m_1,m+m'}+(-1)^{\ell'}\delta_{2m_1,m-m'}) 
J^{4(5)}_{ \ell'2m_1-m;\underline{\ell_1m_1}; \ell_2m-m_1 } J^{4(5)}_{ \underline{\ell m}; \ell_1m_1;\underline{\ell_2m_1-m} } }$} \\ \hline
{\scriptsize Fig.~\ref{Loop7}} & {\scriptsize $ \tilde{\Pi}_{\rm bulk}^{\phi_i}$} & 
{\scriptsize ${\displaystyle   \int^{\Lambda^2}_0 dx \frac{2x}{x+M_{\ell_1}^2} \delta_{mm'}  K^2_{ \underline{\ell_1m_1}; \ell_1m_1; \ell'm; \underline{\ell m} } }$} \\
& {\scriptsize $ \tilde{\Pi}_{\rm bound}^{\phi_i}$} & 
{\scriptsize ${\displaystyle -    2   M_{\ell_1}^2 
(-1)^{\ell_1} \delta_{2m_1,m-m'} K^2_{ \underline{\ell_1,-m_1};\ell_1m_1;\ell' -2m_1+m; \underline{\ell m} }  }$} \\ \hline
{\scriptsize Fig.~\ref{Loop6}} & {\scriptsize $ \Pi_{\rm bulk}^{\phi_i \phi_i (\phi_1 \phi_2)}$} & 
{\scriptsize ${\displaystyle - \int^1_0 d\alpha \int_0^{\Lambda^2} dx \frac{x}{[x+\Delta]^2} \biggl( 1-4 \alpha(1-\alpha) \biggr)
\displaystyle  J^{6(7)}_{ \ell_2m_1-m;\underline{\ell_1m_1};\ell'm } J^{6(7)}_{ \underline{\ell_2m_1-m};\ell_1m_1;\underline{\ell m} } (\delta_{mm'}+(-1)^{\ell'}\delta_{-mm'})}$} \\
& {\scriptsize $ \bar{\Pi}_{\rm bulk}^{\phi_i \phi_i (\phi_1 \phi_2)}$} & 
{\scriptsize ${\displaystyle - \int^1_0 d\alpha \int_0^{\Lambda^2} dx \frac{x}{[x+\Delta]^2} \biggl(x+ [4 \alpha(1-\alpha)-1]p^2 \biggr)}$} 
{\scriptsize ${\displaystyle J^{6(7)}_{ \ell_2m_1-m;\underline{\ell_1m_1};\ell'm } J^{6(7)}_{ \underline{\ell_2m_1-m};\ell_1m_1;\underline{\ell m} } (\delta_{mm'}+(-1)^{\ell'}\delta_{-mm'})}$} \\
& {\scriptsize $ \Pi_{\rm bound}^{\phi_i \phi_i (\phi_1 \phi_2)}$} & 
{\scriptsize ${\displaystyle -  (-1)^{l_2} \frac{1}{3}   
J^{6(7)}_{ \ell_2m-m_1;\underline{\ell_1m_1};\ell'2m_1-m } J^{6(7)}_{ \underline{\ell_2m_1-m};\ell_1m_1;\underline{\ell m} } (\delta_{2m_1,m+m'}+(-1)^{\ell'}\delta_{2m_1,m-m'})}$} \\
& {\scriptsize $ \bar{\Pi}_{\rm bound}^{\phi_i \phi_i (\phi_1 \phi_2)}$} & 
{\scriptsize ${\displaystyle (-1)^{\ell_2}   \Bigl(  M_{\ell_1}^2+M_{\ell_2}^2 \Bigr) }$} 
{\scriptsize ${\displaystyle J^{6(7)}_{ \ell_2m-m_1;\underline{\ell_1m_1};\ell'2m_1-m } J^{6(7)}_{ \underline{\ell_2m_1-m};\ell_1m_1;\underline{\ell m} } (\delta_{2m_1,m+m'}+(-1)^{\ell'}\delta_{2m_1,m-m'})}$}  \\ \hline
\end{tabular}
\vspace{-1ex}
\caption{The contributions from gauge boson loop with self interactions for correction to KK masses of gauge boson. 
The summation symbols and the Log divergence factor  are omitted as in the Table~\ref{OneLoopF}. The overall factor $C_2(G) g_{6a}^2/64 \pi^2 R^2$ is also omitted
where $C_2(G)$ is obtained as $f^{lmi} f^{lmj} = C_2(G) \delta^{ij}$ for corresponding gauge group $G$.\label{OneLoopG-NA}} \vspace{-1ex}
\end{table}

%

\subsection{One loop corrections to KK masses of four dimensional components gauge boson}
For gauge bosons, the corresponding one-loop diagrams are shown in Figs.~\ref{Loop3}-\ref{Loop7}, 
and they are calculated as in the fermion case.
The Fig.~\ref{Loop3} for four dimensional components of gauge bosons can be obtained by using relevant propagators of fermion Eq.~(\ref{propa-F}) and $\bar{\Psi} A_\mu \Psi$ vertices 
in Table~\ref{vertices-F}, 
\begin{align}
i \Pi_{\mu \nu}^{\rm Fig.\ref{Loop3}}(p;\ell m, \ell' m') = & -\frac{1}{4 R^2} \sum_{\ell_1=0}^{\ell_{max}} \sum_{\ell_2=0}^{\ell_{max}} \sum_{m_1=-\ell_1}^{\ell_1} \sum_{m_1'=-\ell_1}^{\ell_1} 
\int_\Lambda \frac{d^4 k}{(2 \pi)^4} \times \nonumber \\
& {\rm Tr} \biggl[ (i g_{6a} T_a \gamma_\mu) \Bigl[ I^\alpha_{\ell_1 m_1'; \ell' m'; \ell_2 m_1' -m'} P_R + I^\beta_{\ell_1 m_1'; \ell' m'; \ell_2 m_1' -m'} P_L \Bigr] \times \nonumber \\
& \frac{i}{\sla{k} + i \gamma_5 M_{\ell_1}} (\delta_{m_1, m_1'} \mp (-1)^{\ell_1+m_1} \delta_{-m_1,m_1'} \gamma_5) \times \nonumber \\
& (i g_{6a} T_a \gamma_\nu) \Bigl[ I^\alpha_{\ell_2 m_1-m; \underline{\ell m}; \ell_1 m_1} P_R + I^\beta_{\ell_2 m_1-m; \underline{\ell m}; \ell_1 m_1} P_L \Bigr] \times  \nonumber \\
& \frac{i}{\sla{k}-\sla{p}+ i\gamma_5 M_{\ell_2}} (\delta_{m_1-m,m_1'-m'} \mp (-1)^{\ell_2 +m_1 +m} \delta_{-(m_1-m),m_1'-m'} \gamma_5) \biggr]
\end{align}
where the sign $\pm$ corresponds to $\Psi_+^{ (\pm \gamma_5 )}$ in the loop.
Taking the sum over $m_1'$, the products of Kronecker delta become $m$-conserving $\{ \delta_{m,m'}, \delta_{-m,m'} \}$ and $m$-violating $\{\delta_{2m_1,m+m'}, \delta_{2m_1,m-m'} \}$, 
and we separate the bulk and the boundary contribution as in the femion case, 
$i \Pi_{\mu \nu}^{\rm Fig.\ref{Loop3}}(p;\ell m, \ell' m')=i \Pi_{\mu \nu}^{\rm Fig.\ref{Loop3} \ bulk}(p;\ell m, \ell' m')+i \Pi_{\mu \nu}^{\rm Fig.\ref{Loop3} \ bound}(p;\ell m, \ell' m')$.
After combining denominators with Feynman parameter and carrying out Wick rotation, a bulk contribution is estimated with Euclidean four momentum integration with cut-off $\Lambda$ and 
a boundary contribution is estimated taking leading log-divergent part as in Eq.~(\ref{logdiv}).
After  some calculations, we can express it in the form of Eqs.~(\ref{OneLoopGSeparate}) and (\ref{OneLoopGBulkBound}) where each coefficients are given in 
the first part of the Table~\ref{OneLoopG}.

The Fig.~\ref{Loop4} is calculated by making use of propagator of scalar boson Eq.~(\ref{propa-S}) and $H^\dagger A_\mu H$ vertex in Table~\ref{vertices-H} such that
\begin{align}
i\Pi_{\mu \nu}^{{\rm Fig.\ref{Loop4} }}(p; \ell m; \ell' m') =& \frac{1}{4R^2} \sum_{\ell_1=0}^{\ell_{max}} \sum_{\ell_2=0}^{\ell_{max}} \sum_{m_1=-\ell_1}^{\ell_1} \sum_{m_1'=-\ell_1}^{\ell_1}
\int \frac{d^4 k}{(2\pi)^4} J^1_{\underline{\ell_1 m_1'};\ell' m'; \ell_2 m_1'-m'} J^1_{\underline{\ell_2 m_1-m};\underline{\ell m};\ell_1,m_1} \nonumber \\
&  \times \biggl[\{ i g_{6a} T_a (2k_{\mu}-p_{\mu}) \} \frac{i}{k^2-M_{\ell_1}^2} 
\bigl( \delta_{m_1,m_1'} + (-1)^{\ell_1} \delta_{-m_1,m_1'}  \bigr) \nonumber \\
&  \quad \times \{ i g_{6a} T_a (2k_{\nu}-p_{\nu}) \} \frac{i}{(k-p)^2-M_{\ell_2}^2}
\bigl(\delta_{m_1-m,m_1'-m'} + (-1)^{\ell_2}\delta_{-(m_1-m),m_1'-m'}  \bigr) \biggr], \nonumber \\   
\end{align}
which is also expressed as the form of Eqs.~(\ref{OneLoopGSeparate}) and (\ref{OneLoopGBulkBound}), and the explicit form of each coefficients are summarized in the 2nd part of Table~\ref{OneLoopG}.

The Fig.~\ref{Loop5} is calculated by making use of propagator of scalar boson Eq.~(\ref{propa-S}) and $H^\dagger HA_\mu A^\mu$ vertex in Table~\ref{vertices-H} such that
\begin{align}
i \Pi_{\mu \nu}^{\rm Fig.~\ref{Loop5}}(\ell m; \ell' m') =& \frac{1}{2R^2} \sum_{\ell_1=0}^{\ell_{max}} \sum_{m_1=-\ell_1}^{\ell_1} \sum_{m_1'=-\ell_1}^{\ell_1}
\int \frac{d^4 k}{(2\pi)^4} (2 i g_{6a}^2 T_a^2 g_{\mu \nu}) \frac{i}{k^2-M_{\ell_1}^2} \nonumber \\
&  \times K^1_{\underline{\ell,m};\ell_1,m_1;\underline{\ell_1,m_1'};\ell',m'} 
[\delta_{m_1,m_1'}+(-1)^{\ell_1} \delta_{-m_1,m_1'}] \delta_{m'-m_1'+m_1-m,0} 
\end{align}
which is expressed as the form of Eqs.~(\ref{OneLoopGSeparate}) and (\ref{OneLoopGBulkBound}), and the explicit form of each coefficients are summarized in the 3rd part of Table~\ref{OneLoopG}.

The Fig.~\ref{Loop6} for two virtual $A_\mu$ is calculated by making use of propagator of $A_\mu$ Eq.~(\ref{propa-G}) and $(A_\mu)^3$ vertex in Table~\ref{vertices-A} such that
\begin{align}
i\Pi_{\mu \nu}^{{\rm Fig.\ref{Loop6}(A_\mu A_\mu) }}(p; \ell m; \ell' m') =& \frac{1}{2} \sum_{\ell_1=0}^{\ell_{max}} \sum_{\ell_2=0}^{\ell_{max}} \sum_{m_1=-\ell_1}^{\ell_1} \sum_{m_1'=-\ell_1}^{\ell_1} 
\int \frac{d^4 k}{(2 \pi)^4} \frac{-i}{k^2-M_{\ell_1}^2} g_{\alpha \beta} \frac{-i}{(k+p)^2-M_{\ell_2}^2} g_{\rho \sigma} \nonumber \\
&\times \frac{1}{2}[\delta_{m_1m_1'}+(-1)^{\ell_1} \delta_{-m_1m_1'} ] \frac{1}{2}[\delta_{m_1-m,m_1'-m'}+(-1)^{\ell_2}\delta_{-(m_1-m),m_1'-m'}] \nonumber \\
& \times \frac{g_{6a}}{R} f^{ikl}[g^{\mu \alpha}(p-k)^{\rho}+g^{\alpha \rho} (p+2k)^{\mu} +g^{\rho \mu}(-2p-k)^{\alpha}] \nonumber \\
& \times \frac{g_{6a}}{R} f^{jkl}[g^{\beta \nu}(p-k)^{\sigma}+g^{\nu \sigma}(-2p-k)^{\beta} +g^{\sigma \beta} (p+2k)^{\nu}] \nonumber  \\
& \times J^1_{\ell'm';\underline{\ell_1m_1'};\ell_2m_1'-m' } J^1_{ \underline{\ell m};\ell_1m_1;\underline{\ell_2m_1-m}}
\end{align}
which is expressed as the form of Eqs.~(\ref{OneLoopGSeparate}) and (\ref{OneLoopGBulkBound}) and the explicit form of each coefficients are summarized in the 1st part of Table~\ref{OneLoopG-NA}.
The Fig.~\ref{Loop6} for one virtual $A_\mu$ and one virtual $\phi_i$ is calculated in the same way by making use of propagators of $A_\mu$ in Eq.~(\ref{propa-G}) 
and $\phi_i$ in Eq.~(\ref{propa-Gex}), and $(A_\mu)^2 \phi_i$ vertex in Table~\ref{vertices-A}.
The results are summarized in the 4th part of Table~\ref{OneLoopG-NA}. 
The Fig.~\ref{Loop6} for two virtual $\phi_i$ is also calculated by making use of  
propagator of $\phi_i$ Eq.~(\ref{propa-Gex}) and $A_\mu (\phi_i)^2 $ vertex in Table~\ref{vertices-A}, and
the results are summarized in the 6th part of Table~\ref{OneLoopG-NA}. 
%
\begin{table}[t] \vspace{1ex}
\begin{tabular}{c||c|l} 
{\scriptsize Diagram} & {\scriptsize Coefficients} &  \\ \hline
{\scriptsize Fig.~\ref{Loop3}} & {\scriptsize $ \Theta_{\rm bulk}^{(i) ff}$} &
{\scriptsize ${\displaystyle  -  \int_0^1 d \alpha \int^{\Lambda^2}_{0} \frac{x dx}{[x + \Delta_M ]^2} (-1)^m
\biggl[   \frac{6x+2 \Delta_M}{x+\Delta_M} \alpha (1-\alpha) \bigl( \delta_{m,m'} + (-1)^{\ell'} \delta_{-m,m'} \bigr)\times}  $} \\
& & {\scriptsize ${\displaystyle \bigl[ C^{\alpha}_{\ell_1m_1;\ell'm;\ell_2m_1-m} C^{\beta}_{\ell_2m_1-m;\ell-m;\ell_1m_1}  
+ C^{\beta}_{\ell_1m_1;\ell'm;\ell_2m_1-m} C^{\alpha}_{\ell_2m_1-m;\ell-m;\ell_1m_1} \bigr] }$} \\
& & {\scriptsize ${\displaystyle - 4 M_{\ell_1} M_{\ell_2} \frac{ \alpha (1- \alpha )}{x + \Delta_M}  \bigl( \delta_{m,m'} - (-1)^{\ell'} \delta_{-m,m'} \bigr) \times}$} \\
& & {\scriptsize ${\displaystyle \bigl[ C^{\alpha}_{\ell_1m_1;\ell'm;\ell_2m_1-m} C^{\alpha}_{\ell_2m_1-m;\ell-m;\ell_1m_1}
+ C^{\beta}_{\ell_1m_1;\ell'm;\ell_2m_1-m} C^{\beta}_{\ell_2m_1-m;\ell-m;\ell_1m_1} \bigr] \biggr]}$} \\
& {\scriptsize $ \tilde{\Theta}_{\rm bulk}^{(i) ff}$} & 
{\scriptsize ${\displaystyle  \int_0^1 d \alpha  \int^{\Lambda^2}_{0} \frac{x dx}{[x + \Delta_M ]^2} (-1)^m} 
\biggl[ 2 x  \bigl( \delta_{m,m'} + (-1)^{\ell'} \delta_{-m,m'} \bigr) \times $}  \\
& & {\scriptsize ${\displaystyle \bigl[ C^{\alpha}_{\ell_1m_1;\ell'm;\ell_2m_1-m} C^{\beta}_{\ell_2m_1-m;\ell-m;\ell_1m_1} 
+ C^{\beta}_{\ell_1m_1;\ell'm;\ell_2m_1-m} C^{\alpha}_{\ell_2m_1-m;\ell-m;\ell_1m_1} \bigr]}$} \\
& & {\scriptsize ${\displaystyle - 2 M_{\ell_1} M_{\ell_2} \bigl( \delta_{m,m'} - (-1)^{\ell'} \delta_{-m,m'} \bigr) \times}$} \\
& & {\scriptsize ${\displaystyle \bigl[ C^{\alpha}_{\ell_1m_1;\ell'm;\ell_2m_1-m} C^{\alpha}_{\ell_2m_1-m;\ell-m;\ell_1m_1}
+ C^{\beta}_{\ell_1m_1;\ell'm;\ell_2m_1-m} C^{\beta}_{\ell_2m_1-m;\ell-m;\ell_1m_1} \bigr] \biggr]}$} \\
& {\scriptsize $\Theta_{\rm bound}^{(i) ff}$} &
{\scriptsize ${\displaystyle -  (-1)^{\ell_2+m_1}    
\Bigl[  \bigl( \delta_{2m_1,m+m'} + (-1)^{\ell'} \delta_{2m_1,m-m'}\bigr) \bigl[ C^{\alpha}_{\ell_1m_1;\ell'2m_1-m;\ell_2m-m_1} C^{\beta}_{\ell_2m_1-m;\ell-m;\ell_1m_1} }$} \\
& & {\scriptsize ${\displaystyle - C^{\beta}_{\ell_1m_1;\ell'2m_1-m;\ell_2m-m_1} C^{\alpha}_{\ell_2m_1-m;\ell-m;\ell_1m_1} \bigr] \Bigr] }$} \\
& {\scriptsize $\tilde{\Theta}_{\rm bound}^{(i) ff}$} & 
{\scriptsize ${\displaystyle  - (-1)^{\ell_2+m_1}  \Bigl[ ( M_{\ell_1}^2 + M_{\ell_2}^2)  \bigl(  \delta_{2m_1,m+m'} + (-1)^{\ell'} \delta_{2m_1,m-m'}\bigr)
\bigl[ C^{\alpha}_{\ell_1m_1;\ell'2m_1-m;\ell_2m-m_1} C^{\beta}_{\ell_2m_1-m;\ell-m;\ell_1m_1} }$} \\
& & {\scriptsize ${  - C^{\beta}_{\ell_1m_1;\ell'2m_1-m;\ell_2m-m_1} C^{\alpha}_{\ell_2m_1-m;\ell-m;\ell_1m_1} \bigr]}$} \\
& & {\scriptsize ${\displaystyle + 2 M_{\ell_1} M_{\ell_2}  \bigl( \delta_{2m_1,m+m'} - (-1)^{\ell'} \delta_{2m_1,m-m'}\bigr) 
\bigl[ C^{\alpha}_{\ell_1m_1;\ell'2m_1-m;\ell_2m-m_1} C^{\alpha}_{\ell_2m_1-m;\ell-m;\ell_1m_1} } $} \\
& & {\scriptsize ${\displaystyle  - C^{\beta}_{\ell_1m_1;\ell'2m_1-m;\ell_2m-m_1} C^{\beta}_{\ell_2m_1-m;\ell-m;\ell_1m_1} \bigr] \Bigr]}$} \\ \hline 
{\scriptsize Fig.~\ref{Loop4}} & {\scriptsize ${\Theta_{\rm bulk}^{(1(2)) HH }}$} & 
{\scriptsize ${\displaystyle - \frac{1}{R^2} \int_0^1 d \alpha \int^{\Lambda^2}_{0} \frac{x dx}{[x + \Delta_M ]^2} \frac{2 \alpha (1-\alpha)}{x+\Delta_M} 
\bigl( \delta_{m,m'} + (-1)^{\ell'} \delta_{-m,m'} \bigr) \times }$} \\
& & {\scriptsize ${\displaystyle \{ J^{2(3)}_{\ell_1-m_1;\ell'm;\ell_2m_1-m} - J^{2(3)}_{\ell_2m_1-m; \ell'm; \ell_1-m_1} \} \{ J^{2(3)}_{\ell_2m-m_1;\ell-m;\ell_1m_1} - J^{2(3)}_{\ell_1m_1;\ell-m;\ell_2m-m_1} \} }$} \\
& {\scriptsize ${\tilde{\Theta}_{\rm bulk}^{(1(2))HH}}$} & 
{\scriptsize ${\displaystyle  \frac{1}{R^2} \int_0^1 d \alpha \int^{\Lambda^2}_{0} \frac{ x dx}{[x + \Delta_M ]^2} 
\bigl( \delta_{m,m'} + (-1)^{\ell'} \delta_{-m,m'} \bigr) \times}$} \\ 
& & {\scriptsize ${\displaystyle \{ J^{2(3)}_{\ell_1-m_1;\ell'm;\ell_2m_1-m}- J^{2(3)}_{\ell_2m_1-m; \ell'm; \ell_1-m_1} \} \{ J^{2(3)}_{\ell_2m-m_1;\ell-m;\ell_1m_1}- J^{2(3)}_{\ell_1m_1;\ell-m;\ell_2m-m_1 } \}}$} \\
& {\scriptsize $\tilde{\Theta}_{\rm bound}^{(1(2)) HH}$} & 
{\scriptsize ${\displaystyle   \frac{(-1)^{\ell_2} }{R^2}
\bigl( \delta_{2m_1,m+m'} + (-1)^{\ell'}\delta_{2 m_1,m-m'}  \bigr) \times}$} \\
& & {\scriptsize ${\displaystyle \{ J^{2(3)}_{\ell_1 -m_1;\ell' 2m_1-m;\ell_2 m-m_1} - J^{2(3)}_{\ell_2m-m_1;\ell'2m_1-m;\ell_1-m_1} \} }$} 
 {\scriptsize ${\displaystyle \{ J^{2(3)}_{\ell_2 m-m_1;\ell-m;\ell_1m_1} -  J^{2(3)}_{ \ell_1 m_1;\ell -m;\ell_2 m-m_1} \}}$} \\ \hline
{\scriptsize Fig.~\ref{Loop5}} & {\scriptsize $\tilde{\Theta}_{\rm bulk}^{(i) H}$} & 
{\scriptsize ${\displaystyle   \int^{\Lambda^2}_0  \frac{4 x dx}{x+M_{\ell_1}^2} 
(-1)^{m+m_1} \delta_{m,m'}K^2_{\ell -m; \ell'm ; \ell_1-m_1;\ell_1m_1} }$} \\
& {\scriptsize $\tilde{\Theta}_{\rm bound}^{(i) H}$} & 
{\scriptsize ${\displaystyle - 4 M_{\ell_1}^2  
(-1)^{\ell_1+m-m_1} \delta_{2m_1,m-m'} K^2_{ \ell-m;\ell'm' ; \ell_1 m_1;\ell_1 -m_1} }$} \\ \hline
\end{tabular}
\vspace{-1ex}
\caption{The contributions from fermions and Higgs boson loop  for correction to KK masses of extra components of gauge bosons. 
The summation symbols and the Log divergence are omitted as in the Table~\ref{OneLoopF}. The overall factor $Tr[T^a_iT^a_j] g_{6a}^2/64 \pi^2 R^2$ is also omitted for each expression. \label{OneLoopGex}} \vspace{-1ex}
\end{table}

%
The Fig.~\ref{Loop7} for virtual $A_\mu$ is calculated by making use of propagator of $A_\mu$ in Eq.~(\ref{propa-G}) and $(A_\mu)^4$ vertex in Table~\ref{vertices-A} such that
\begin{align}
i\Pi_{\mu \nu}^{{\rm Fig.\ref{Loop7}(A_\mu) }}(p; \ell m; \ell' m')=& \frac{1}{2} \sum_{\ell_1=0}^{\ell_{max}} \sum_{m_1=-\ell_1}^{\ell_1} \sum_{m_1'=-\ell_1}^{\ell_1}
\int \frac{d^4k}{(2\pi)^4} \frac{-i}{k^2-M_{\ell_1}^2} g_{\sigma \rho} \delta^{kl} \nonumber \\
& \biggl( -i \frac{g_{6a}^2}{R^2} \biggr) [ f^{ijm}f^{klm}(g^{\mu \rho}g^{\nu \sigma}-g^{\mu \sigma}g^{\nu \rho}) 
+ f^{ikm}f^{jlm}(g^{\mu \nu}g^{\rho \sigma}-g^{\mu \sigma}g^{\nu \rho}) \nonumber \\ 
& \qquad \qquad  +  f^{ilm}f^{jkm}(g^{\mu \nu}g^{\rho \sigma}-g^{\mu \rho}g^{\nu \sigma})] \nonumber \\
& \times \frac{1}{2}[\delta_{m_1m_1'}+(-1)^{\ell_1} \delta_{-m_1m_1'} ] \delta_{m'-m_1'+m_1-m,0}  K^1_{\ell'm';\underline{\ell m}; \ell_1m_1; \underline{\ell_1m_1'}}
\end{align}
which is expressed as the form of Eqs.~(\ref{OneLoopGSeparate}) and (\ref{OneLoopGBulkBound}) and the explicit form of each coefficients are summarized in the 2nd part of Table~\ref{OneLoopG-NA}.
The Fig.~\ref{Loop7} for virtual $\phi_i$ is calculated in the same way by making use of propagator of $\phi_i$ in Eq.~(\ref{propa-Gex}) and $(A_\mu)^2(\phi_i)^2$ vertex in Table~\ref{vertices-A}.
The results are summarized in the 5th part of Table~\ref{OneLoopG-NA}. 

The Fig.~\ref{Loop4} for Ghost contribution is calculated by making use of propagator of scalar boson in Eq.~(\ref{propa-S}) and $\bar{c} c A_\mu$ vertex in Table~\ref{vertices-A} such that
\begin{align}
i\Pi_{\mu \nu}^{{\rm Fig.\ref{Loop4}(cc) }}(p; \ell m; \ell' m')=& (-1) \sum_{\ell_1=0}^{\ell_{max}} \sum_{\ell_2=0}^{\ell_{max}} \sum_{m_1=-\ell_1}^{\ell_1} \sum_{m_1'=-\ell_1}^{\ell_1} 
\int \frac{d^4 k}{(2 \pi)^4} \frac{i}{k^2-M_{\ell_1}^2}  \frac{i}{(k+p)^2-M_{\ell_2}^2}  \nonumber \\
& \times \frac{g_{6a}^2}{R^2} f^{ilk} (k+p)^{\mu} f^{kjl}k^{\nu} J^1_{\ell'm';\underline{\ell_1m_1'};\ell_2m_1'-m'} J^1_{ \underline{\ell m};\ell_1m_1;\underline{\ell_2m_1-m} } \nonumber \\
&\times \frac{1}{2}[\delta_{m_1m_1'}+(-1)^{\ell_1} \delta_{-m_1m_1'} ] \frac{1}{2}[\delta_{m_1-m,m_1'-m'}+(-1)^{\ell_2}\delta_{-(m_1-m),m_1'-m'}] \nonumber \\
\end{align}
which is expressed as the form of Eqs.~(\ref{OneLoopGSeparate}) and (\ref{OneLoopGBulkBound}) and the explicit form of each coefficients are summarized in the 3rd part of Table~\ref{OneLoopG-NA}.

\begin{table}[t] \vspace{1ex}
\begin{tabular}{c||c|l} 
{\scriptsize Diagram} & {\scriptsize Coefficients} &  \\ \hline
{\scriptsize Fig.~\ref{Loop6} } & {\scriptsize $ \Theta^{(1) A_\mu \phi_1(\phi_2)}_{\rm bulk}$} & 
{\scriptsize ${\displaystyle  \int^1_0 d\alpha \int^{\Lambda^2}_0 dx  \frac{x[ (3 \alpha^2-4\alpha+1)x +(\alpha-1)^2 \Delta_M ]}{[x+\Delta_M]^2(x+\Delta_M)} (\delta_{mm'}+(-1)^{\ell'}\delta_{-mm'})   }$} 
{\scriptsize ${\displaystyle  J^{6(7)}_{ \ell'm;\underline{\ell_1m_1};\ell_2m_1-m} J^{6(7)}_{\underline{\ell m};\ell_1m_1;\underline{\ell_2m_1-m} }  }$} \\
& {\scriptsize $ \tilde{\Theta}^{(1) A_\mu \phi_1(\phi_2)}_{\rm bulk}$} & 
{\scriptsize ${\displaystyle  \int^1_0 d\alpha \int^{\Lambda^2}_0 dx  \frac{x^2}{[x+\Delta_M]^2} 
(\delta_{mm'}+(-1)^{\ell'}\delta_{-mm'}) J^{6(7)}_{\ell'm;\underline{\ell_1m_1};\ell_2m_1-m} J^{6(7)}_{ \underline{\ell m};\ell_1m_1;\underline{\ell_2m_1-m} } }$} \\
& {\scriptsize $ \tilde{\Theta}^{(1) A_\mu \phi_1 (\phi_2)}_{\rm bound}$} & 
{\scriptsize ${\displaystyle    - ( M_{l_1}^2+M_{l_2}^2 ) (-1)^{\ell_2}(\delta_{2m_1,m+m'}+(-1)^{\ell'}\delta_{2m_1,m-m'}) 
J^{6(7)}_{ \ell'2m_1-m;\underline{\ell_1m_1};\ell_2m-m_1 } J^{6(7)}_{ \underline{\ell m};\ell_1m_1;\underline{\ell_2m_1-m} } }$} \\ \hline
{\scriptsize Fig.~\ref{Loop7}} & {\scriptsize $ \tilde{\Theta}^{(1) A_\mu}_{\rm bulk}$} & 
{\scriptsize ${\displaystyle  \int^{\Lambda^2}_0 dx \frac{8 x}{x+M_{\ell_1}^2} \delta_{mm'}  K^2_{ \underline{\ell m};\ell'm;\underline{\ell_1m_1};\ell_1m_1}  }$} \\ 
& {\scriptsize $ \tilde{\Theta}^{(1) A_\mu}_{\rm bound}$} & 
{\scriptsize ${\displaystyle - 8   M_{l_1}^2 (-1)^{l_1} \delta_{2m_1,m-m'} K^2_{ \underline{lm};l'-2m_1+m; \underline{l_1-m_1};l_1m_1} }$} \\ \hline
{\scriptsize Fig.~\ref{Loop6}} & {\scriptsize $ \Theta^{(1) A_\mu A_\mu}_{\rm bulk} $} & 
{\scriptsize ${\displaystyle \frac{1}{R^2}  \int^1_0 d\alpha \int_0^{\Lambda^2} dx \frac{4 x}{[x+\Delta_M]^2} \frac{2\alpha(1-\alpha)}{x+\Delta_M}
(\delta_{mm'}+(-1)^{\ell'}\delta_{-mm'}) J^4_{ \ell_2m_1-m;\underline{\ell_1m_1};\ell'm } J^4_{ \underline{\ell_2m_1-m};\ell_1m_1;\underline{\ell m} } }$} \\
& {\scriptsize $ \bar{\Theta}^{(1) A_\mu A_\mu}_{\rm bulk}$} & 
{\scriptsize ${\displaystyle - \frac{1}{R^2} \int^1_0 d\alpha \int_0^{\Lambda^2} dx \frac{4 x}{[x+\Delta_M]^2} 
(\delta_{mm'}+(-1)^{\ell'}\delta_{-mm'}) J^4_{ \ell_2m_1-m;;\underline{\ell_1m_1};\ell'm } J^4_{ \underline{\ell_2m_1-m};\ell_1m_1;\underline{\ell m} } }$} \\
& {\scriptsize $ \bar{\Theta}^{(1) A_\mu A_\mu}_{\rm bound}$} & 
{\scriptsize ${\displaystyle   -  \frac{4}{R^2}  (-1)^{\ell_2}(\delta_{2m_1,m+m'}+(-1)^{\ell'}\delta_{2m_1,m-m'}) 
J^4_{ \ell_2m-m_1;\underline{\ell_1m_1};\ell'2m_1-m } J^4_{ \underline{\ell_2m_1-m};\ell_1m_1;\underline{\ell m} }  }$} \\ \hline
{\scriptsize Fig.~\ref{Loop6}} & {\scriptsize $ \Theta^{(1) \phi_1 \phi_1}_{\rm bulk}$} & 
{\scriptsize ${\displaystyle \frac{1}{R^2} \int^1_0 d\alpha \int^{\Lambda^2}_0 dx \frac{x}{[x+\Delta_M]^2} \frac{ \alpha(1-\alpha)}{x+\Delta_M} 
(\delta_{mm'}+(-1)^{\ell'}\delta_{-mm'}) J^{10}_{ \ell'm;\ell_2m_1-m;\underline{\ell_1m_1} } J^{10}_{ \underline{\ell_2m_1-m};\ell_1m_1;\underline{\ell m} } }$} \\
& {\scriptsize $  \tilde{\Theta}^{(1) \phi_1 \phi_1}_{\rm bulk}$} & 
{\scriptsize ${\displaystyle  - \frac{1}{R^2} \int^1_0 d\alpha \int^{\Lambda^2}_0 dx \frac{x}{2[x+\Delta_M]^2}  
(\delta_{mm'}+(-1)^{\ell'}\delta_{-mm'}) J^{10}_{ \ell'm;\ell_2m_1-m;\underline{\ell_1m_1} } J^{10}_{ \underline{\ell_2m_1-m};\ell_1m_1;\underline{\ell m} } }$} \\
& {\scriptsize $ \tilde{\Theta}^{(1) \phi_1 \phi_1}_{\rm bound}$} & 
{\scriptsize ${\displaystyle - \frac{1}{2 R^2}   (-1)^{\ell_2}(\delta_{2m_1,m+m'}+(-1)^{\ell'}\delta_{2m_1,m-m'}) 
 J^{10}_{ \ell'2m_1-m;\ell_2m-m_1;\underline{\ell_1m_1} } J^{10}_{ \underline{\ell_2m_1-m};\ell_1m_1;\underline{\ell m} } }$} \\ \hline
{\scriptsize Fig.~\ref{Loop6}} & {\scriptsize $ \Theta^{(1) \phi_2 \phi_2}_{\rm bulk}$} & 
{\scriptsize ${\displaystyle \frac{1}{R^2}  \int^1_0 d\alpha \int^{\Lambda^2}_0 dx \frac{x}{[x+\Delta_M]^2} \frac{ \alpha(1-\alpha)}{x+\Delta_M} 
(\delta_{mm'}+(-1)^{\ell'}\delta_{-mm'}) J^9_{ \ell_2m_1-m;\underline{\ell_1m_1};\ell'm } J^9_{ \underline{\ell_2m_1-m};\ell_1m_1;\underline{\ell m} } }$} \\
& {\scriptsize $ \tilde{\Theta}^{(1) \phi_2 \phi_2}_{\rm bulk}$} & 
{\scriptsize ${\displaystyle - \frac{1}{R^2} \int^1_0 d\alpha \int^{\Lambda^2}_0 dx \frac{x}{2[x+\Delta_M]^2}  
(\delta_{mm'}+(-1)^{\ell'}\delta_{-mm'}) J^9_{ \ell_2m_1-m;\underline{\ell_1m_1};\ell'm } J^9_{ \underline{\ell_2m_1-m};\ell_1m_1;\underline{\ell m} } }$} \\
& {\scriptsize $ \tilde{\Theta}^{(1) \phi_2 \phi_2}_{\rm bound}$} & 
{\scriptsize ${\displaystyle - \frac{1}{2 R^2}   (-1)^{\ell_2}(\delta_{2m_1,m+m'}+(-1)^{\ell'}\delta_{2m_1,m-m'}) 
 J^9_{ \ell_2m-m_1;\underline{\ell_1m_1};\ell'2m_1-m } J^9_{ \underline{\ell_2m_1-m};\ell_1m_1;\underline{\ell m} }  }$} \\ \hline
{\scriptsize Fig.~\ref{Loop6}} & {\scriptsize $ \Theta^{(1) \phi_1 \phi_2}_{\rm bulk}$} & 
{\scriptsize ${\displaystyle \frac{1}{R^2} \int^1_0 d\alpha \int^{\Lambda^2}_0 dx \frac{x}{[x+\Delta_M]^2} \frac{ \alpha(1-\alpha)}{x+\Delta_M} 
(\delta_{mm'}+(-1)^{\ell'}\delta_{-mm'}) J^8_{ \ell'm;\underline{\ell_1m_1};\ell_2m_1-m } J^8_{ \underline{\ell m};\ell_1m_1;\underline{\ell_2m_1-m} } }$} \\
& {\scriptsize $ \tilde{\Theta}^{(1) \phi_1 \phi_2}_{\rm bulk}$} & 
{\scriptsize ${\displaystyle - \frac{1}{R^2} \int^1_0 d\alpha \int^{\Lambda^2}_0 dx \frac{x}{2[x+\Delta_M]^2}  
(\delta_{mm'}+(-1)^{\ell'}\delta_{-mm'}) J^8_{ \ell'm;\underline{\ell_1m_1};\ell_2m_1-m } J^8_{ \underline{\ell m};\ell_1m_1;\underline{\ell_2m_1-m} } }$} \\
& {\scriptsize $ \tilde{\Theta}^{(1) \phi_1 \phi_2}_{\rm bound}$} & 
{\scriptsize ${\displaystyle - \frac{1}{2 R^2}   (-1)^{\ell_2}(\delta_{2m_1,m+m'}+(-1)^{\ell'}\delta_{2m_1,m-m'}) 
 J^8_{ \ell'2m_1-m;\underline{\ell_1m_1};\ell_2m-m_1 } J^8_{ \underline{\ell m};\ell_1m_1;\underline{\ell_2m_1-m} } }$} \\ \hline
{\scriptsize Fig.~\ref{Loop7}} & {\scriptsize $ \tilde{\Theta}^{(1) \phi_1}_{\rm bulk}$} &
{\scriptsize ${\displaystyle  \int_0^{\Lambda^2} dx \frac{x}{x+M_{\ell_1}^2} 
\delta_{mm'} \left( 2 K^{4}_{ \ell'm;\underline{\ell_1m_1}; \underline{\ell m} ;\ell_1m_1 }-K^{4}_{\underline{\ell m};\ell'm;\underline{\ell_1m_1} ;\ell_1m_1}
-K^{4}_{\ell'm;\underline{\ell m};\underline{\ell_1m_1} ;\ell_1m_1}  \right) }$} \\
& {\scriptsize $ \tilde{\Theta}^{(1) \phi_1}_{\rm bound}$} & 
{\scriptsize ${\displaystyle     
-  M_{\ell_1}^2 \delta_{2m_1,m-m'} \left( 2 K^{4}_{ \ell'-2m_1+m;\underline{\ell_1-m_1};\underline{\ell m}; \ell_1m_1}-K^{4}_{\underline{\ell m};\ell'-2m_1+m; \underline{\ell_1-m_1}; \ell_1m_1 } 
  -K^{4}_{ \ell'-2m_1+m;\underline{\ell m}; \underline{\ell_1-m_1}; \ell_1m_1} \right) }$}  \\  \hline
{\scriptsize Fig.~\ref{Loop7}} & {\scriptsize $ \tilde{\Theta}^{(1) \phi_2}_{\rm bulk}$} &
{\scriptsize ${\displaystyle - \int_0^{\Lambda^2} dx \frac{x}{x+M_{\ell_1}^2} 
\delta_{mm'} \left( 2 K^{6}_{ \ell'm;\underline{\ell m}; \underline{\ell_1m_1};\ell_1m_1 } - K^{4}_{\underline{\ell m};\ell'm; \underline{\ell_1m_1};\ell_1m_1 } 
-K^{4}_{ \ell'm;\underline{\ell m}; \underline{\ell_1m_1};\ell_1m_1} \right)  }$} \\
& {\scriptsize $ \tilde{\Theta}^{(1) \phi_2}_{\rm bound}$} & 
{\scriptsize ${\displaystyle    
  M_{\ell_1}^2 \delta_{2m_1,m-m'} \left( 2 K^{6}_{ \ell'-2m_1+m;\underline{\ell m}; \underline{\ell_1-m_1}; \ell_1m_1 } - K^{4}_{ \underline{\ell m};\ell'-2m_1+m; \underline{\ell_1-m_1}; \ell_1m_1} 
  -K^{4}_{ \ell'-2m_1+m;\underline{\ell m}; \underline{\ell_1-m_1}; \ell_1m_1} \right) }$}  \\  \hline
\end{tabular}
\vspace{-1ex}
\caption{The contributions from gauge boson loop with self interactions for correction to $\phi_1 \phi_1$ terms for KK masses of extra components of gauge bosons. 
The summation symbols and the Log divergence factor  are omitted as in the Table~\ref{OneLoopF}. The overall factor $C_2(G) g_{6a}^2/64 \pi^2 R^2$ is also omitted.
\label{OneLoopGexNA11}} \vspace{-1ex}
\end{table}

\subsection{One loop corrections to KK masses of extra dimensional components gauge boson}
Here we calculate one loop corrections to $\phi_1 \phi_1$ term. 
%
The Fig.~\ref{Loop3} for extra dimensional components of gauge bosons $\phi_1$ can be obtained by making use of regarding propagator of fermion in Eq.~(\ref{propa-F}) and $\bar{\Psi} \phi_i \Psi$ 
vertex in Table~\ref{vertices-F}, such that
\begin{align}
\label{GexLoop}
i \Theta^{(i)}_{\rm Fig.\ref{Loop3}}(p;\ell m, \ell' m') = & -\frac{1}{4 R^2} \sum_{\ell_1=0}^{\ell_{max}} \sum_{\ell_2=0}^{\ell_{max}} \sum_{m_1=-\ell_1}^{\ell_1} \sum_{m_1'=-\ell_1}^{\ell_1} 
\int_{\Lambda} \frac{d^4 k}{(2\pi)^4} \times \nonumber \\
& {\rm Tr} \biggl[ (i g_{6a} T^a \gamma_5) \Bigl[ C^\alpha_{\ell_1 m_1'; \ell' m'; \ell_2 m_1' -m'} P_L + C^\beta_{\ell_1 m_1'; \ell' m'; \ell_2 m_1' -m'} P_R \Bigr] \times \nonumber \\
& \frac{i}{\sla{k} + i \gamma_5 M_{\ell_1}} (\delta_{m_1, m_1'} \mp (-1)^{\ell_1+m_1} \delta_{-m_1,m_1'} \gamma_5) \times \nonumber \\
& (i g_{6a} T^a \gamma_5) \Bigl[ C^\alpha_{\ell_2 m_1-m; \underline{\ell m}; \ell_1 m_1} P_L + C^\beta_{\ell_2 m_1-m; \underline{\ell m}; \ell_1 m_1} P_R \Bigr] \times  \nonumber \\
& \frac{i}{\sla{k}-\sla{p}+ i\gamma_5 M_{\ell_2}} (\delta_{m_1-m,m_1'-m'} \mp (-1)^{\ell_2 +m_1 +m} \delta_{-(m_1-m),m_1'-m'} \gamma_5) \biggr]
\end{align}
where the sign $\pm$ corresponds to $\Psi_+^{ (\pm \gamma_5 )}$ in the loop.
After taking the sum over $m_1'$, it is separated into bulk and boundary contribution as in the previous cases.
We then arrange the terms in the form of Eqs.~(\ref{OneLoopPhiSeparate}) and (\ref{OneLoopPhiBulkBound}) and each coefficients are given in 
the first line of Table~\ref{OneLoopGex}.
In arranging the terms, we expanded the denominator of Eq.~(\ref{GexLoop}) to extract the terms proportional to $p^2$ such as 
\begin{align}
\label{Expansion}
\frac{1}{[\sla{k} + i \gamma_5 M_{\ell_1}][\sla{k}-\sla{p}+ i\gamma_5 M_{\ell_2}]} \rightarrow & \int_0^1 d \alpha \frac{1}{[k_E^2 +(1-\alpha)M_{\ell_1}^2 + \alpha M_{\ell_2}^2 -\alpha(1-\alpha)p^2]^2} \nonumber \\
& = \frac{1}{k_E^2 + \Delta_M} \Bigl[ 1 + \frac{2 \alpha (1-\alpha)}{k_E^2 + \Delta_M} p^2 + O(p^4) \Bigr]
\end{align}
where $\Delta_M = (1-\alpha)M_{\ell_1}^2 + \alpha M_{\ell_2}^2$, and $k_E$ is a Euclidean momentum.

The Fig.~\ref{Loop4} is calculated in the same manner by  by making use of propagator of scalar boson in Eq.~(\ref{propa-S}) and $H^\dagger H \phi_i$ vertex in Table~\ref{vertices-H} such that
\begin{align}
i\Theta^{\rm Fig.\ref{Loop4}}(p;\ell m;\ell' m') =& \frac{1}{4R^4} \sum_{\ell_1=0}^{\ell_{max}} \sum_{\ell_2=0}^{\ell_{max}} \sum_{m_1=-\ell_1}^{\ell_1} \sum_{m_1'=-\ell_1}^{\ell_1}
\int \frac{d^4 k}{(2\pi)^4} g_{6a} T_a \frac{-i}{k^2-M_{\ell_1}^2} \bigl( \delta_{m_1,m_1'} + (-1)^{\ell_1} \delta_{-m_1,m_1'}  \bigr) \nonumber \\
&  \quad \times g_{6a} T_a \frac{-i}{(k-p)^2-M_{\ell_2}^2}
\bigl(\delta_{m_1-m,m_1'-m'} + (-1)^{\ell_2}\delta_{-(m_1-m),m_1'-m'}  \bigr)  \nonumber \\ 
&  \quad \times [J^2_{\underline{\ell_1,m_1'};\ell',m';\ell_2,m_1'-m' }-J^2_{\ell_2,m_1'-m';\ell',m';\underline{\ell_1,m_1'}}] \nonumber \\
& \quad \times  [J^2_{\underline{\ell_2,m_1-m};\underline{\ell,m}; \ell_1,m_1}-J^2_{(\ell_1,m_1);\underline{\ell,m}; \underline{\ell_2,m_1-m}}],
\end{align}
which is expressed as the form of Eqs.~(\ref{OneLoopPhiSeparate}) and (\ref{OneLoopPhiBulkBound}) and each coefficients are summarized in the 2nd part of Table~\ref{OneLoopGex}.

The Fig.~\ref{Loop5} is calculated in the same manner by making use of propagator of scalar boson in Eq.~(\ref{propa-S}) and $H^\dagger H (\phi_i)^2$ vertex in Table~\ref{vertices-H} such that
\begin{align}
i \Theta^{\rm Fig. \ref{Loop5}}(\ell,m,\ell',m') =& \frac{1}{2 R^2} \sum_{\ell_1=0}^{\ell_{max}} \sum_{m_1=-\ell_1}^{\ell_1} \sum_{m_1'=- \ell_1}^{\ell_1}
\int \frac{d^4 k}{(2\pi)^4} 2i(g_{6a} T_a)^2  \frac{i}{k^2-M_{\ell_1}^2}  \nonumber \\
&  \times K^2_{\underline{\ell_1,m_1'};\ell',m';\underline{\ell,m};\ell_1,m_1} 
[\delta_{m_1,m_1'}+(-1)^{\ell_1} \delta_{-m_1,m_1'}] \delta_{m'-m_1'+m_1-m,0}
\end{align}
which is expressed as the form of Eqs.~(\ref{OneLoopPhiSeparate}) and (\ref{OneLoopPhiBulkBound}), and each coefficients are summarized in the 3rd part of Table~\ref{OneLoopGex}.
%

\begin{table}[t] \vspace{1ex}
\begin{tabular}{c||c|l} 
{\scriptsize Diagram} & {\scriptsize Coefficients} &  \\ \hline
{\scriptsize Fig.~\ref{Loop6}} & {\scriptsize $ \Theta^{(2) A_\mu  \phi_1}_{\rm bulk}$} & 
{\scriptsize ${\displaystyle  \int^1_0 d\alpha \int^{\Lambda^2}_0 dx  \frac{x[ (3 \alpha^2-4\alpha+1)x +(\alpha-1)^2 \Delta_M] }{[x+\Delta_M]^2 (x+\Delta_M) } }$} 
 {\scriptsize ${\displaystyle (\delta_{mm'}+(-1)^{\ell'}\delta_{-mm'}) J^7_{ \underline{\ell_1m_1};\ell'm; \ell_2m_1-m } J^7_{ \ell_1m_1;\underline{\ell m};\underline{\ell_2m_1-m} }  }$} \\
 & {\scriptsize $ \tilde{\Theta}^{(2) A_\mu \phi_1}_{\rm bulk}$} & 
{\scriptsize ${\displaystyle  \int^1_0 d\alpha \int^{\Lambda^2}_0 dx  \frac{-x^2}{[x+\Delta_M]^2}  
(\delta_{mm'}+(-1)^{\ell'}\delta_{-mm'}) J^7_{ \underline{\ell_1m_1};\ell'm; \ell_2m_1-m } J^7_{ \ell_1m_1;\underline{\ell m};\underline{\ell_2m_1-m} } }$} \\
& {\scriptsize $ \tilde{\Theta}^{(2) A_\mu \phi_1}_{\rm bound}$} & 
{\scriptsize ${\displaystyle   -( M_{l_1}^2+M_{l_2}^2 )  (-1)^{l_2}(\delta_{2m_1,m+m'}+(-1)^{l'}\delta_{2m_1,m-m'})
J^7_{ \underline{\ell_1m_1};\ell'2m_1-m;\ell_2m-m_1 } J^7_{ \ell_1m_1;\underline{\ell m};\underline{\ell_2m_1-m} } }$} \\ \hline
{\scriptsize  Fig.~\ref{Loop6}} & {\scriptsize $ \Theta^{(2) A_\mu \phi_2}_{\rm bulk}$} & 
{\scriptsize ${\displaystyle  \int^1_0 d\alpha \int^{\Lambda^2}_0 dx  \frac{x[ (3 \alpha^2-4\alpha+1)x +(\alpha-1)^2 \Delta_M ]}{[x+\Delta_M]^2 (x+\Delta_M) } }$} 
 {\scriptsize ${\displaystyle (\delta_{mm'}+(-1)^{\ell'}\delta_{-mm'}) J^6_{ \ell'm;\underline{\ell_1m_1}; \ell_2m_1-m } J^6_{ \underline{\ell m};\ell_1m_1;\underline{\ell_2m_1-m} } }$} \\
 & {\scriptsize $ \tilde{\Theta}^{(2) A_\mu \phi_2}_{\rm bulk}$} & 
{\scriptsize ${\displaystyle  \int^1_0 d\alpha \int^{\Lambda^2}_0 dx  \frac{-x^2}{[x+\Delta_M]^2}  
(\delta_{mm'}+(-1)^{\ell'}\delta_{-mm'}) J^6_{ \ell'm;\underline{\ell_1m_1}; \ell_2m_1-m } J^6_{ \underline{\ell m};\ell_1m_1;\underline{\ell_2m_1-m} } }$} \\
& {\scriptsize $ \tilde{\Theta}^{(2) A_\mu \phi_2}_{\rm bound}$} & 
{\scriptsize ${\displaystyle   -( M_{l_1}^2+M_{l_2}^2 )  (-1)^{l_2}(\delta_{2m_1,m+m'}+(-1)^{l'}\delta_{2m_1,m-m'})
J^6_{ \ell'2m_1-m;\underline{\ell_1m_1};\ell_2m-m_1 } J^6_{ \underline{\ell m};\ell_1m_1;\underline{\ell_2m_1-m} } }$} \\ \hline
{\scriptsize Fig.~\ref{Loop7}} & {\scriptsize $ \tilde{\Theta}^{(2) A_\mu }_{\rm bulk}$} & 
{\scriptsize ${\displaystyle   \int^{\Lambda^2}_0 dx \frac{8 x}{x+M_{\ell_1}^2} \delta_{mm'}  K^2_{ \underline{\ell m};\ell'm;\underline{\ell_1m_1};\ell_1m_1 } }$} \\
& {\scriptsize $ \tilde{\Theta}^{(2) A_\mu}_{\rm bound}$} & 
{\scriptsize ${\displaystyle   - 8   M_{\ell_1}^2 (-1)^{\ell_1} \delta_{2m_1,m-m'} K^2_{ \underline{\ell m};\ell'-2m_1+m; \underline{\ell_1-m_1}; \ell_1m_1 } }$} \\ \hline
{\scriptsize Fig.~\ref{Loop6}} & {\scriptsize $ \Theta^{(2) A_\mu A_\mu}_{\rm bulk}$} & 
{\scriptsize ${\displaystyle \frac{1}{R^2} \int^1_0 d\alpha \int_0^{\Lambda^2} dx \frac{4 x}{[x+\Delta_M]^2} \frac{2\alpha(1-\alpha)}{x+\Delta_M}
  (\delta_{mm'}+(-1)^{\ell'}\delta_{-mm'}) J^5_{ \ell_2m_1-m;\underline{\ell_1m_1};\ell'm } J^5_{ \underline{\ell_2m_1-m};\ell_1m_1;\underline{\ell m} }  }$} \\
 & {\scriptsize $ \tilde{\Theta}^{(2) A_\mu A_\mu}_{\rm bulk}$} & 
{\scriptsize ${\displaystyle - \frac{1}{R^2} \int^1_0 d\alpha \int_0^{\Lambda^2} dx \frac{4 x}{[x+\Delta_M]^2} 
 (\delta_{mm'}+(-1)^{\ell'}\delta_{-mm'}) J^5_{ \ell_2m_1-m;\underline{\ell_1m_1};\ell'm } J^5_{ \underline{\ell_2m_1-m};\ell_1m_1;\underline{\ell m} }  }$} \\
& {\scriptsize $ \tilde{\Theta}^{(2) A_\mu A_\mu}_{\rm bound}$} & 
{\scriptsize ${\displaystyle   - \frac{4}{R^2}     (-1)^{\ell_2}(\delta_{2m_1,m+m'}+(-1)^{\ell'}\delta_{2m_1,m-m'}) 
J^5_{ \ell_2m-m_1;\underline{\ell_1m_1};\ell'2m_1-m } J^5_{ \underline{\ell_2m_1-m};\ell_1m_1;\underline{\ell m} }  }$} \\ \hline
{\scriptsize Fig.~\ref{Loop6}} & {\scriptsize $ \Theta^{(2) \phi_1 \phi_1}_{\rm bulk}$} & 
{\scriptsize ${\displaystyle \frac{1}{R^2} \int^1_0 d\alpha \int_0^{\Lambda^2} dx \frac{ x}{[x+\Delta_M]^2} \frac{ \alpha(1-\alpha)}{x+\Delta_M}
  (\delta_{mm'}+(-1)^{\ell'}\delta_{-mm'}) J^8_{ \ell_2m_1-m;\underline{\ell_1m_1};\ell'm } J^8_{ \underline{\ell_2m_1-m};\ell_1m_1;\underline{\ell m} }  }$} \\
 & {\scriptsize $ \tilde{\Theta}^{(2) \phi_1 \phi_1}_{\rm bulk}$} & 
{\scriptsize ${\displaystyle - \frac{1}{R^2} \int^1_0 d\alpha \int_0^{\Lambda^2} dx \frac{ x}{2[x+\Delta_M]^2} 
 (\delta_{mm'}+(-1)^{\ell'}\delta_{-mm'}) J^8_{ \ell_2m_1-m;\underline{\ell_1m_1};\ell'm } J^8_{ \underline{\ell_2m_1-m};\ell_1m_1;\underline{\ell m} }  }$} \\
& {\scriptsize $ \tilde{\Theta}^{\phi_1 \phi_1}_{\rm bound}$} & 
{\scriptsize ${\displaystyle  - \frac{1}{2 R^2}    (-1)^{\ell_2}(\delta_{2m_1,m+m'}+(-1)^{\ell'}\delta_{2m_1,m-m'}) 
J^8_{ \ell_2m-m_1;\underline{\ell_1m_1};\ell'2m_1-m } J^8_{ \underline{\ell_2m_1-m};\ell_1m_1;\underline{\ell m} }  }$} \\ \hline
{\scriptsize Fig.~\ref{Loop6}} & {\scriptsize $i \Theta^{(2) \phi_2 \phi_2}_{\rm bulk}$} & 
{\scriptsize ${\displaystyle \frac{1}{R^2} \int^1_0 d\alpha \int_0^{\Lambda^2} dx \frac{ x}{[x+\Delta_M]^2} \frac{ \alpha(1-\alpha)}{x+\Delta_M}
  (\delta_{mm'}+(-1)^{\ell'}\delta_{-mm'}) J^9_{ \ell'm;\underline{\ell_1m_1};\ell_2m_1-m } J^9_{ \underline{\ell m};\ell_1m_1;\underline{\ell_2m_1-m} }  }$} \\
 & {\scriptsize $ \tilde{\Theta}^{(2) \phi_2 \phi_2}_{\rm bulk}$} & 
{\scriptsize ${\displaystyle - \frac{1}{R^2} \int^1_0 d\alpha \int_0^{\Lambda^2} dx \frac{ x}{2[x+\Delta_M]^2} 
  (\delta_{mm'}+(-1)^{\ell'}\delta_{-mm'}) J^9_{ \ell'm;\underline{\ell_1m_1};\ell_2m_1-m } J^9_{ \underline{\ell m};\ell_1m_1;\underline{\ell_2m_1-m} }  }$} \\
& {\scriptsize $ \tilde{\Theta}^{(2) \phi_2 \phi_2}_{\rm bound}$} & 
{\scriptsize ${\displaystyle -  \frac{1}{2 R^2}    (-1)^{\ell_2}(\delta_{2m_1,m+m'}+(-1)^{\ell'}\delta_{2m_1,m-m'}) 
J^9_{ \ell'2m_1-m;\underline{\ell_1m_1};\ell_2m-m_1}  J^9_{ \underline{\ell m};\ell_1m_1;\underline{\ell_2m_1-m} }  }$} \\ \hline
{\scriptsize Fig.~\ref{Loop7}} & {\scriptsize $ \tilde{\Theta}^{(2) \phi_1 }_{\rm bulk}$} & 
{\scriptsize ${\displaystyle  \int_0^{\Lambda^2} dx \frac{x}{x+M_{\ell_1}^2} 
\delta_{mm'} \left( 2 K^{4}_{\ell'm;\underline{\ell_1m_1};\underline{\ell m}; \ell_1m_1}-K^{4}_{\underline{\ell m};\ell'm; \underline{\ell_1m_1};\ell_1m_1} 
-K^{4}_{\ell'm;\underline{\ell m}; \underline{\ell_1m_1};\ell_1m_1} \right)}$} \\ 
& {\scriptsize $ \tilde{\Theta}^{(2) \phi_1}_{\rm bound}$} & 
{\scriptsize ${\displaystyle   
-  M_{\ell_1}^2 \delta_{2m_1,m-m'} \left( 2 K^{4}_{ \ell'-2m_1+m;\underline{\ell_1-m_1};\underline{\ell m}; \ell_1m_1 }-K^{4}_{\underline{\ell m};\ell'-2m_1+m; \underline{\ell_1-m_1};\ell_1m_1}
 -K^{4}_{ \ell'-2m_1+m;\underline{\ell m}; \underline{\ell_1-m_1};\ell_1m_1} \right)}$}\\ \hline 
{\scriptsize Fig.~\ref{Loop7}} & {\scriptsize $ \tilde{\Theta}^{(2) \phi_2}_{\rm bulk}$} & 
{\scriptsize ${\displaystyle - \int_0^{\Lambda^2} dx \frac{x}{x+M_{l_1}^2}  \delta_{mm'} \left( 2 K^{6}_{ \underline{\ell_1m_1};\ell_1m_1;\ell'm;\underline{\ell m} }
- K^{4}_{\ell_1m_1; \underline{\ell_1m_1};\ell'm;\underline{\ell m} }-K^{4}_{ \underline{\ell_1m_1};\ell_1m_1;\ell'm;\underline{\ell m}} \right) }$} \\
& {\scriptsize $ \tilde{\Theta}^{(2) \phi_2}_{\rm bound}$} & 
{\scriptsize ${\displaystyle 
M_{\ell_1}^2 \delta_{2m_1,m-m'} \left( 2 K^{6}_{ \underline{\ell_1-m_1};  \ell_1m_1; \ell'-2m_1+m;\underline{\ell m} } 
-K^{4}_{ \ell_1m_1;\underline{\ell_1-m_1};  \ell'-2m_1+m;\underline{\ell m} }-K^{4}_{ \underline{\ell_1-m_1};  \ell_1m_1; \ell'-2m_1+m;\underline{\ell m}} \right)}$} \\ \hline
\end{tabular}
\vspace{-1ex}
\caption{The contributions from gauge boson loop with self interactions for correction to $\phi_2 \phi_2$ terms for KK masses of extra components of gauge bosons. 
The summation symbols and the Log divergence factor  are omitted as in the Table~\ref{OneLoopF}. The overall factor $C_2(G) g_{6a}^2/64 \pi^2 R^2$ is also omitted .\label{OneLoopGexNA22}} \vspace{-1ex}
\end{table}

%
The Fig.~\ref{Loop6} for two virtual $A_\mu$ is calculated by making use of propagator of $A_\mu$ in Eq.~(\ref{propa-G}) and $(A_\mu)^2 \phi_1$ vertex in Table~\ref{vertices-A} such that
\begin{align}
i\Theta^{{\rm Fig.\ref{Loop6}(A_\mu A_\mu) }}(p; \ell m; \ell' m') =& \sum_{\ell_1=0}^{\ell_{max}} \sum_{\ell_2=0}^{\ell_{max}} \sum_{m_1=-\ell_1}^{\ell_1} \sum_{m_1'=-\ell_1}^{\ell_1} 
\int \frac{d^4 k}{(2 \pi)^4} \frac{-i}{k^2-M_{\ell_1}^2} \frac{-i}{(k+p)^2-M_{\ell_2}^2} g_{\mu \nu} \nonumber \\
& \times \frac{1}{2}[\delta_{m_1m_1'}+(-1)^{\ell_1} \delta_{-m_1m_1'} ] \frac{1}{2}[\delta_{m_1-m,m_1'-m'}+(-1)^{\ell_2}\delta_{-(m_1-m),m_1'-m'}] \nonumber \\
& \times \frac{g_{6a}}{R} f^{kil} (-k-p)^{\mu} \frac{g_{6a}}{R} f^{kjl} (-k-p)^{\nu} J^6_{ \ell'm';\underline{\ell_1m_1'};\ell_2m_1'-m' } J^6_{ \underline{\ell m};\ell_1m_1;\underline{\ell_2m_1-m}}  
\end{align}
which is expressed as the form of Eqs.~(\ref{OneLoopPhiSeparate}) and (\ref{OneLoopPhiBulkBound}) and each coefficients are summarized in the 3rd part of Table~\ref{OneLoopGexNA11}.
The Fig.~\ref{Loop6} for one virtual $A_\mu$ and one virtual $\phi_i$ is calculated in the same way by making use of propagator of $A_\mu$ in Eq.~(\ref{propa-G}), 
that of $\phi_i$ Eq.~(\ref{propa-Gex}) and $(A_\mu)^2 \phi_i$ vertex in Table~\ref{vertices-A}.
The results are summarized in the 1st part of Table~\ref{OneLoopGexNA11}. 
Fig.~\ref{Loop6} for two virtual $\phi_i$ is calculated by making use of  
propagator of $\phi_i$ in Eq.~(\ref{propa-Gex}) and $ (\phi_i)^4 $ vertex in Table~\ref{vertices-A}, 
and the results are summarized in the 4th, the 5th and the 6th part of Table~\ref{OneLoopGexNA11}. 

Fig.~\ref{Loop7} for virtual $A_\mu$ is calculated by applying propagator of $A_\mu$ Eq.~(\ref{propa-G}) and $(A_\mu)^2 (\phi_i)^2$ vertex in Table~\ref{vertices-A} such that
\begin{align}
i\Theta^{{\rm Fig.\ref{Loop7}(A_\mu) }}(p; \ell m; \ell' m')=& \frac{1}{2} \sum_{\ell_1=0}^{\ell_{max}} \sum_{\ell_2=0}^{\ell_{max}} \sum_{m_1=-\ell_1}^{\ell_1} \sum_{m_1'=-\ell_1}^{\ell_1}
\int \frac{d^4k}{(2\pi)^4} \frac{-i}{k^2-M_{\ell_1}^2} g_{\sigma \rho} \delta^{kl} \nonumber \\
& \times i \frac{g_{6a}^2}{R^2}  g^{\rho \sigma} [ f^{ikm}f^{jlm} + f^{ilm}f^{jkm} ] \nonumber \\
& \times \frac{1}{2}[\delta_{m_1m_1'}+(-1)^{\ell_1} \delta_{-m_1m_1'} ] \delta_{m'-m_1'+m_1-m,0}  K^2_{ \underline{\ell m};\ell'm'; \underline{\ell_1m_1'};\ell_1m_1}
\end{align}
which is expressed as the form of Eqs.~(\ref{OneLoopPhiSeparate}) and (\ref{OneLoopPhiBulkBound}) and each coefficients are summarized in the 2nd part of Table~\ref{OneLoopGexNA11}.
Fig.~\ref{Loop7} for virtual $\phi_i$ is calculated in the same way by  making use of propagator of $\phi_i$ in Eq.~(\ref{propa-Gex}) and $(\phi_i)^4$ vertex in Table~\ref{vertices-A},
and the results are summarized in the 7th and the 8th parts of Table~\ref{OneLoopGexNA11}. 

The one loop corrections to $\phi_2 \phi_2$ term and off-diagonal $\phi_1 \phi_2$ term are carried out in the same way, 
and the results are summarized in Table~\ref{OneLoopGexNA22} and \ref{OneLoopGexNA12} respectively.


\begin{table}[t] \vspace{1ex}
\begin{tabular}{c||c|l} 
{\scriptsize Diagram} & {\scriptsize Coefficients} &  \\ \hline
{\scriptsize Fig.~\ref{Loop6}} & {\scriptsize $ \Theta^{(12) \phi_1 \phi_1}_{\rm bulk} $ } & 
{\scriptsize ${\displaystyle \frac{1}{R^2} \int^1_0 d\alpha \int^{\Lambda^2}_0 dx \frac{x}{[x+\Delta_M]^2}  \frac{\alpha(1-\alpha)}{x+\Delta_M}
 (\delta_{mm'}+(-1)^{\ell'}\delta_{-mm'}) J^{10}_{ \ell_2m_1-m;\underline{\ell_1m_1};\ell'm } J^8_{ \underline{\ell_2m_1-m};\ell_1m_1;\underline{\ell m} } }$} \\
 & {\scriptsize $ \tilde{\Theta}^{(12) \phi_1 \phi_1}_{\rm bulk} $ } & 
{\scriptsize ${\displaystyle - \frac{1}{R^2} \int^1_0 d\alpha \int^{\Lambda^2}_0 dx \frac{x}{2[x+\Delta_M]^2}  
(\delta_{mm'}+(-1)^{\ell'}\delta_{-mm'}) J^{10}_{ \ell_2m_1-m;\underline{\ell_1m_1};\ell'm } J^8_{ \underline{\ell_2m_1-m};\ell_1m_1;\underline{\ell m} }  }$} \\ 
& {\scriptsize $ \tilde{\Theta}^{(12) \phi_1 \phi_1}_{\rm bound}$} & 
{\scriptsize ${\displaystyle  - \frac{1}{2 R^2}  (-1)^{\ell_2}(\delta_{2m_1,m+m'}+(-1)^{l'}\delta_{2m_1,m-m'}) 
J^{10}_{ \ell_2m-m_1;\underline{\ell_1m_1}; \ell'2m_1-m } J^8_{ \underline{\ell_2m_1-m}; \ell_1m_1;\underline{\ell m} }  }$} \\ \hline
{\scriptsize Fig.~\ref{Loop6}} & {\scriptsize $ \Theta^{(12) \phi_1 \phi_2}_{\rm bulk} $ } & 
{\scriptsize ${\displaystyle - \frac{1}{R^2} \int^1_0 d\alpha \int^{\Lambda^2}_0 dx \frac{x}{[x+\Delta_M]^2}  \frac{\alpha(1-\alpha)}{x+\Delta_M}
 (\delta_{mm'}+(-1)^{\ell'}\delta_{-mm'}) J^8_{ \ell'm; \ell_2m_1-m;\underline{\ell_1m_1} } J^9_{ \underline{\ell m};\ell_1m_1;\underline{\ell_2m_1-m} } }$} \\
 & {\scriptsize $ \tilde{\Theta}^{(12) \phi_1 \phi_2}_{\rm bulk} $ } & 
{\scriptsize ${\displaystyle  \frac{1}{R^2} \int^1_0 d\alpha \int^{\Lambda^2}_0 dx \frac{x}{2[x+\Delta_M]^2}  
 (\delta_{mm'}+(-1)^{\ell'}\delta_{-mm'}) J^8_{ \ell'm; \ell_2m_1-m;\underline{\ell_1m_1} } J^9_{ \underline{\ell m};\ell_1m_1;\underline{\ell_2m_1-m} }  }$} \\
& {\scriptsize $ \tilde{\Theta}^{(12) \phi_1 \phi_2}_{\rm bound}$} & 
{\scriptsize ${\displaystyle    \frac{1}{2 R^2}  (-1)^{\ell_2}(\delta_{2m_1,m+m'}+(-1)^{l'}\delta_{2m_1,m-m'}) 
J^8_{ \ell'2m_1-m; \ell_2m-m_1;\underline{\ell_1m_1} } J^9_{ \underline{\ell m}; \ell_1m_1;\underline{\ell_2m_1-m} } }$} \\ \hline
{\scriptsize Fig.~\ref{Loop7}} & {\scriptsize $ \tilde{\Theta}^{(12) \phi_1 }_{\rm bulk}$} & 
{\scriptsize ${\displaystyle  - \frac{1}{R^2} \int^1_0 d\alpha \int^{\Lambda^2}_0 dx \frac{1}{x+M_{\ell_1}^2} 
\Bigl[ K^{5}_{ \underline{\ell m};\ell'm;\ell_1 m_1; \underline{\ell_1 m_1} }+K^{5}_{ \underline{\ell m};\ell'm; \underline{\ell_1 m_1};\ell_1 m_1 } 
-2 K^{5}_{ \underline{\ell m};\underline{\ell_1 m_1};\ell_1 m_1; \ell'm } \Bigr] \delta_{mm'} }$} \\ 
& {\scriptsize $ \tilde{\Theta}^{(12) \phi_1 }_{\rm bound}$}  & 
{\scriptsize ${\displaystyle    \frac{1}{R^2}
\Bigl[ K^{5}_{ \underline{\ell m};\ell' m-2m_1 ;\ell_1 m_1; \underline{\ell_1 -m_1} }+K^{5}_{ \underline{\ell m};\ell' m-2m_1 ; \underline{\ell_1 -m_1};\ell_1 m_1 } 
-2 K^{5}_{ \underline{\ell m}; \underline{\ell_1 -m_1} ;\ell_1 m_1;\ell' m-2m_1} \Bigr] \delta_{2m_1, m-m'}}$} \\ \hline
{\scriptsize Fig.~\ref{Loop7}} & {\scriptsize $ \tilde{\Theta}^{(12)  \phi_2}_{\rm bulk}$} & 
{\scriptsize ${\displaystyle \frac{1}{R^2} \int^1_0 d\alpha \int^{\Lambda^2}_0 dx \frac{1}{x+M_{\ell_1}^2} 
\Bigl[ K^{5}_{ \underline{\ell m};\ell'm;\ell_1 m_1; \underline{\ell_1 m_1} }+K^{5}_{ \underline{\ell m};\ell'm; \underline{\ell_1 m_1};\ell_1 m_1 }
-2 K^{5}_{ \underline{\ell m}; \underline{\ell_1 m_1};\ell_1 m_1;\ell'm } \Bigr] \delta_{mm'} }$} \\ 
& {\scriptsize $ \tilde{\Theta}^{(12) \phi_2}_{\rm bound}$}  & 
{\scriptsize ${\displaystyle -    \frac{1}{R^2}
\Bigl[ K^{5}_{ \underline{\ell m};\ell' m-2m_1 ;\ell_1 m_1; \underline{\ell_1 -m_1} } + K^{5}_{ \underline{\ell m};\ell' m-2m_1; \underline{\ell_1 -m_1} ;\ell_1 m_1} 
-2 K^{5}_{ \underline{\ell m}; \underline{\ell_1 -m_1};\ell_1 m_1;\ell' m-2m_1 } \Bigr] \delta_{2m_1, m-m'}}$} \\ \hline
\end{tabular}
 \vspace{-1ex}
\caption{The contributions from gauge boson loop with self interactions for correction to $\phi_1 \phi_2$ terms for KK masses of extra components of gauge bosons. 
The summation symbols and the Log divergence factor  are omitted as in the Table~\ref{OneLoopF}. The overall factor $C_2(G) g_{6a}^2/64 \pi^2 R^2$ is also omitted.
\label{OneLoopGexNA12}} \vspace{-1ex}
\end{table}


\begin{table}[t] 
\begin{tabular}{c||c|l} 
{\scriptsize Diagram} & {\scriptsize Coefficients} &  \\ \hline
 {\scriptsize Fig.~\ref{Loop11}} & {\scriptsize $ \Xi_{\rm bulk}^{A_\mu H}$} & 
 {\scriptsize ${\displaystyle    \int_0^1 d\alpha \int_0^{\Lambda^2}   \frac{-x dx}{[x+\Delta_M]^2} 
  \Bigl( (\alpha+1)^2 - \frac{2 \alpha(1-\alpha) x}{x+\Delta_M} \Bigr) [\delta_{m, m'}  + (-1)^{\ell'} \delta_{m,-m'} ] 
  J^1_{ \underline{\ell_1 m_1}; \ell_2 m_1-m;\ell' m } J^1_{ \underline{\ell m};\underline{\ell_2 m_1-m};\ell_1 m_1} }$} \\
& {\scriptsize $ \tilde{\Xi}_{\rm bulk}^{A_\mu H}$} & 
{\scriptsize ${\displaystyle - \int_0^1 d\alpha \int_0^{\Lambda^2} dx  \frac{x^2}{[x+\Delta_M]^2}  [\delta_{m, m'}  + (-1)^{\ell'} \delta_{m,-m'} ]   
J^1_{ \underline{\ell_1 m_1};\ell_2 m_1-m; \ell' m } J^1_{ \underline{\ell m};\underline{\ell_2 m_1-m};\ell_1 m_1} }$} \\ 
& {\scriptsize $ {\Xi}_{\rm bound}^{A_\mu H}$} & 
{\scriptsize ${\displaystyle -2   (-1)^{\ell_2} [\delta_{2m_1,m+m'}  + (-1)^{\ell'} \delta_{2m_1,m-m'}   ] 
J^1_{ \underline{\ell_1  m_1};\ell_2 m-m_1; \ell' 2m_1-m } J^1_{ \underline{\ell m};\underline{\ell_2 m_1-m};\ell_1 m_1 } }$} \\ 
& {\scriptsize $ \tilde{\Xi}_{\rm bound}^{A_\mu H}$} &  
{\scriptsize ${\displaystyle    (M_{\ell_1}^2+M_{\ell_2}^2)   (-1)^{\ell_2} [\delta_{2m_1,m+m'}  + (-1)^{\ell'} \delta_{2m_1,m-m'}   ] 
J^1_{ \underline{\ell_1 m_1};\ell_2 m-m_1;\ell'2m_1-m } J^1_{ \underline{\ell m};\underline{\ell_2 m_1-m};\ell_1 m_1 }  }$} \\ \hline
 {\scriptsize Fig.~\ref{Loop11}} & {\scriptsize $ \Xi_{\rm bulk}^{\phi_{1(2) H}}$} & 
{\scriptsize ${\displaystyle - \frac{1}{R^2}  \int_0^1 d\alpha \int_0^{\Lambda^2} dx \frac{x}{[x+\Delta_M]^2} \frac{2 \alpha (1-\alpha)}{x+\Delta_M}  [\delta_{m, m'}   - (-1)^{\ell'} \delta_{-m,m'} ] }$} \\
& & {\scriptsize ${\displaystyle [ J^{2(3)}_{ \underline{\ell_1m_1};\ell_2m_1-m; \ell'm }-J^{2(3)}_{ \ell'm;\ell_2 m_1-m;\underline{\ell_1m_1} } ] 
[J^{2(3)}_{ \underline{\ell m};\underline{\ell_2 m_1-m};\ell_1 m_1} - J^{2(3)}_{ \ell_1 m_1;\underline{\ell_2 m_1-m};\underline{\ell m}}  ]  }$} \\
& {\scriptsize $\bar{\Xi}_{\rm bulk}^{\phi_{1(2)} H}$} & 
{\scriptsize ${\displaystyle \frac{1}{R^2}  \int_0^1 d\alpha \int_0^{\Lambda^2} dx \frac{x}{[x+\Delta_M]^2}   [\delta_{m, m'}   - (-1)^{l'} \delta_{-m,m'} ] }$} \\
& & {\scriptsize ${\displaystyle [ J^2_{ \underline{\ell_1m_1};\ell_2m_1-m; \ell'm }-J^2_{ \ell'm;\ell_2 m_1-m;\underline{\ell_1m_1} } ] 
[J^{2(3)}_{ \underline{\ell m};\underline{\ell_2 m_1-m};\ell_1 m_1 } - J^{2(3)}_{ \ell_1 m_1;\underline{\ell_2 m_1-m};\underline{\ell m} }  ]}$} \\
& {\scriptsize $ \bar{\Xi}_{\rm bound}^{\phi_{1(2)} H} $} &
{\scriptsize ${\displaystyle    \frac{(-1)^{\ell_2}}{R^2}  [\delta_{2m_1,m+m'} - (-1)^{\ell'} \delta_{2m_1,m-m'}   ]}$} \\
& & {\scriptsize ${\displaystyle [ J^{2(3)}_{ \underline{\ell_1m_1};\ell_2m-m_1; \ell'2m_1-m } -J^{2(3)}_{ \ell'2m_1-m;\ell_2m-m_1;\underline{\ell_1m_1} }] 
[J^{2(3)}_{ \underline{\ell m};\underline{\ell_2 m_1-m}; \ell_1 m_1 } - J^{2(3)}_{ \ell_1 m_1;\underline{\ell_2 m_1-m};\underline{\ell m} } ]  }$} \\ \hline 
 {\scriptsize Fig.~\ref{Loop10}} & {\scriptsize $\tilde{\Xi}_{\rm bulk}^{A_\mu}$} & 
 {\scriptsize ${\displaystyle  \int_0^{\Lambda^2} \frac{8 x dx}{x+M_{\ell_1}^2} K^1_{ \underline{\ell_1m_1}; \ell_1m_1;\underline{\ell m};\ell'm } \delta_{m,m'}}$} \\ 
 & {\scriptsize $\tilde{\Xi}^{A_\mu}_{\rm bound}$} & 
 {\scriptsize ${\displaystyle -(-1)^{\ell_1} 8 M_{\ell_1}^2  K^1_{ \underline{\ell_1m_1}; \ell_1-m_1;\underline{\ell m};\ell'm' } \delta_{2m_1,m'-m} }$} \\ \hline 
 {\scriptsize Fig.~\ref{Loop10}} & {\scriptsize $\tilde{\Xi}_{\rm bulk}^{\phi_i}  $} & 
 {\scriptsize ${\displaystyle  \int_0^{\Lambda^2} \frac{2 x dx}{x+M_{\ell_1}^2} K^2_{ \underline{\ell_1m_1}; \ell_1m_1;\underline{\ell m};\ell'm } \delta_{m,m'}}$} \\
 & {\scriptsize $\tilde{\Xi}_{\rm bound}^{\phi_i}$} & 
 {\scriptsize ${\displaystyle - (-1)^{\ell_1} 2 M_{\ell_1}^2  K^2_{ \underline{\ell_1m_1}; \ell_1-m_1;\underline{\ell m};\ell'm' } \delta_{2m_1,m'-m} }$} \\ \hline
 {\scriptsize Fig.~\ref{Loop8}} & {\scriptsize $\tilde{\Xi}^{H}_{\rm bulk}$} & 
{\scriptsize ${\displaystyle  \int_0^{\Lambda^2} \frac{2 x dx}{x+M_{\ell_1}^2} K^1_{ \underline{\ell_1m_1}; \ell_1m_1;\underline{\ell m};\ell'm } \delta_{m,m'}}$} \\
& {\scriptsize $\tilde{\Xi}^{H}_{\rm bound}$} & 
{\scriptsize ${\displaystyle  -  (-1)^{\ell_1} 2 M_{\ell_1}^2  K^1_{ \underline{\ell_1m_1}; \ell_1-m_1;\underline{\ell m};\ell'm' } \delta_{2m_1,m'-m}}$} \\ \hline
\end{tabular}
 \vspace{-1ex}
\caption{The contributions from Higgs boson and gauge boson loop for correction to KK masses of Higgs boson. 
The summation symbols and the Log divergence factor  are omitted as in the Table~\ref{OneLoopF}. The overall factor $(T_a)^2 g_{6a}^2/64 \pi^2 R^2$ and $ \lambda_6/64 \pi^2 R^2$
are also omitted for loops with gauge interaction and Higgs-self interaction respectively.\label{OneLoopS}} \vspace{-1ex}
\end{table}


\begin{table}[t] 
\begin{tabular}{c||c|l} 
{\scriptsize Diagram} & {\scriptsize Coefficients} &  \\ \hline
{\scriptsize Fig.~\ref{Loop9}} & {\scriptsize $ \Xi_{\rm bulk}^{ff}$} & 
{\scriptsize ${\displaystyle   \int_0^1 d \alpha \int_0^{\Lambda^2} dx \frac{2x}{[x+\Delta_M]^2} 
\biggl[  \Bigl( \alpha(1-\alpha) +  \frac{2 \alpha(1-\alpha)x}{x+\Delta_M} \Bigr) [\delta_{m,m'} + (-1)^{l'} \delta_{-m,m' }] \times}$} \\
& & {\scriptsize ${\displaystyle (I^{\alpha}_{\ell_1m_1;\ell'm;\ell_2m_1-m} I^{\alpha}_{ \ell_2 m_1-m;\underline{ \ell m}; \ell_1m_1}
+ I^{\beta}_{ \ell_1m_1;\ell'm;\ell_2m_1-m } I^{\beta}_{ \ell_2m_1-m;\underline{ \ell m}; \ell_1m_1}  ) }$} \\ 
& & {\scriptsize ${\displaystyle  + M_{l_1}M_{l_2}  \frac{2 \alpha(1-\alpha)}{x+\Delta_M}  [\delta_{m,m'} + (-1)^{\ell'} \delta_{-m,m' }] \times }$}   \\
& & {\scriptsize ${\displaystyle (I^{\alpha}_{ \ell_1m_1;\ell'm;\ell_2m_1-m } I^{\beta}_{ \ell_2m_1-m;\underline{\ell m};\ell_1m_1}
+ I^{\beta}_{ \ell_1m_1;\ell'm;\ell_2m_1-m } I^{\alpha}_{ \ell_2m_1-m;\underline{\ell m}; \ell_1m_1} )  \biggr]}$} \\
& {\scriptsize $ \tilde{\Xi}_{\rm bulk}^{ff}$} & 
{\scriptsize ${\displaystyle -  \int_0^1 d \alpha \int_0^{\Lambda^2} dx \frac{2x}{[x+\Delta_M]^2} 
\bigl[  x [\delta_{m,m'} + (-1)^{l'} \delta_{-m,m' }] \times}$} \\
& & {\scriptsize ${\displaystyle (I^{\alpha}_{ \ell_1m_1;\ell'm;\ell_2m_1-m } I^{\alpha}_{ \ell_2 m_1-m;\underline{ \ell m}; \ell_1m_1 }
+ I^{\beta}_{ \ell_1m_1;\ell'm;\ell_2m_1-m } I^{\beta}_{ \ell_2m_1-m;\underline{ \ell m}; \ell_1m_1}  ) }$} \\ 
& & {\scriptsize ${\displaystyle  + M_{\ell_1}M_{\ell_2}    [\delta_{m,m'} + (-1)^{\ell'} \delta_{-m,m' }] \times }$}   \\
& & {\scriptsize ${\displaystyle (I^{\alpha}_{ \ell_1m_1;\ell'm;\ell_2m_1-m } I^{\beta}_{ \ell_2m_1-m;\underline{\ell m};\ell_1m_1 }  
+ I^{\beta}_{ \ell_1m_1;\ell'm;\ell_2m_1-m } I^{\alpha}_{ \ell_2m_1-m;\underline{\ell m}; \ell_1m_1} )  \bigr]}$} \\
& {\scriptsize $ \Xi_{\rm bound}^{ff}$} & 
{\scriptsize ${\displaystyle -   (-1)^{\ell_2+m_1+m} [\delta_{2m_1,m+m' }+(-1)^{\ell'}\delta_{2m_1,m-m'} ] \times}$} \\
& & {\scriptsize ${\displaystyle (I^{\alpha}_{ \ell_1m_1;\ell'2m_1-m;\ell_2m-m_1} I^{\alpha}_{ \ell_2m_1-m;\underline{\ell m}; \ell_1m_1} 
- I^{\beta}_{ \ell_1m_1;\ell'2m_1-m;\ell_2m-m_1 } I^{\beta}_{ \ell_2m_1-m;\underline{\ell m}; \ell_1m_1}  ) }$} \\
& {\scriptsize $ \tilde{\Xi}_{\rm bound}^{ff}$} & 
{\scriptsize ${\displaystyle - 2  \bigl[ (M_{\ell_1}^2+M_{\ell_2}^2 ) (-1)^{\ell_2+m_1+m} [  \delta_{2m_1,m+m' }+(-1)^{\ell'}\delta_{2m_1,m-m'}] \times }$} \\
& & {\scriptsize ${\displaystyle \bigl( I^{\alpha}_{ \ell_1m_1;\ell'2m_1-m; \ell_2m-m_1 } I^{\alpha}_{ \ell_2m_1-m;\underline{\ell m}; \ell_1m_1}  
- I^{\beta}_{ \ell_1m_1;\ell'2m_1-m;\ell_2m-m_1 } I^{\beta}_{ \ell_2m_1-m;\underline{\ell m}; \ell_1m_1 } \bigr)}$} \\ 
& & {\scriptsize ${\displaystyle -M_{ \ell_1}M_{ \ell_2} (-1)^{ \ell_2+m_1+m}  [  \delta_{2m_1,m+m' } - (-1)^{ \ell'}\delta_{2m_1,m-m'}] \times }$} \\
& & {\scriptsize ${\displaystyle \bigl(I^{\alpha}_{ \ell_1m_1;\ell'2m_1-m;\ell_2m-m_1 } I^{\beta}_{ \ell_2m_1-m;\underline{\ell m}; \ell_1m_1 } 
- I^{\beta}_{ \ell_1m_1;\ell'2m_1-m;\ell_2m-m_1 } I^{\alpha}_{ \ell_2m_1-m;\underline{\ell m}; \ell_1m_1} \bigr) \bigr] }$} \\ \hline
\end{tabular}
 \vspace{-1ex}
\caption{The contributions from fermion loop for correction to KK masses of Higgs boson. 
The summation symbols and the Log divergence factor  are omitted as in the Table~\ref{OneLoopF}. The overall factor $ Y_f^2/64 \pi^2 R^2$ is also omitted. \label{OneLoopS-F}} \vspace{-1ex}
\end{table}


\subsection{One loop corrections to KK mass of Higgs boson}
For Higgs boson, the corresponding one-loop diagrams are shown in Fig.~\ref{Loop11}-\ref{Loop8}.
These diagrams are calculated as in the previous cases.
Fig.~\ref{Loop8} for virtual $A_\mu$ can be obtained by making use of propagators in Eq.~(\ref{propa-S}) and $H^\dagger HA_\mu$ vertex in Table~\ref{vertices-H}, such that
\begin{align}
i \Xi_{\rm Fig.\ref{Loop11}}(p;\ell m, \ell' m')  =& \frac{1}{4 R^2} \sum_{\ell_1=0}^{\ell_{max}} \sum_{\ell_2=0}^{\ell_{max}} \sum_{m_1 = -\ell_1}^{\ell_1} \sum_{m_1' = -\ell_1}^{\ell_1} \int \frac{d^4 k}{(2 \pi)^4} \times \nonumber \\
&  ig_{6a} T^a (p+k)^\mu J^1_{\underline{\ell_1 m_1'}; \ell_2 m_1'-m';\ell' m' } \times  \nonumber \\
& \frac{i}{k^2-M_{\ell_1}^2} [\delta_{m_1-m,m_1'-m' }+(-1)^{\ell_2} \delta_{-m_1+m,m_1'-m'}] \times \nonumber \\
&  ig_{6a} T^a (p+k)_\mu J^1_{\underline{\ell m}; \underline{\ell_2 m_1-m} ;\ell_1 m_1} \times  \nonumber \\
& \frac{-i}{(p-k)^2-M_{\ell_2}^2} [\delta_{m_1,m_1' }+(-1)^{\ell_1} \delta_{-m_1,m_1'}].
\end{align}
After summing over $m_1'$, it is separated into the bulk and the boundary contribution as  in the previous cases. 
We then arrange the terms in the form of Eqs.~(\ref{OneLoopHSeparate}) and (\ref{OneLoopHBulkBound}) using Eq.~(\ref{Expansion}) to extract terms proportional to $p^2$, and  each coefficients are given in 
the first part of the Table~\ref{OneLoopS}.
Fig.~\ref{Loop8} for one virtual $\phi_i$ is calculated in the same way by applying propagator of $\phi_i$ in Eq.~(\ref{propa-Gex}), 
propagator of scalar boson in Eq.~(\ref{propa-S}) and $H^\dagger H \phi_i$ vertex in Table~\ref{vertices-H},
and the results are summarized in the 2nd part of Table~\ref{OneLoopS}. 

The Fig.~\ref{Loop9} for virtual $A_\mu$ is calculated by making use of propagator of $A_\mu$  in Eq.~(\ref{propa-G}) and $ H^\dagger H (A_\mu)^2 $ vertex in Table~\ref{vertices-H} such that
\begin{align}
i\Xi^{{\rm Fig.\ref{Loop9}(A_\mu) }}(p; \ell m; \ell' m')=& i\frac{g_{6a}^2}{R^2} T_a^2 g_{\mu\nu}g^{\mu\nu} 
\sum_{\ell_1=0}^{\ell_{max}}  \sum_{m_1=-\ell_1}^{\ell_1} \sum_{m_1'=-\ell_1}^{\ell_1} \int \frac{d^4k}{(2\pi)^4} K^1_{ \underline{\ell_1m_1}; \ell_1m_1'; \underline{\ell m}; \ell'm'} \nonumber \\
& \times \frac{1}{2} \frac{-i}{k^2-M^2_{\ell_1}} [\delta_{m_1,m_1'} + (-1)^{\ell_1} \delta_{-m_1,m_1'}] \delta_{m+m_1,-m,m_1'} 
\end{align}
which is expressed as the form of Eqs.~(\ref{OneLoopHSeparate}) and (\ref{OneLoopHBulkBound}) and each coefficients are summarized in the 3rd part of Table~\ref{OneLoopS}.
The Fig.~\ref{Loop9} for virtual $\phi_i$ is calculated in the same way by applying propagator of $\phi_i$ in Eq.~(\ref{propa-Gex}) and $H^\dagger H (\phi_i)^2$ vertex in Table~\ref{vertices-H},
and the results are summarized in the 4th part of Table~\ref{OneLoopS}. 

The Fig.~\ref{Loop11} is calculated in the same manner by making use of propagator of scalar boson in Eq.~(\ref{propa-S}) and $ (H^\dagger H )^2 $ vertex in Table~\ref{vertices-H} such that
\begin{align}
i\Xi^{{\rm Fig.\ref{Loop11} }}(p; \ell m; \ell' m')=& \frac{-i \lambda_6}{2R^2} \sum_{\ell_1=0}^{\ell_{max}} \sum_{m_1=-\ell_1}^{\ell_1} \sum_{m_1'=-\ell_1}^{\ell_1}
\int \frac{d^4k}{(2\pi)^4} K^1_{\underline{\ell_1m_1}; \ell_1m_1';\underline{\ell m}; \ell'm'} \nonumber \\
& \times \frac{i \delta_{m+m_1,m'+m_1'} }{k^2-M_{\ell_1}^2}[\delta_{m_1,m_1'} +(-1)^{\ell_1} \delta_{-m_1,m_1'} ]
\end{align}
which is expressed as the form of Eqs.~(\ref{OneLoopHSeparate}) and (\ref{OneLoopHBulkBound}) and each coefficients are summarized in the 5th part of Table~\ref{OneLoopS}.

Fig.~\ref{Loop10} is calculated by making use of propagator of scalar boson in Eq.~(\ref{propa-S}) and $ \Psi \Psi H $ vertex in Table~\ref{vertices-H} such that
\begin{align}
i\Xi^{{\rm Fig.\ref{Loop10} }}(p; \ell m; \ell' m')=& -\frac{(Y_{f})^2}{4R^4} \sum_{\ell_1=0}^{\ell_{max}} \sum_{\ell_2=0}^{\ell_{max}} \sum_{m_1=-\ell_1}^{\ell_1} \sum_{m_1'=-\ell_1}^{\ell_1}
\int \frac{d^4k}{(2\pi)^4} \nonumber \\
&  {\rm Tr} 
\Bigl[
( I^{\alpha}_{\ell_1 m_1'; \ell'm'; \ell_2m_1'-m' }P_L + I^{\beta}_{\ell_1m_1'; \ell'm'; \ell_2m_1'-m'}P_R )  \frac{i}{\gamma^\mu k_\mu +i\gamma^5 M_{\ell_1}}  \times \nonumber \\
&   (\delta_{m_1,m_1'} \mp (-1)^{\ell_1+m_1} \delta_{-m_1,m_1'} \gamma^5)  ( I^{\alpha}_{ \ell_2 m_1-m; \underline{ \ell m}; \ell_1m_1}P_L + I^{\beta}_{\ell_2 m_1-m; \underline{ \ell m }; \ell_1m_1} P_R ) \nonumber \\
&  \frac{i}{\gamma^\mu (k_\mu-p_\mu) +i\gamma^5 M_{\ell_2}} 
(\delta_{m_1-m,m_1'-m'} \mp (-1)^{\ell_2+m_1+m} \delta_{-(m_1-m),m_1'-m'} \gamma^5)
\Bigr] 
\end{align}
which is expressed as in the previous case and the result is summarized in Table~\ref{OneLoopS-F}.

%



\end{document}